\documentclass[12pt,a4paper]{article}
\usepackage[utf8]{inputenc}
\usepackage{authblk}
\usepackage{graphicx}
\usepackage{float}
\usepackage{color}
\usepackage{xcolor}
\usepackage{caption}
\usepackage{subcaption}
\usepackage{makeidx}
\usepackage{array}
\usepackage[T1]{fontenc}
\usepackage[utf8]{inputenc}
\usepackage{cite}
\usepackage[font=small,labelfont=bf]{caption}
\usepackage{graphicx}
\newcolumntype{P}[1]{>{\centering\arraybackslash}p{#1}}
\newcolumntype{M}[1]{>{\centering\arraybackslash}m{#1}}
\author[]{A. Arul Anne Elden}
\author[]{M. Ponmurugan {\footnote{email correspondence: ponphy@cutn.ac.in}}}
\affil[]{School of Basic and Applied Sciences, Central University of Tamil Nadu, \\Thiruvarur - 610 005, Tamil Nadu, India.}
\date{}
\usepackage[left=2cm,right=2cm,top=2cm,bottom=2cm]{geometry}

\begin{document}
\title{Monte Carlo investigation of phase changes and the order of transition of Ising modeled single-walled Nanotube}

\maketitle
\section*{Abstract} 
The Monte Carlo analysis for the magnetic response of a single-walled nanotube using the Metropolis and Wang Landau algorithms is reported in the present paper. The  nanotube architecture used in the present study utilizes the spin half Ising model with nearest neighbors interaction and obtained various magnetic orderings namely, ferromagnetic, G-type anti-ferromagnetic, A-type anti-ferromagnetic, and C-type anti-ferromagnetic.  It is also found that the phase changes from ferromagnetic/anti-ferromagnetic to paramagnetic with the modificiation of system's control parameters. The transition temperatures is determined for various interaction strength in the absence of magnetic field and for fixed interaction strength with the inclusion of external magnetic field.  The present study confirms the transition from ferromagnetic /anti-ferromagnetic to paramagnetic is a second order transition.\\
\\
Key words: Nanotube, Ising model, Monte-Carlo techniques, Phase changes, transition temperature.

\section{{Introduction}}
The study of magnetic properties in low-dimensional systems has attracted a lot of theoretical and experimental interest recent years \cite{Parente2008Spin, Hachem2021Phase}. The low dimensional systems such as the magnetic nano-materials have wider range of technical applications, viz., sensors \cite{Kurlyand2003Gaint}, drug delivery \cite{salem2003Multi}, bio-medical applications \cite{Pank2003Application}, magnetic resonance imaging \cite{Lopez2014Applica}, magnetic recording media \cite{Ray2010High}, memory devices \cite{Fuhrer2002High} and  permanent magnets \cite{Lopez2014Applica}. The nano-structured materials can be produced in a variety of shapes, such as nano films \cite {Sultan2009Magnetiz}, nano rings \cite {Kho2009Single}, nanotubes \cite{Xu2008Synthesis}, nano wires\cite{Yang2019Single} and nano particles \cite {Tadic20151Manetic}. Among other shapes, the magnetic nanotube  has interesting uses in  high performance battery \cite{Xie2020Hierarchical}, humidity sensors \cite{Zhu2020Cellulose}, tactile sensors \cite{Chen2020Self} and energy storage applications \cite{Xie2019In}. Thus, analyzing the mechanical and thermal properties of a  nanotube is important in enhancing its applications in various fileds. The simple interaction model known as the Ising model \cite{Taroni2015Years, Wolf2000The, Strecka2015Brief} is well-suited to study the magnetic behavior of nanotubes using various theoretical techniques.  

The phase transition of a hexagonal cylindrical magnetic super-lattice nanotube is investigated using the molecular field theory approximation \cite{Tanriver2021The}.  The effective field theory with correlations is used to analyze the phase diagram and magnetization of a cylindrical nanotube as a transverse Ising model \cite{kaneoshi2012Some}. Using the differential operator technique and effective field theory, the magnetic properties of a six-legged spin half and spin one nanotubes are studied in the presence of an applied magnetic field \cite{Maddahi2017Magnetic, ElMaddahi2019Magnetic}. The phase transition and magnetization of Ising nanotubes are analyzed by cellular automata approach with ferromagnetic and anti-ferromagnetic interactions \cite{Astaraki2018Investig}. Magnon thermodynamic properties of nanotubes are investigated using the many-body green function approach and studied the effect of the spin correlation  on the magnetic properties of the single walled nanotubes \cite{Bin2016Magnon}. The thermal and magnetic properties of low dimensional spin related systems are also reported in detail in Refs. \cite{Avetisyan2016Magneti, Castano2018A_compara, Boyacioglu2012Dia, Castano2019Impact, Atoyan2006Interband, Castano2020Super, Castano2021Super, Ourabah2017Quantum}. The Monte-Carlo algorithm is used to study the dielectric properties along with the polarization and hysteresis behavior of mixed spin in a nanotube system \cite{Sahdane2020Dielectric}. Several studies also discussed the magnetic responses of Ising modeled nanotubes \cite{Salazar2012Influence,Salazar2012,Konsta2008Theoretical,Zheng2018Zheng} using  Metropolis Monte Carlo algorithm. In the present paper, both the Boltzmann (Metropolis) and non-Boltzmann (Wang Landau) Monte-Carlo algorithms are used to  analyze the phase changes and the order of the transition of the single-walled nanotube which is modeled as spin half Ising system.

The Metropolis algorithm \cite{Metro1953Equations} is particularly useful to compute the mechanical properties and the hysteresis behavior by generating the micro-states that correspond to the Boltzmann distribution \cite{Masrour2016Critical}. However, the Metropolis algorithm is not suitable for obtaining the  thermal properties of nanotube by estimating its density of states.  This is due to the fact that the  individual micro-states cannot be assigned an entropy value since entropy is a collective property  \cite{Suman2019Non-}. Hence,  in the present study,  the Wang Landau (WL) algorithm is used to calculate the  density of states and thereby the temperature-dependent observables can be determined with un-weighting and re-weighting techniques \cite{Murthy2000Monte, Jayasri2005Wang}. 

In this paper,  a systematic analysis of the magnetic properties of  a single walled Ising nanotube is explored in detail.  The influence of the control parameters (the interaction strength and the external magnetic field) on the properties of the Ising nanotube are also investigated. The manuscript is organised as follows: In section. 2, the detailed description of the model and the simulation techniques are explained.  The results and its discussions are eloborated in section.3. The observations and its conclusions are given in the conclusion section of the manuscript. 
 
\section{Model description and the Monte Carlo simulation techniques}

\subsection{Model}  
\begin{figure}[H]
	\centering
	\includegraphics[scale=0.5]{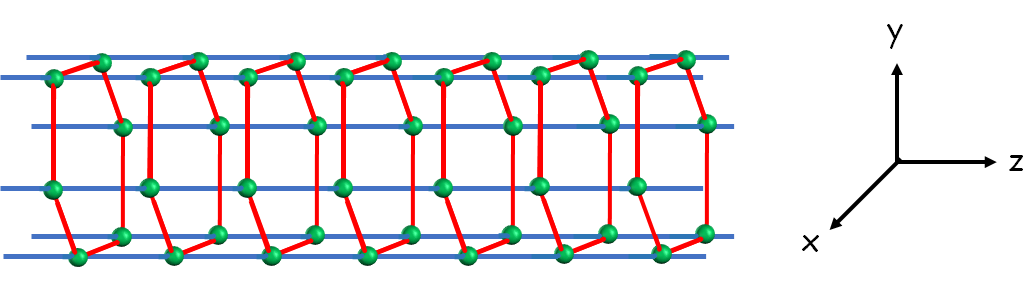}
	\caption{Illustration of Ising nanotube. Green balls represents lattice points, red lines represents $J_1$ interactions and blue line represents $J_2$ interactions.}
	\label{f1}
\end{figure}

In the present work, the spin half Ising model is examined and decorated as a single wall nanotube which consists of a layered graphite-like structure with magnetic moments ${S_i}$, occupied in its respective lattice sites. Each layer would be composed of $6$ lattice sites ${S_i}$. Every lattice site can take a spin value of ($+1$)/($-1$) for spin up/spin down, respectively. The Hamiltonian of the configuration can be described as,
\begin{equation}
\label{eq:1}
H ~=~-J_1\sum_{<i,j>} S_i S_j ~- J_2\sum_{<i,k>} S_i S_k ~- B\sum_{i} S_i.
\end{equation}
Where $<i, j>$, $<i, k>$ represents the nearest neighbour sites, $J_1$ stands for the exchange interaction strength between the first neighbouring sites within every layer, $J_2$ stands for the exchange interaction strength between the layers and $B$ is the external magnetic field. These interactions ($J_1$ \&  $J_2$ ) will favour the spin to align either parallel (Ferro Magnetic: FM) or anti-parallel (Anti Ferro Magnetic: AFM). If $J_1$ and $J_2$ are greater than zero, then the system follows ferromagnetic order, whereas the system follows anti-ferromagnetic order if either of the interaction ($J_1$ or $J_2$) is less than zero. The detailed picture of the Ising nanotube is depicted in Fig. \ref{f1}. In the figure, each site (green ball) is connected with $4$ neighbouring sites. The $J_1$, i.e., the intralayer interaction where each site is connected with two neighbouring sites are shown in red lines and the blue line that connects the two neighbouring sites of different layers represents the inter-layer interaction ($J_2$).  The periodic boundary condition is applied in the axis of the system's length $(z)$ whereas the closed boundary conditions are given in the $x$ and $y$ axes. 
The following sub section reviews  the Metropolis algorithm for canonical ensemble by considering the system in equilibrium with the reservoir  and further considers WL algorithm with un-weighting and re-weighting to calculate the thermal properties.

\subsection{Simulation technique 1: Boltzmann technique}

Ising nanotube can be analyzed in simulation by using the temperature dependent Metropolis algorithm \cite{Metro1953Equations, Landau2000A_guide}, called the Boltzmann technique. The spin configurations (Markov chain) are generated according to the Boltzmann distribution ($C_0\rightarrow C_1\rightarrow C_2\rightarrow ... C_i\rightarrow C_{i+1}\rightarrow ... C_N$). The simulation begins with the initial random configuration $(C_0)$ followed by the trial configurations $(C_t)$ which are generated by flipping a randomly selected spin. This trial configuration is accepted with transition probability $p=\mbox{min}\{1,\exp(-\beta\Delta E)\}$, where $\Delta E=E_t-E_i$ is the energy difference between the trial and the initial spin configurations, $\beta=\frac{1}{k_{B}T}$ ($k_B$ is the Boltzmann constant and it is set to unity for the whole simulation) is the inverse temperature and $T$ is the temperature of the reservoir.  About $10^6$ Monte Carlo Sweeps (MCS) is performed for each temperature in which the first $10^5$ MCS are discarded for the system equilibration. The remaining MCS part is taken into account for calculating the average values. 
The canonical ensemble average of a desired physical quantity $O$ is given by,
\begin{eqnarray}
\label{eq:2}
\langle O \rangle_T = \frac{1}{M_s} \sum\limits_{C} O(C),
\end{eqnarray}
where $M_s$ is the total number of MCS after equilibration. The average energy and average magnetization can be obtained from the equation \ref{eq:2}. The specific heat capacity $(C_V)$ and magnetic susceptibility $(\chi)$ were calculated using the following expressions,
\begin{eqnarray}
\label{eq:3} 
C_V(T)&=&\frac{1}{k_{B}T^2} ~(\langle E^2 \rangle_T - \langle E \rangle^2_T),
\end{eqnarray}

\begin{eqnarray}
\label{eq:4} 
\chi(T)&=&\frac{1}{k_{B}T} ~(\langle M^2 \rangle_T - \langle M \rangle^2_T),
\end{eqnarray}
where $\langle E \rangle$ is the average energy and $\langle M \rangle$ is the average magnetization of the system.  The Magnetization per spin  ($M$) is calculated by summing all magnetic moments $S_i$ and is given by $ M  = \frac{1}{N}\sum_{i} S_i$, where, $N = 6L$ is the total number of spins and  $L$ represents the total number of layers (each layer consist of six spins). In general (otherwise specified),  $N = 180$ spins with $L = 30$ layers are used in the entire simulation.  As mentioned earlier, the Metropolis algorithm is not suitable for analyzing the thermal properties in terms of the system density of states, hence the WL algorithm is used to simulate the Ising nanotube for obtaining the other thermal properties in addition to the above properties (calculated from the estimated density of states). 

\subsection{Simulation technique 2: Non-Boltzmann technique} 

The conventional density of states $g(E)$ is calculated using the temperature-independent WL algorithm, called non-Boltzmann Monte-Carlo technique \cite{Wang2001Efficient, Wang2001Determining}. This approach aims to perform a random walk on the energy space by randomly choosing a lattice site and changing its magnetic moment with an appropriate probability, which is proportional to the inverse of the density of states $g(E)$, i.e. $P(E)\propto1/g(E)$. The density of states (DOS) is not known as \textit{a priori}. Thus, the algorithm is initiated by assuming $g(E)=1$ and the energy histogram, $H(E)=0$ for all energy levels. The simulation is started with a random configuration of energy $E_{i}$. The trial configuration is made by flipping a randomly chosen spin and its energy is given by $E_{t}$. The trial configuration is accepted with the transition probability,
\begin{eqnarray}
	\label{eq:5}
	P(E_i\rightarrow E_t)&=& \mbox{min}\Bigg\{1,\frac{g(E_i)}{g(E_t)}\Bigg\}.
\end{eqnarray}
Then, the corresponding $g(E)$ is updated by multiplying with a modification factor $f$ $(f>1)$ and the histogram is updated with unity as,
\begin{equation}
	\label{eq:6}
	g(E)= g(E) \times f,
\end{equation}
\begin{equation}
	\label{eq:7}
	H(E)= H(E)+1.
\end{equation}
The logarithm of DOS is updated as,
\begin{equation}
	\label{eq:8}
	\ln~[g(E)] = \ln~[g(E)] + \ln f.
\end{equation}
If the configuration gets rejected, the initial configuration is taken as a trial configuration and update it accordingly. Once the histogram is flat, the modification factor is reduced monotonically by multiplying with $0.5$ $(\ln f\Leftarrow \ln f\times 0.5)$ and the histogram resets to $H(E)=0$ for all energies. Typically, the flatness $(80\%)$ condition for the energy histogram is checked for every $10,000$ MCS. The algorithm repeats the iteration with a new $\ln f$ until it reaches a small enough value (e.g., $\ln f<10^{-8}$). The entire process results in the converged DOS of the system. The thermo-dynamical quantities can be calculated by finding the partition function of the system from the estimated $g(E)$. The canonical partition function $Z(\beta)$ at any finite temperature can be obtained by, $Z(\beta)=\sum_{E}g(E)~\exp[-\beta E]$. The production run is performed with the WL acceptance rule as in the equation (\ref{eq:5}) with converged DOS. It begins with an arbitrary configuration $C_0$ and generates the micro-state sequence that constitutes the Markov chain,

\begin{eqnarray*}
	C_0\rightarrow C_1\rightarrow C_2\rightarrow ...~ C_i\rightarrow C_{i+1}\rightarrow ... ~C_M.
\end{eqnarray*}   
The average of any macroscopic observable $O$ can be computed for the canonical distribution using the estimated DOS by un-weighting and re-weighting with the corresponding weight factor.  Un-weighting is the process of dividing the value $O(C)$ by the probability weight factor $[g(E(C))]^{-1}$ for each configuration while, re-weighting is the process of multiplying  the value $O(C)$ with the Boltzmann weight factor $\exp[-\beta E]$ for each configuration. Thus, the weight factor \cite{Murthy2000Monte, Jayasri2005Wang},

\begin{eqnarray}
	\label{eq:9}
	W(C,\beta)&=&g(E(C))~\exp[-\beta E(C)],
\end{eqnarray}
such that,  the canonical ensemble average of $O$ can be calculated as,
\begin{eqnarray} 
	\label{eq:10}
	\langle O \rangle_\beta &=& \frac{\sum\limits_{C} O(C)~ W(C,\beta)}{\sum\limits_{C}W(C,\beta)},
\end{eqnarray}

\begin{eqnarray}
	\label{eq:11}
	\langle O \rangle_\beta &=& \frac{\sum\limits_{C} O(C)~ g(E(C))~\exp[-\beta E(C)]}{\sum\limits_{C}g(E(C))~\exp[-\beta E(C)]},
\end{eqnarray}
where $C$ represents the configuration generated by the Monte-Carlo production run. The average energy and average magnetization are calculated using the above equation (\ref{eq:11}). Further, the equations (\ref{eq:3}) and (\ref{eq:4}) are used to calculate the specific heat capacity $C_V(T)$ and susceptibility $\chi(T)$ of the system. In addition to that, the Gibbs free energy  $(F)$ and Canonical entropy $(S)$  are calculated as,

\begin{eqnarray}
	\label{eq:14} 
	F(T)=-k_{B}T\ln(Z)=-k_{B}T\ln\Bigg(\sum\limits_{E}g(E)~\exp[-\beta E]\Bigg),
\end{eqnarray}
\begin{eqnarray}
	\label{eq:15} 
	S(T)&=&\frac{U(T)-F(T)}{T}
\end{eqnarray}
where $ U(T)=\langle E \rangle_T$ is the internal energy. The Monte Carlo simulation of Ising nanotube using the above explained techniques is carried out and the results are discussed in the following section.

\section{Simulation results and its discussions}

\begin{figure}[H]			
	\centering
	\begin{subfigure}[b]{0.22\textwidth}
		\includegraphics[width=\textwidth]{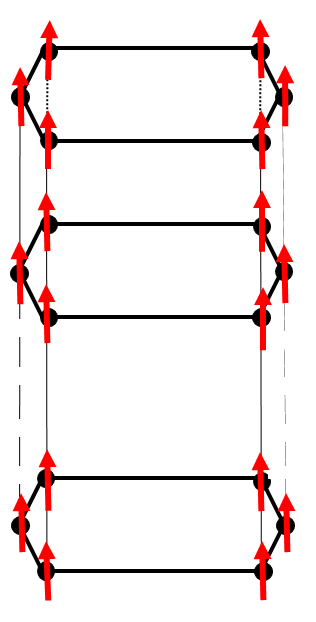}
		\caption{$ $}
		\label{f2a}
	\end{subfigure}
~
	\begin{subfigure}[b]{0.22\textwidth}
		\includegraphics[width=\textwidth]{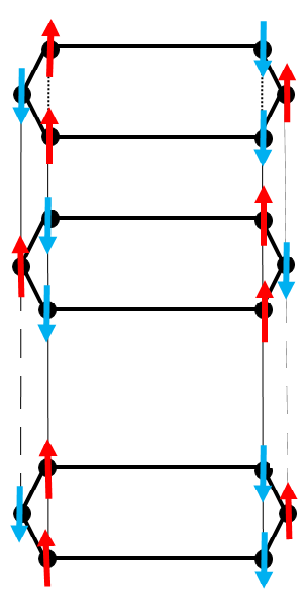}
		\caption{$ $}
		\label{f2b}
	\end{subfigure}
~
	\begin{subfigure}[b]{0.22\textwidth}
		\includegraphics[width=\textwidth]{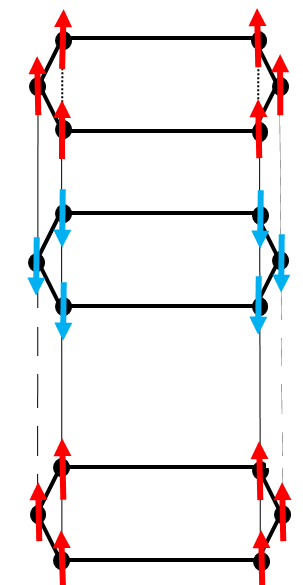}
		\caption{$ $}
		\label{f2c}
	\end{subfigure}
~
	\begin{subfigure}[b]{0.22\textwidth}
		\includegraphics[width=\textwidth]{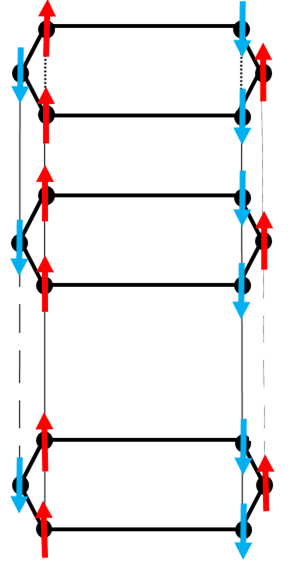}
		\caption{$ $}
		\label{f2d}
	\end{subfigure}
~
	\caption{Graphical representation of ground state ferromagnetic (FM) and anti-ferromagnetic (G-AFM, A-AFM and C-AFM) spin order for various interaction strength
(a) FM: $J_1=1; ~J_2=1$, (b) G-AFM: $J_1=-1; ~J_2=-1$, (c) A-AFM: $J_1=+1; ~J_2=-1$ and (d) C-AFM:  $J_1=-1; ~J_2=+1$  in the absence of external magnetic field $B=0$.}
\label{f2}
\end{figure}

The simulations are carried out for the Ising nanotube with different control parameters over a finite range of temperatures. The Fig. \ref{f2} shows the ground state spin orientation for various magnetic orderings namely, FM (Fig. \ref{f2a}), G-AFM (Fig. \ref{f2b}), A-AFM (Fig. \ref{f2c}) and C-AFM (Fig. \ref{f2d}) ordering \cite{Wqllan1955Neutron, Dagotto2003Nanoscale}.  In FM ordering, all the magnetic spins are aligned parallel to each other by the ferromagnetic interaction with the interaction strength  $J_1=+1$ and $J_2=+1$.  Whereas  in G-type AFM,  all the spins are aligned anti-parallel among the nearest neighbouring sites with the interaction strength $J_1=-1$ and $J_2=-1$. While the other two, in which  A-type AFM magnetic ordering is characterized with parallel alignment due to FM interaction and the spins between the layers are anti-parallel due to AFM interaction which is viceversa for C-type. The interaction strengths are $(J_1, J_2)=(+1,-1)$ and $(J_1, J_2)=(-1,+1)$, respectively, for A-type and C-type ordereing.

The present simulation is performed by varying either of the control parmeters $J_1$ and B thereby keeping the other to be constant. While the control parameter $J_2$ is fixed for every set of simulations.  The result analysis of the above mentioned framework is discussed in the subsequent sections.

\subsection{Metropolis results}
\subsubsection*{In the absence of magnetic field}

\begin{figure}[H]
	\centering
	\begin{subfigure}[b]{0.45\textwidth}
		\includegraphics[width=\textwidth]{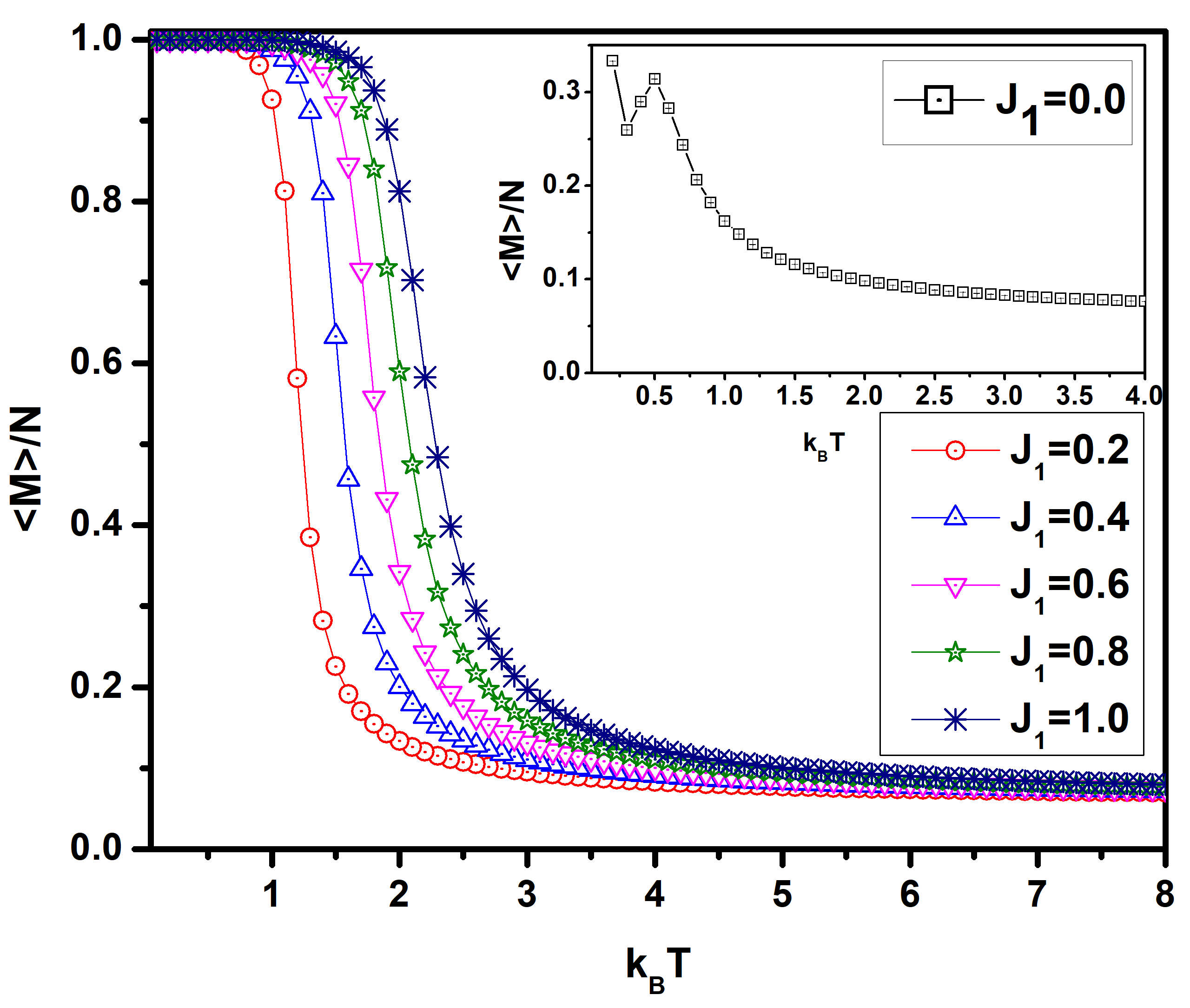}
		\caption{$ $}
		\label{f3a}
	\end{subfigure}
	~
	\begin{subfigure}[b]{0.45\textwidth}
		\includegraphics[width=\textwidth]{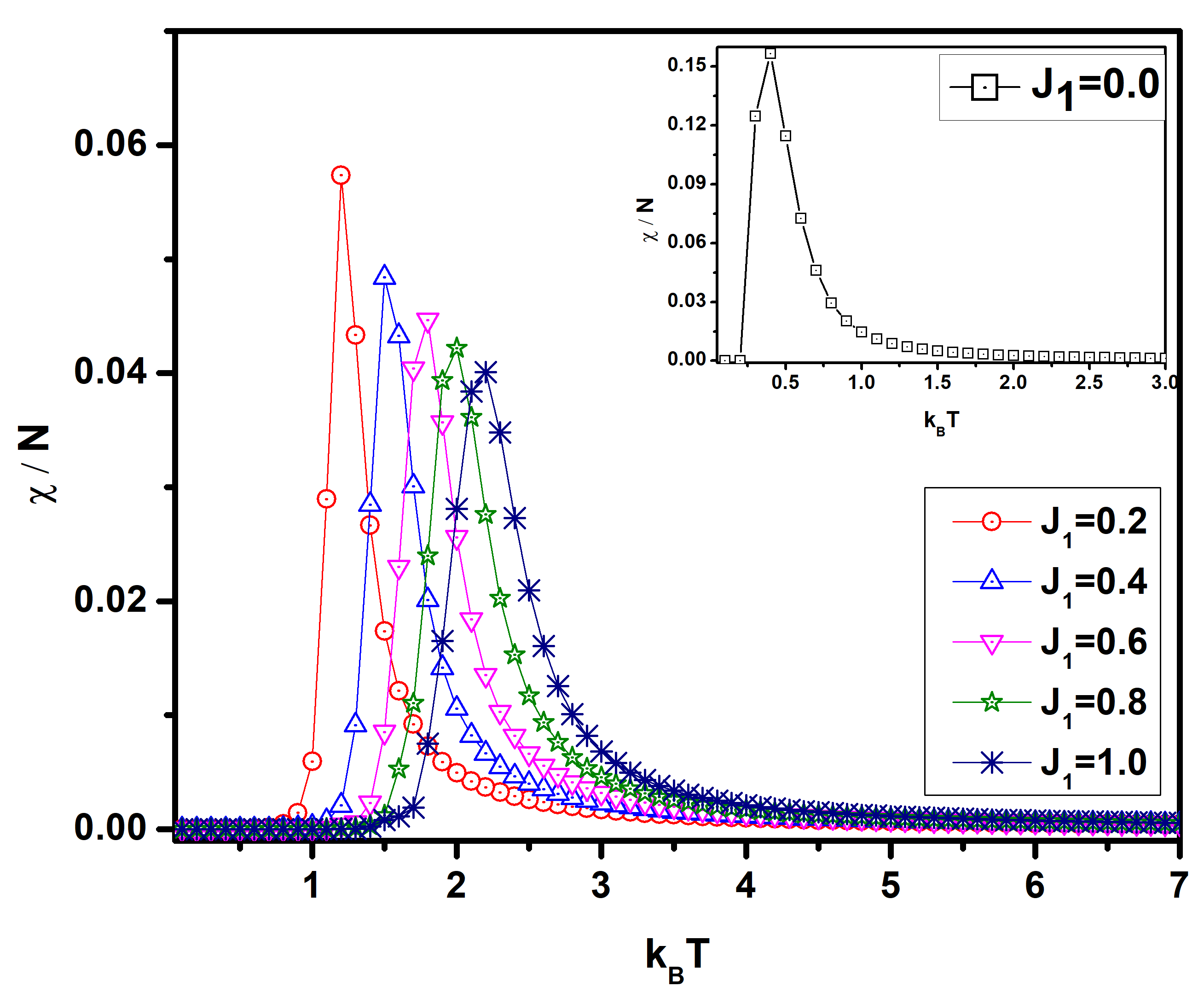}
		\caption{$ $}
		\label{f3b}
	\end{subfigure}
	~
	\caption{Temperature dependence of observables for ferromagnetic interactions in the absence of magnetic field $B=0$, (a) average magnetization and (b) susceptibility. The interactions are $0.0 \leq J_1 \leq 1.0$ and $J_2=1.0$. The inset plot shows the observables for $J_1=0.0$.}
	\label{f3}
\end{figure}

The system evolves under the Metropolis algorithm without  the external magnetic field $(B=0)$ in this part of the simulation. Initially, the interaction strength between the adjacent layers, $J_2$, is set to $1.0$ and  $J_1$, the interaction strength within the layers is varied  from $0.0$ to $1.0$ to simulate the ferromagnetic system. The temperature-dependent magnetization and magnetic susceptibility graphs are plotted in Fig. \ref{f3}. Fig. \ref{f3a} shows that the system exhibits spontaneous magnetization at lower temperatures. The saturation of magnetization occurs due to the parallel ordering of the spins. The spontaneous magnetization breaks at a certain temperature and turns into random ordering. Thus, the net (absolute) magnetization curve smoothly settles down to the minimum magnetization at higher temperatures. When $J_1=0$,  the interaction of spins within the layer is zero and hence only the interactions of adjacent layer spins will contribute to the FM ordering. So the total magnetization is reduced (inset Fig. \ref{f3a}). The Monte Carlo errors for all simulation data are calculated and it is  found to be smaller than the size of the data points in the entire analysis. The transition temperature $(T_{C})$ is positioned by identifying the peak value of the susceptibility curve (Fig. \ref{f3b}). The susceptibility curve for different values of interaction strength are plotted and the peak value of the susceptibility is found to be decreasing (as a result of the decrease in the fluctuations) and shifting towards the higher temperatures with increasing interaction strength $J_1$. The maximum value of susceptibility is found to be higher for $J_1=0$ (inset Fig. \ref{f3b}).

\begin{figure}[H]
	\centering
	\begin{subfigure}[b]{0.45\textwidth}
		\includegraphics[width=\textwidth]{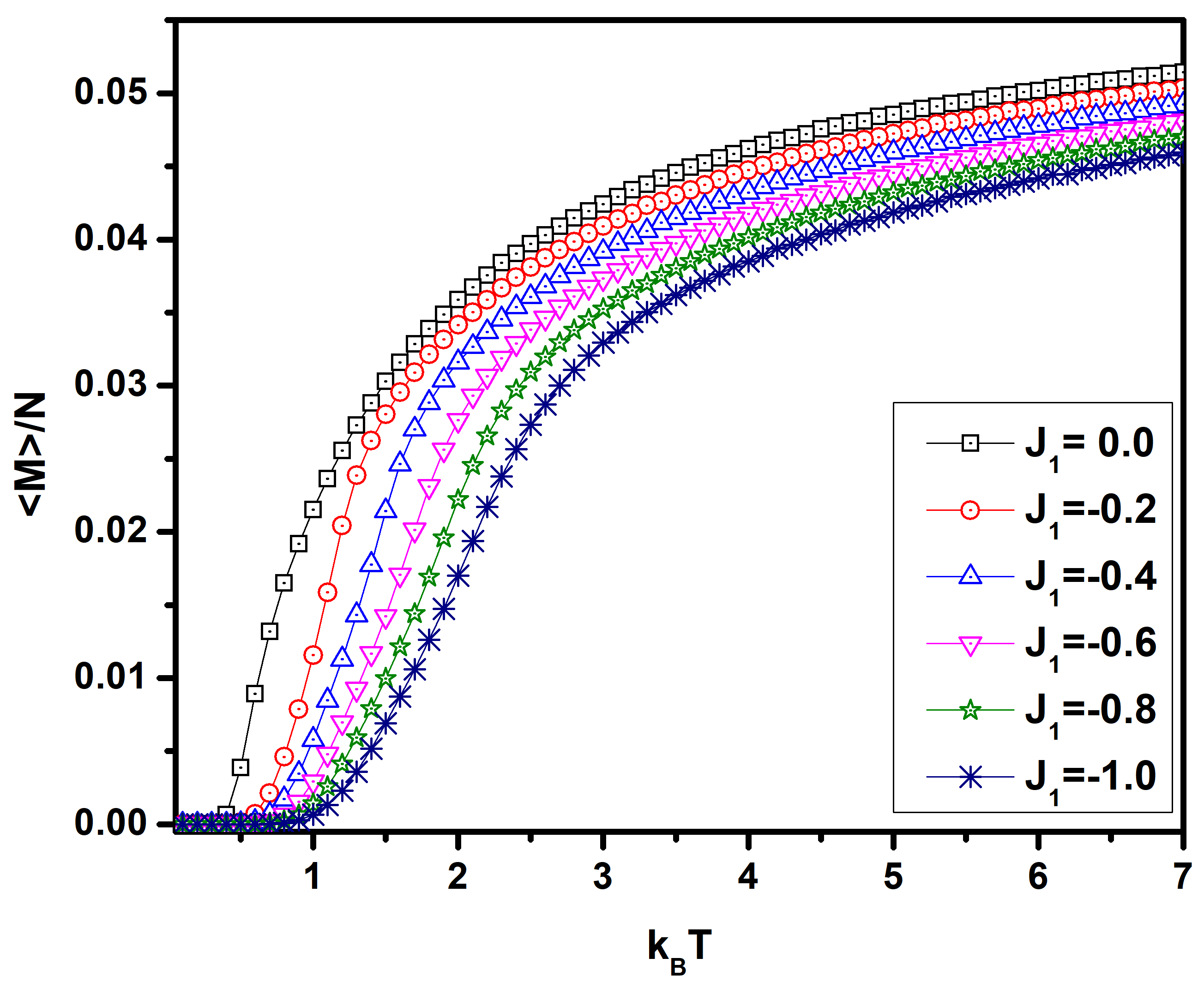}
		\caption{$ $}
		\label{f4a}
	\end{subfigure}
	~
	\begin{subfigure}[b]{0.45\textwidth}
		\includegraphics[width=\textwidth]{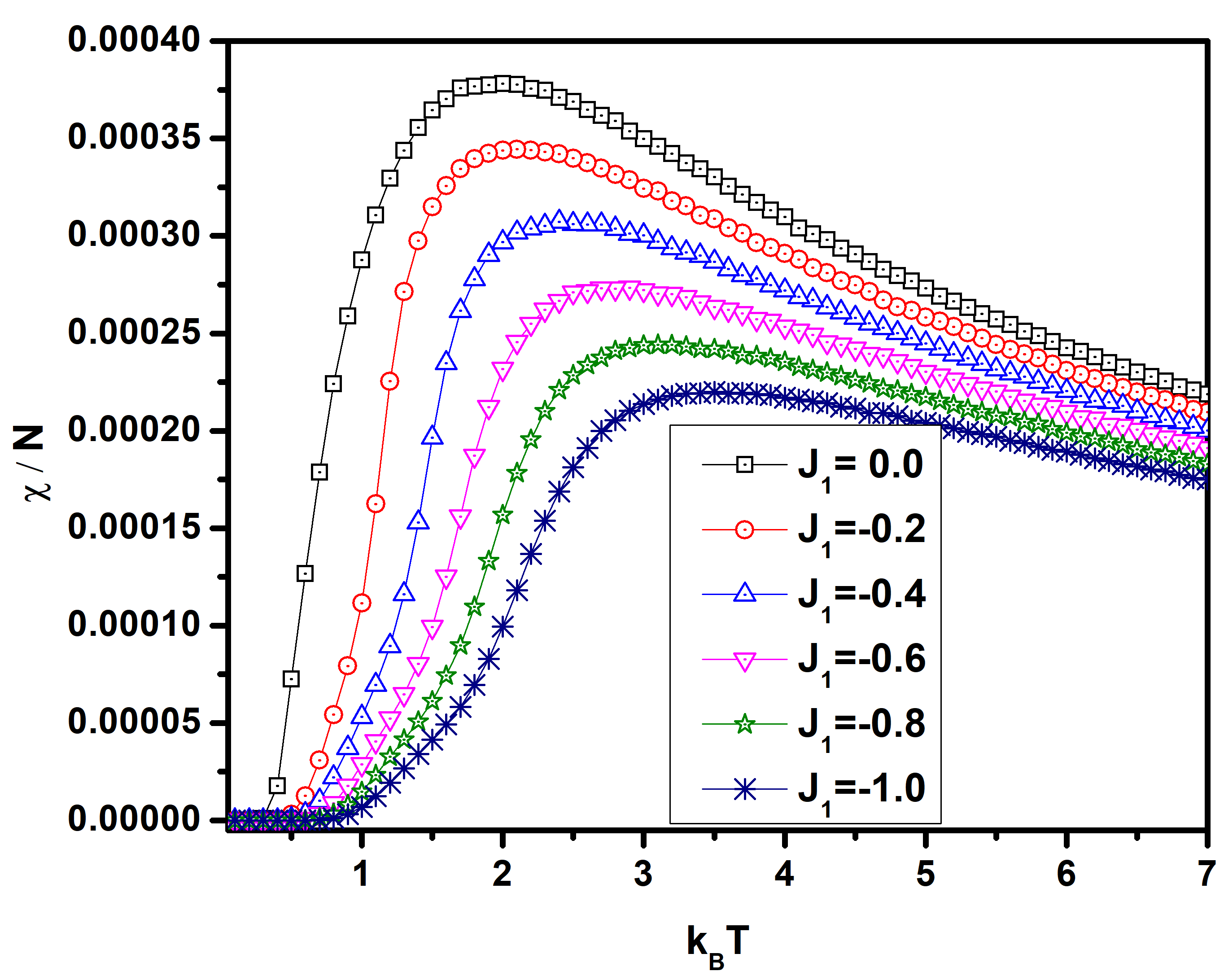}
		\caption{$ $}
		\label{f4b}
	\end{subfigure}
	~
	\caption{Temperature dependence of observables for G-type anti-ferromagnetic interactions excluding the magnetic field, i.e.,  $B=0$, (a) average magnetization and (b) susceptibility. The interactions are $-1.0 \leq J_1 \leq 0.0$ and $J_2=-1.0$.}
	\label{f4}
\end{figure}

In a similar way, the interaction strength between the layers, $J_2$, is fixed as $-1.0$ and the interaction strength within the layer, $J_1$, is varied from $-1.0$ to $0.0$, which brings the system into G-type AFM ordering. The corresponding average magnetization and magnetic susceptibility plots are shown in Fig. \ref{f4}. The temperature-dependent magnetization for G-AFM in Fig. \ref{f4a} is zero at the lower temperature region, as all the spins are aligned in the anti-parallel pattern. The average magnetization curve increases steadily and reaches its saturation point, evidencing the transition to a random order state. The transition point is determined from the maximum value of susceptibility (Fig. \ref{f4b}). The transition temperature is shifted to the  lower values as the interaction strength, $J_1$, increases. The maximum value of susceptibility also increases with increase in $J_1$. Though the interaction strength, $J_1=0$, the system follows the G-type AFM order.

\begin{figure}[H]
	\centering
	\begin{subfigure}[b]{0.45\textwidth}
		\includegraphics[width=\textwidth]{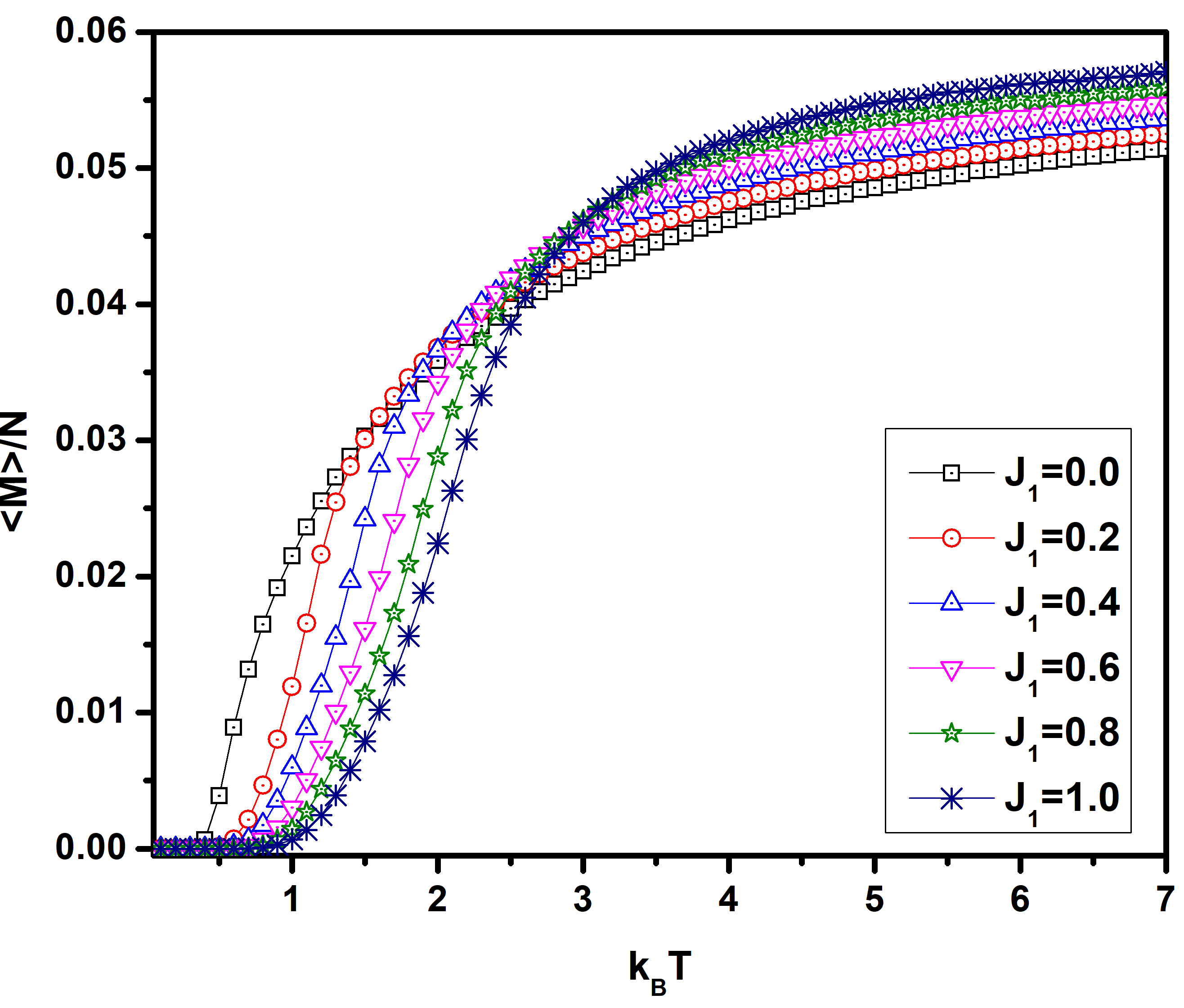}
		\caption{$ $}
		\label{f5a}
	\end{subfigure}
	~
	\begin{subfigure}[b]{0.45\textwidth}
	\includegraphics[width=\textwidth]{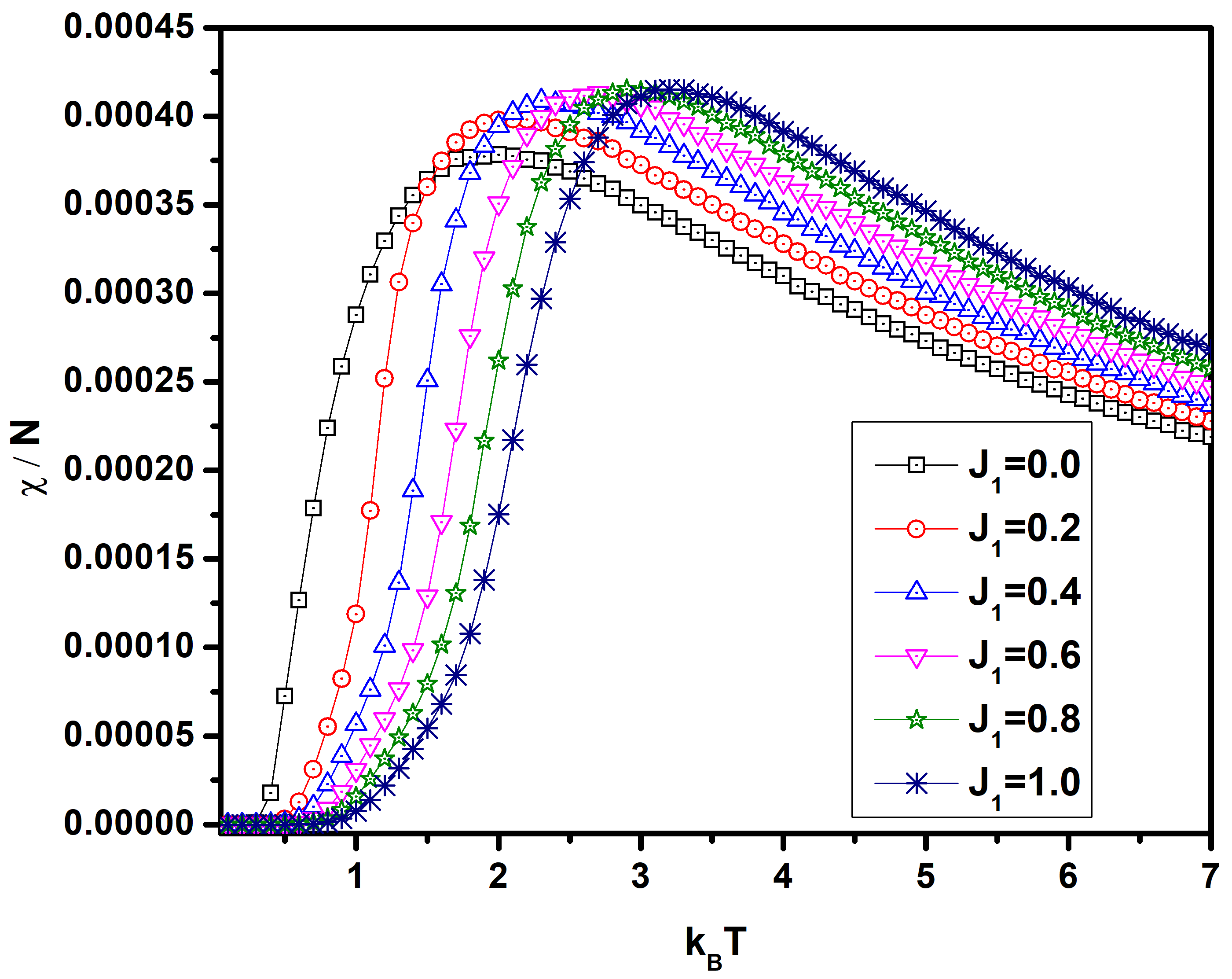}
		\caption{$ $}
		\label{f5b}
	\end{subfigure}
	~
	\caption{Temperature dependence of observables in the absence of magnetic field $B=0$, (a) average magnetization and (b) susceptibility for the interactions $0.0 \leq J_1 \leq 1.0$ and $J_2=-1.0$.} 
	\label{f5}
\end{figure}

\begin{figure}[H]
	\centering
	\begin{subfigure}[b]{0.45\textwidth}
		\includegraphics[width=\textwidth]{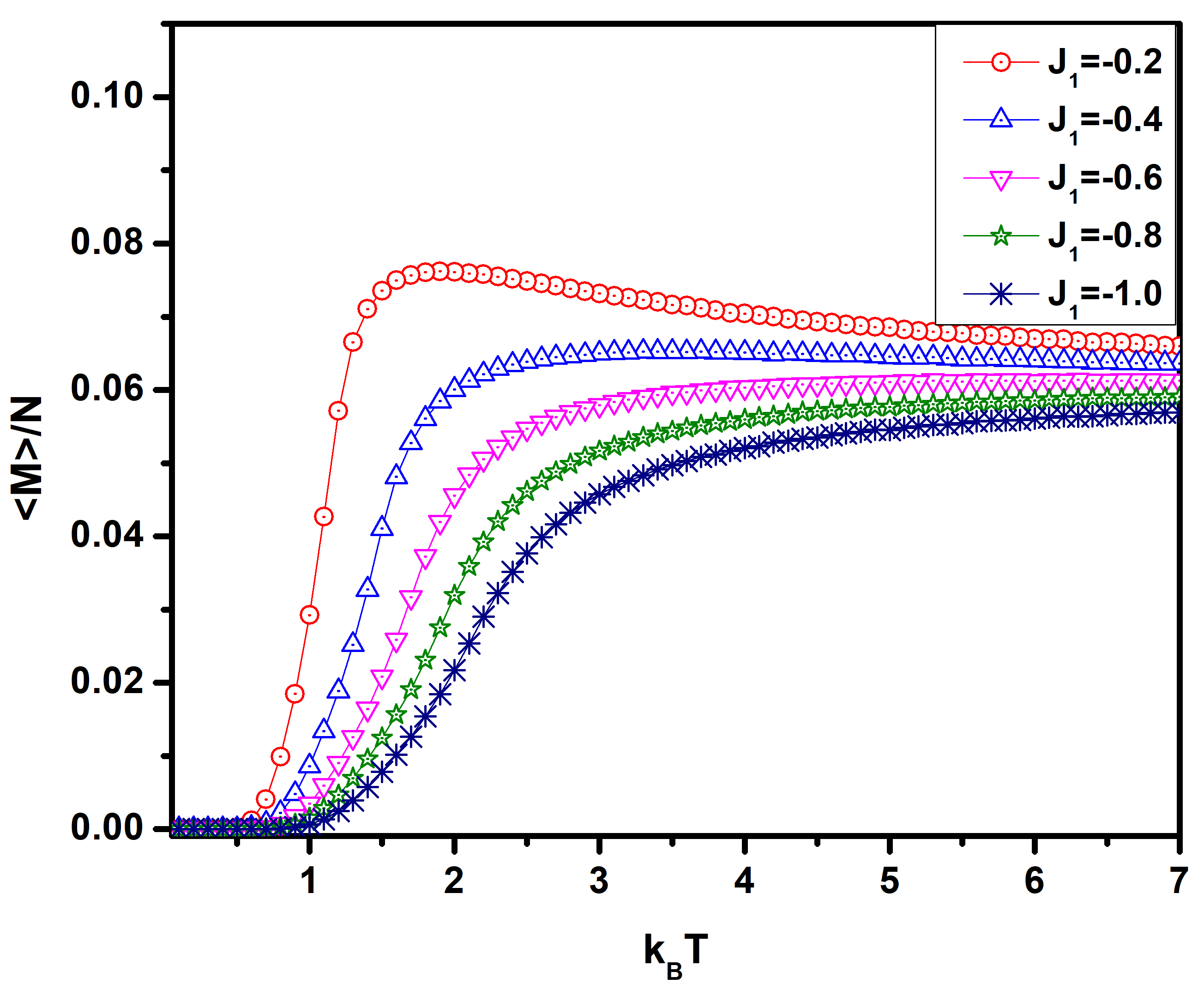}
		\caption{$ $}
		\label{f6a}
	\end{subfigure}
	~
	\begin{subfigure}[b]{0.45\textwidth}
		\includegraphics[width=\textwidth]{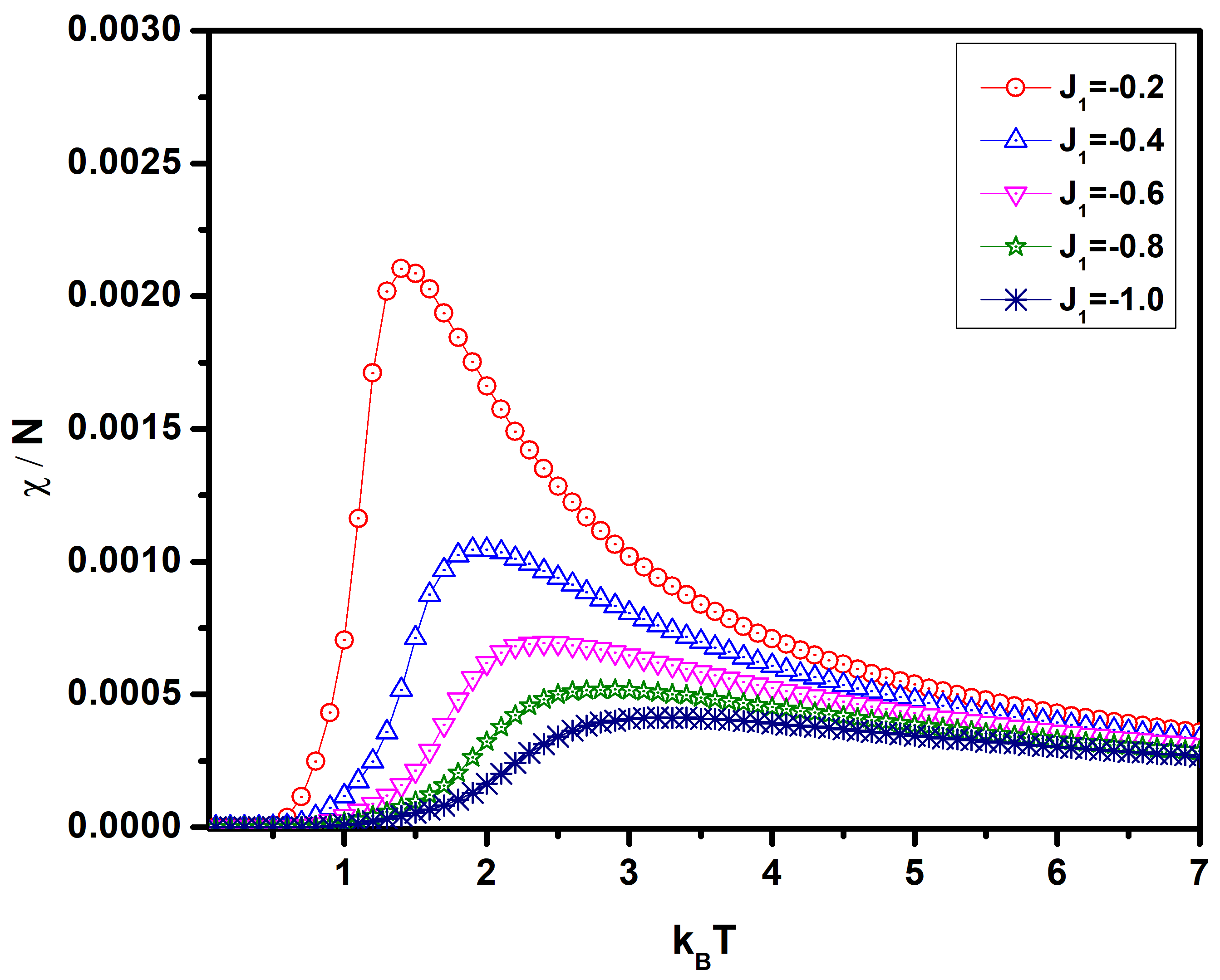}
		\caption{$ $}
		\label{f6b}
	\end{subfigure}
	~
	\caption{Temperature dependence of observables in the absence of magnetic field $B=0$, (a) average magnetization and (b) susceptibility for the interactions $-1.0 \leq J_1 \leq -0.2$ and $J_2=+1.0$.}
	\label{f6}
\end{figure}

The system evolving with opposite interaction (A type: $J_1=+1$, $J_2=-1$) and (C type: $J_1=-1$, $J_2=+1$) are respectively shown in Figs. \ref{f5} and \ref{f6}. The average magnetic observables for the A-type interaction namely, the average magnetization and susceptibility are shown in Figs. \ref{f5a} and \ref{f5b}. The magnetization curve (Fig. \ref{f5a}) clearly shows that the system exhibits AFM dominance despite the presence of FM interaction within the layers. The susceptibility plot (Fig. \ref{f5b}) shows a transition from the anti-ferromagnetic to the paramagnetic phase though the FM interaction is present. The transition temperature rises as $J_1$ interaction increases.

Despite having a high FM interaction of $J_2=+1$, the magnetization plot in Fig. \ref{f6a} shows that the system follows C-AFM ordering.  While for the interaction $J_1=0.0$, $J_2=+1$, 
system follows FM ordering which is already shown earlier in the inset plot of Fig. \ref{f3a}.  The susceptibility curves are shown  in Fig. \ref{f6b}. As the interaction $J_1$ is reduced towards $-1.0$, the value of $T_C$ changes towards higher temperature. The susceptibility peak is higher for $J_1=0.0$, $J_2=1.0$ and it is already highlighted in the inset plot of Fig. \ref{f3b}.

\begin{figure}[H]
	\centering
	\begin{subfigure}[b]{0.45\textwidth}
		\includegraphics[width=\textwidth]{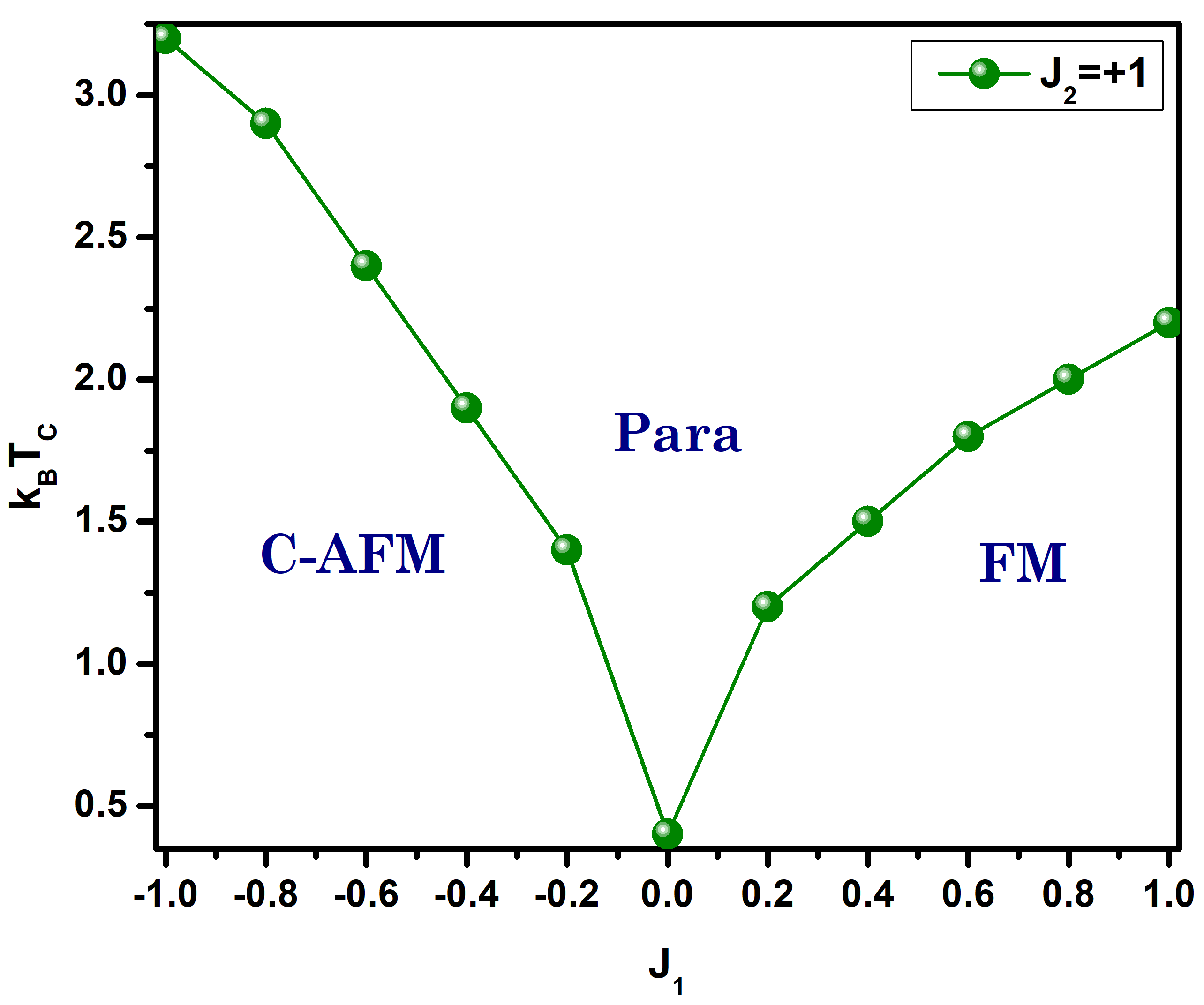}
		\caption{$ $}
		\label{f7a}
	\end{subfigure}
	~
	\begin{subfigure}[b]{0.45\textwidth}
		\includegraphics[width=\textwidth]{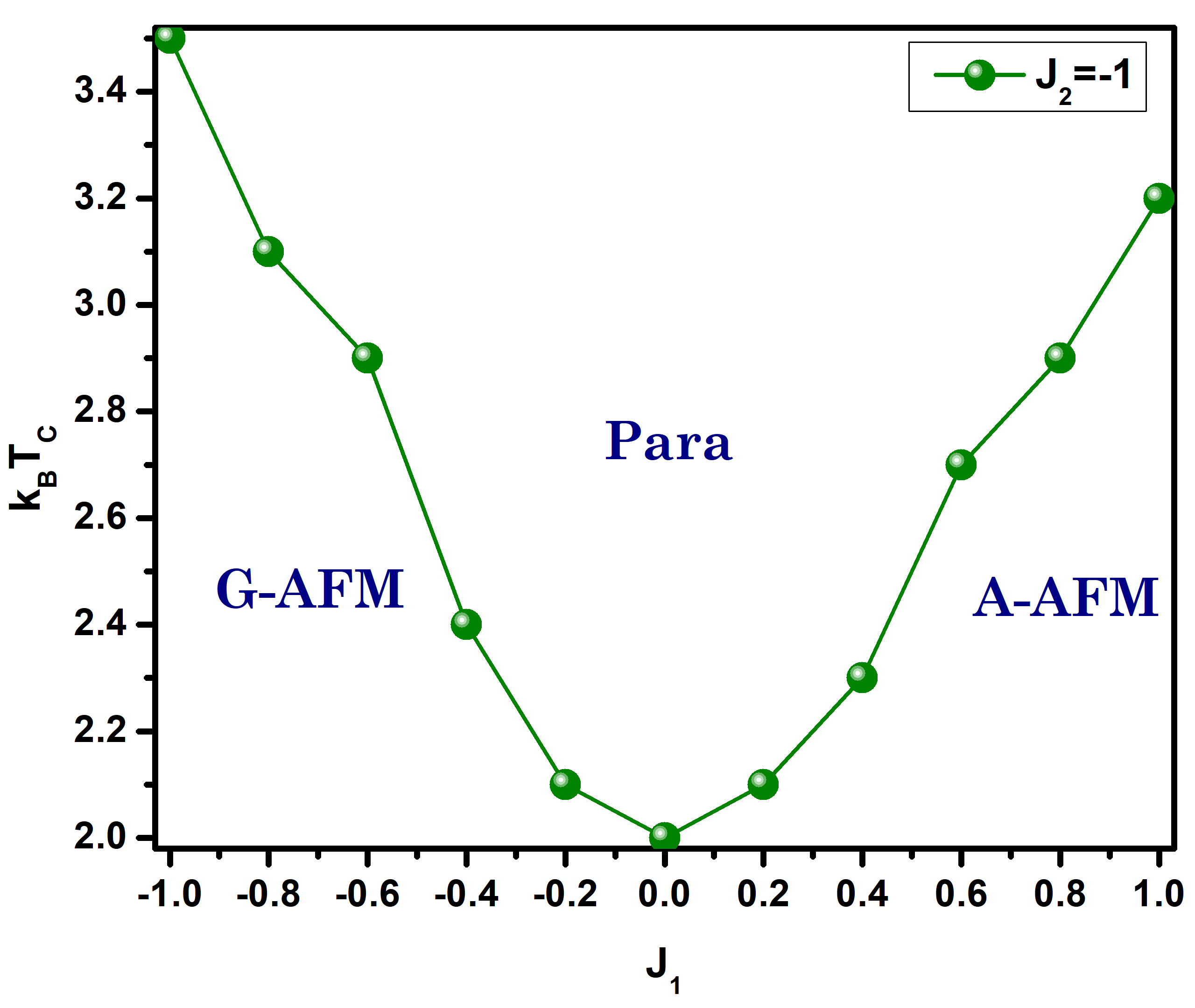}
		\caption{$ $}
		\label{f7b}
	\end{subfigure}
	~
	\caption{Phase diagram: $J_1$ vs $T_C$. (a) $J_2=+1$; $-1.0 \leq J_1 \leq 1.0$, and (b) $J_2=-1$; $-1.0 \leq J_1 \leq 1.0$.}
	\label{f7}
\end{figure}

The phase diagram is plotted for $T_C$ with various values of interaction strength, $J_1$ and is shown in Fig. \ref{f7}. The interaction $J_2$ is set to $+1$  and the interaction $J_1$ is raised from $-1.0$ to $1.0$ as in Fig. \ref{f7a}. Therein, the phase line separates the phases into C-AFM, FM and paramagnetic regimes. At $J_1=0.0$, the C-type AFM turns in to FM ordering due to the fixed value of $J_2=1.0$. The transition temperature for the interactions $J_2=-1$ and $-1.0 \leq J_1 \leq +1$ are plotted and shown in Fig. \ref{f7b}.  $T_C$ is found to decreases with increase in $J_1$ till $J_1=0.0$ and above that it increases. The phase line separates the region of G-AFM, A-AFM and the Paramagnetic order. The system will follow the G-AFM ordering when $J_1=0.0$ and $T_C=2.0$ due to the fixed value of $J_2=-1.0$. The observations from these results suggest that though the system has a ferromagnetic interaction, a small anti-ferromagnetic interaction would lead the system to possess the anti-ferromagnetic ordering.

The following part explores the system with the inclusion of magnetic field of various field strengths for different types of interaction. 

\subsubsection*{In the presence of magnetic field}

\begin{figure}[H]
	\centering
	\begin{subfigure}[b]{0.45\textwidth}
		\includegraphics[width=\textwidth]{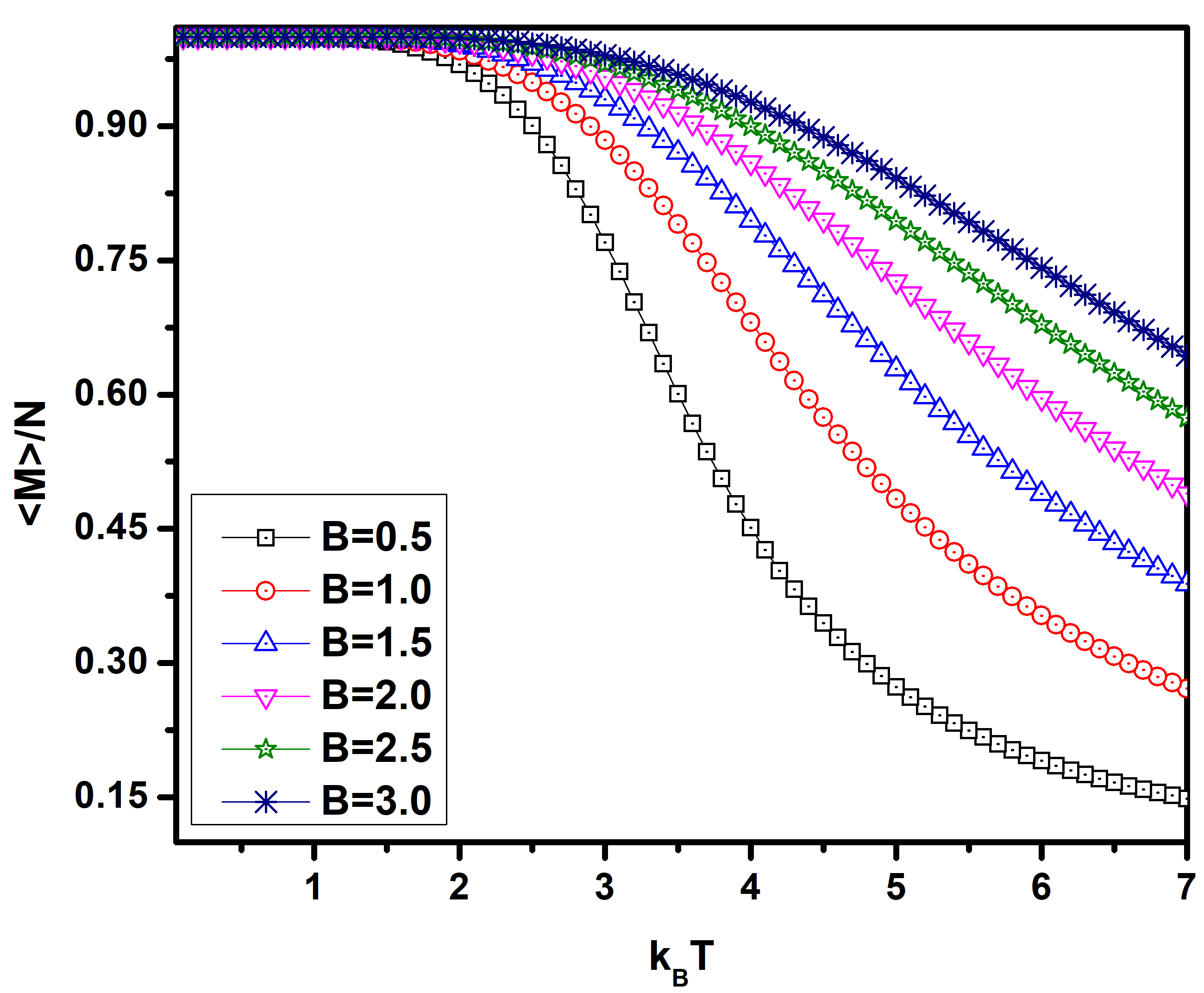}
		\caption{$ $}
		\label{f8a}
	\end{subfigure}
	~
	\begin{subfigure}[b]{0.45\textwidth}
		\includegraphics[width=\textwidth]{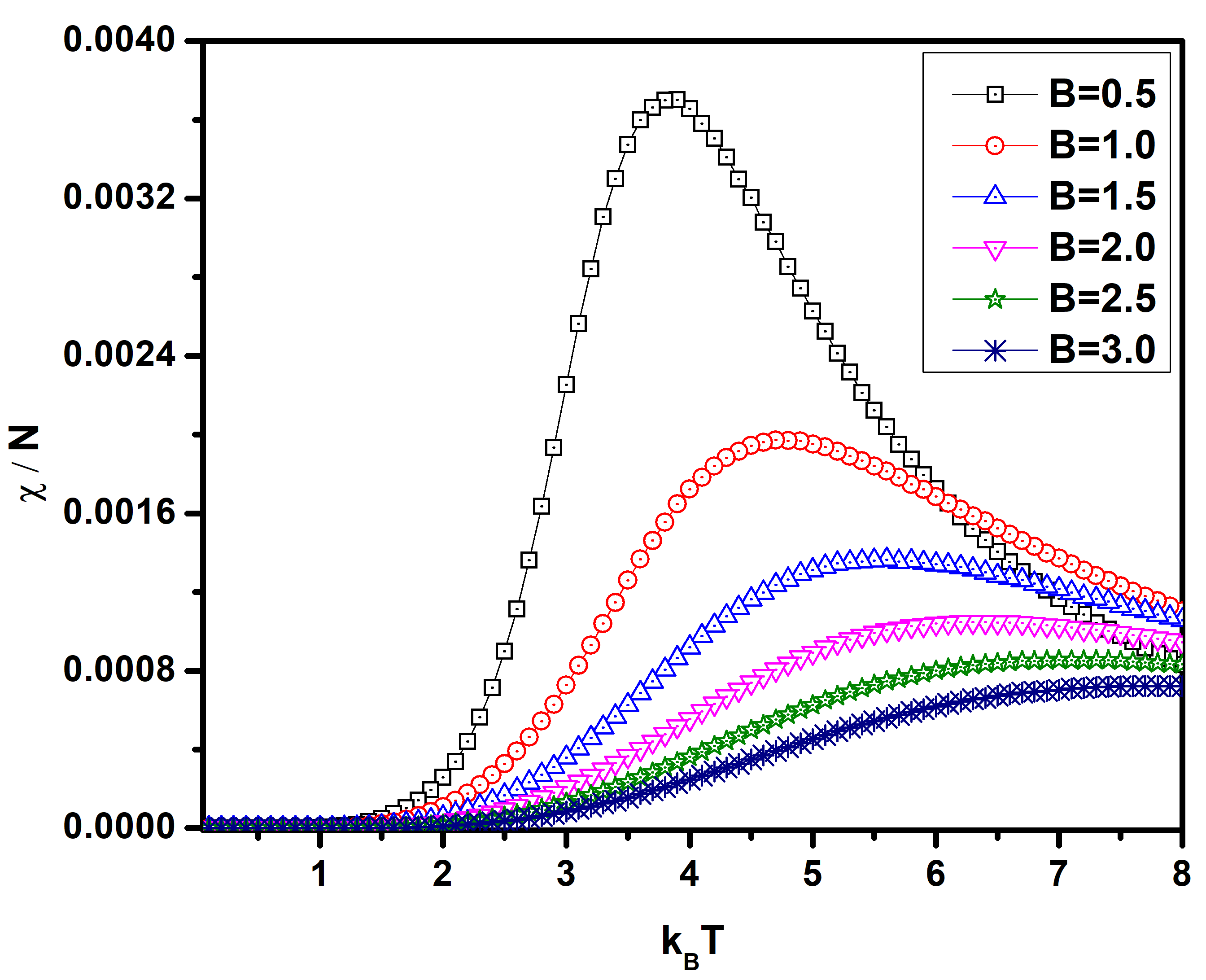}
		\caption{$ $}
		\label{f8b}
	\end{subfigure}
	~
	\caption{Field dependence of observables for ferromagnetic interactions with magnetic field, (a) average magnetization and (b) susceptibility. The interactions are $J_1=+1.0$ and $J_2=+1.0$.}
		\label{f8}	
\end{figure}

The magnetic field is applied along the $Z$ axis. The interaction strengths, $J_1=+1$ and $J_2=+1$ are applied to set the system's state in the ferromagnetic order. The external magnetic field is varied from $0.5$ to $3.0$ and the average observables are plotted in Fig. \ref{f8}. Fig. \ref{f8a} depicts the average magnetization and the influence of field in the spin ordering. At lower temperature, the average magnetization attains its saturation value and the system is found to maintain FM ordering at those lower temperatures. Due to the presence of magnetic field, the spins are favoured to orient in the direction of the field. Increasing the temperature breaks that orientation into a random spin order. This breaking point (the point at which random spin order occurs) increases while increasing the field, since, the system is favoured to sustain ferromagnetic ordering at the higher magnetic field and it needs more temperature to break this orientation. Field dependent magnetic susceptibility is shown in Fig. \ref{f8b} and the transition temperature is obtained from the peak value of susceptibility, which is observed to be shifting and decreasing towards the higher temperatures with increasing magnetic field.

\begin{figure}[H]
	\centering
	\begin{subfigure}[b]{0.45\textwidth}
		\includegraphics[width=\textwidth]{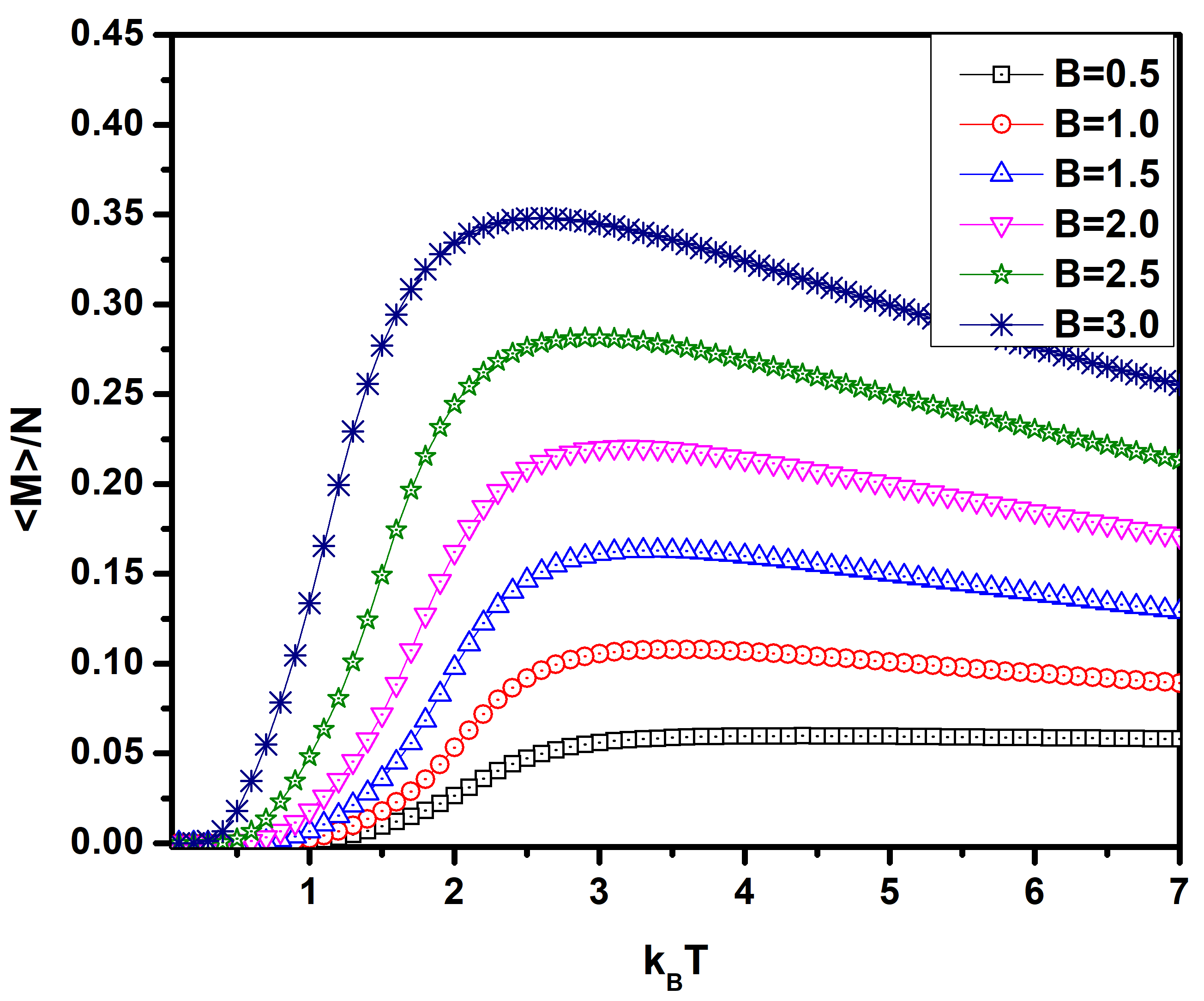}
		\caption{$ $}
		\label{f9a}
	\end{subfigure}
	~
	\begin{subfigure}[b]{0.45\textwidth}
		\includegraphics[width=\textwidth]{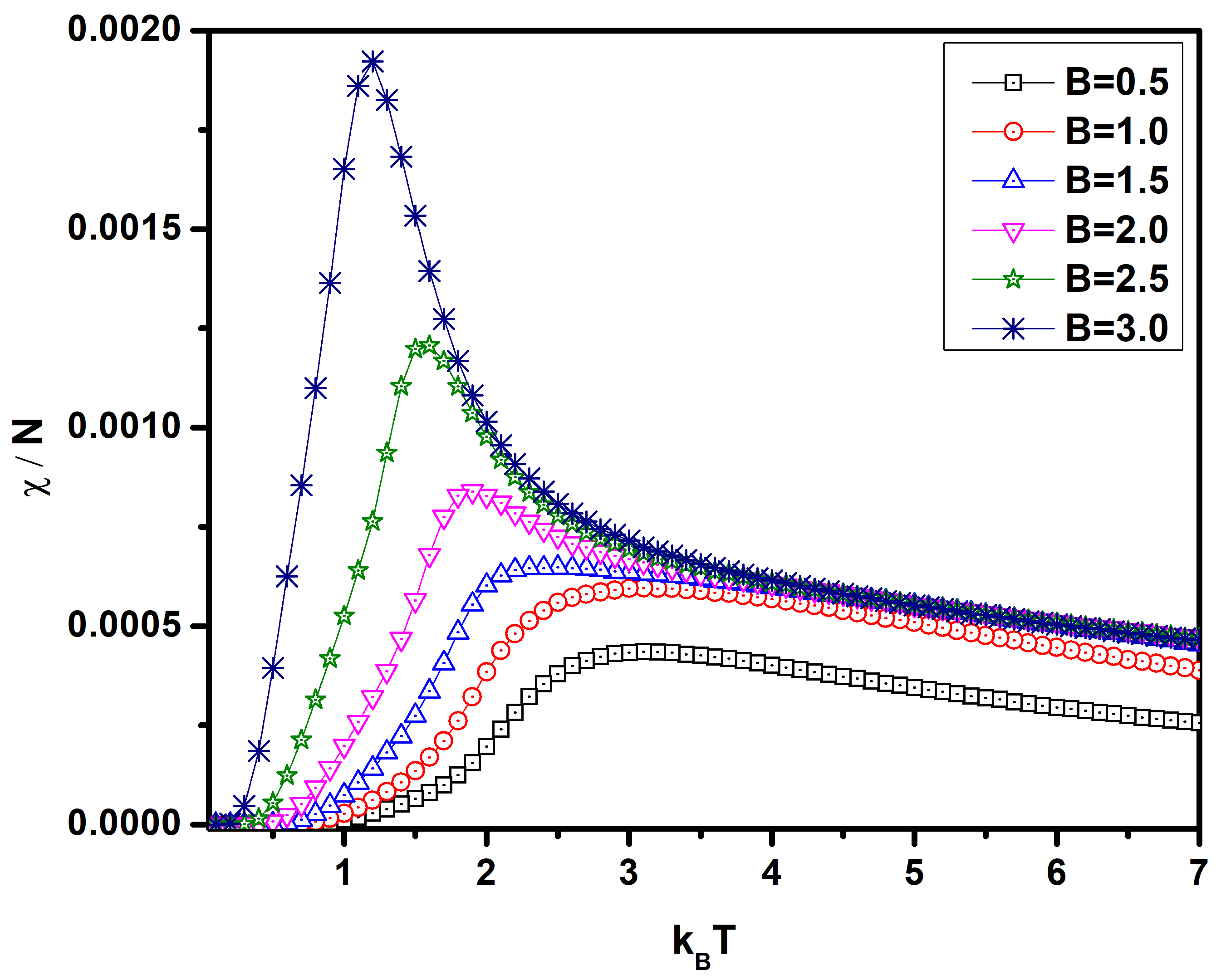}
		\caption{$ $}
		\label{f9b}
	\end{subfigure}
	~
	\caption{Field dependence of observables for G-type anti-ferromagnetic interactions  with magnetic field, (a) average magnetization and (b) susceptibility for the interactions $J_1=-1.0$ and $J_2=-1.0$.}
	\label{f9}
\end{figure}

The magnetic observables for anti-ferromagnetic interactions with the inclusion of magnetic field is plotted in Fig. \ref{f9}. The average magnetization at lower temperatures is zero due to the anti-parallel alignment of the spins that cancels each other. The value of average magnetization increases as the field increases (see the  Fig. \ref{f9a}). The critical temperature is found from the susceptibility plot Fig. \ref{f9b}. The peak value of susceptibility increases and its position is shifted towards the lower temperatures while increasing the magnetic field value. 

\begin{figure}[H]
	\centering
	\begin{subfigure}[b]{0.45\textwidth}
		\includegraphics[width=\textwidth]{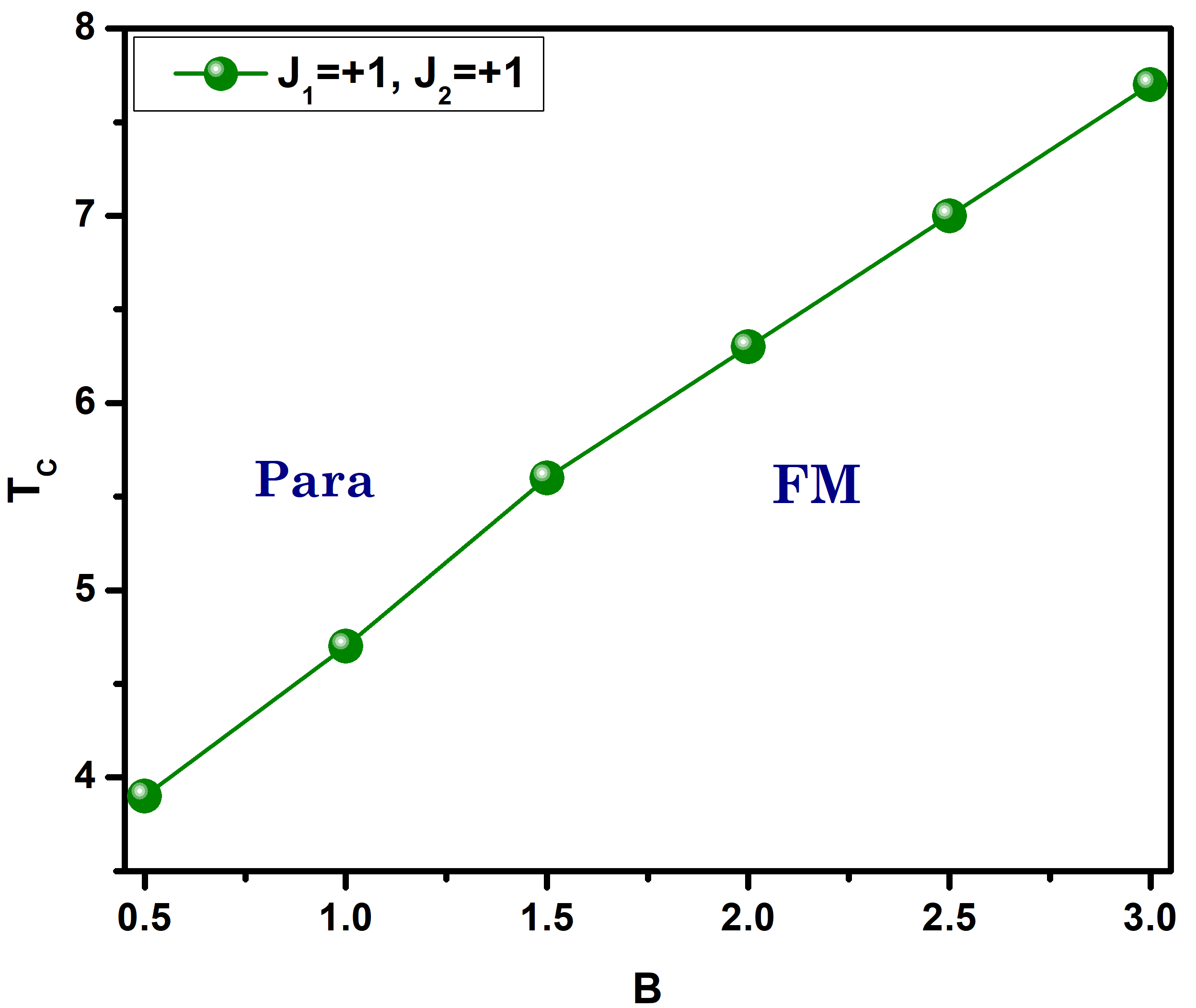}
		\caption{$ $}
		\label{f10a}
	\end{subfigure}
	~
	\begin{subfigure}[b]{0.45\textwidth}
		\includegraphics[width=\textwidth]{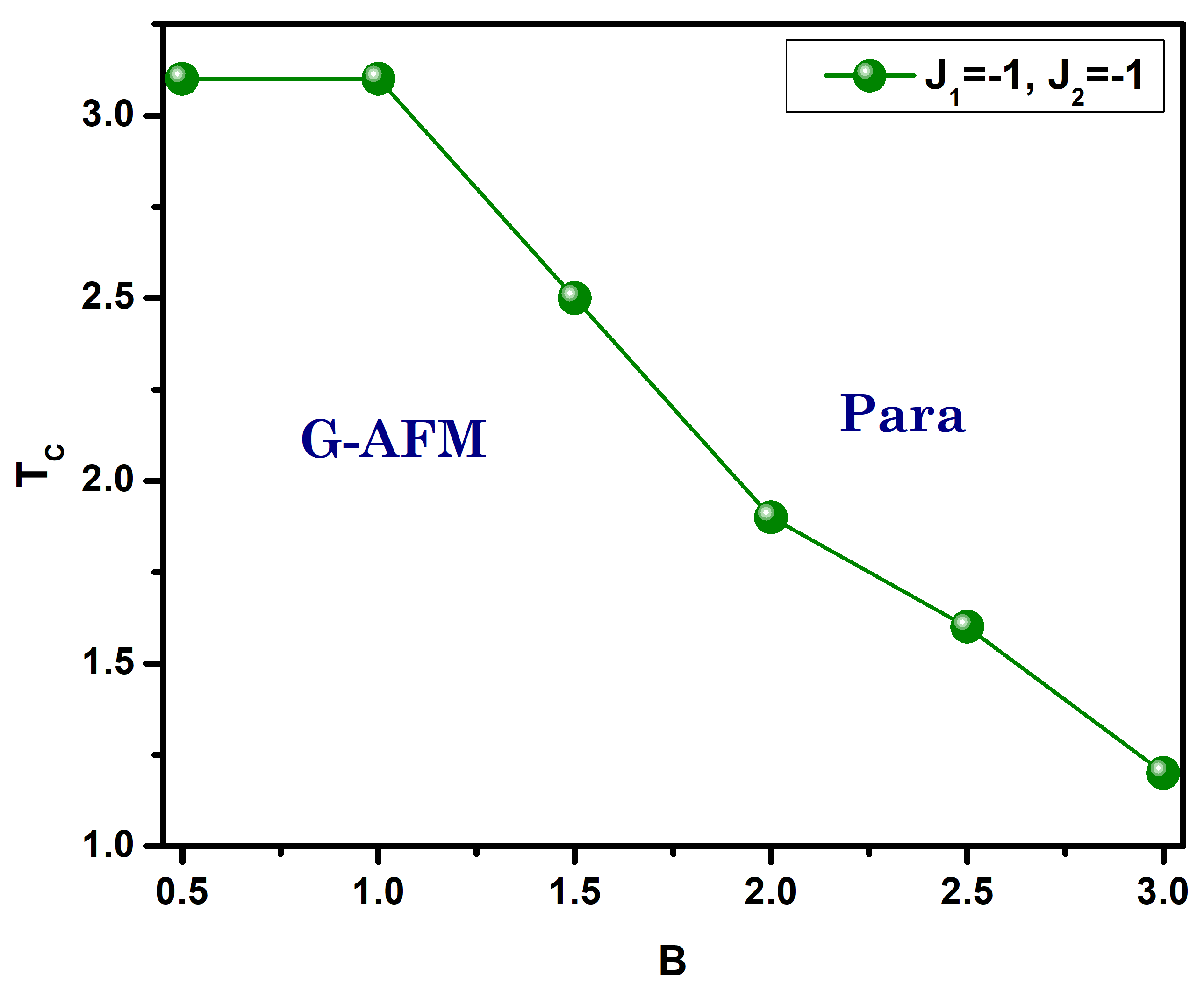}
		\caption{$ $}
		\label{f10b}
	\end{subfigure}
	~
	\caption{Phase diagram: $B$ vs $T_C$. (a) $J_1=1.0$ \& $J_2=1.0$; $0.5\leq B\leq 3.0$, and (b) $J_1=-1.0$ \& $J_2=-1.0$; $0.5\leq B\leq 3.0$. }
	\label{f10}
\end{figure}

The variation of transition temperature with field intensity for FM and G-AFM interactions are plotted as phase diagram and is shown in Fig. \ref{f10}.  It is observed that the transition temperature shifted towards higher temperature for FM interactions (Fig. \ref{f10a}) and  towards lower temperature for G-AFM interactions with the increase in the field intensity (Fig. \ref{f10b}). From the figure (Fig. \ref{f10}), it is also observed that the system transits from FM/G-AFM to paramagnetic phase with respect to the values of interactions $J_{1}$ and $J_{2}$.
                                                                   
\begin{figure}[H]
	\centering
	\begin{subfigure}[b]{0.45\textwidth}
		\includegraphics[width=\textwidth]{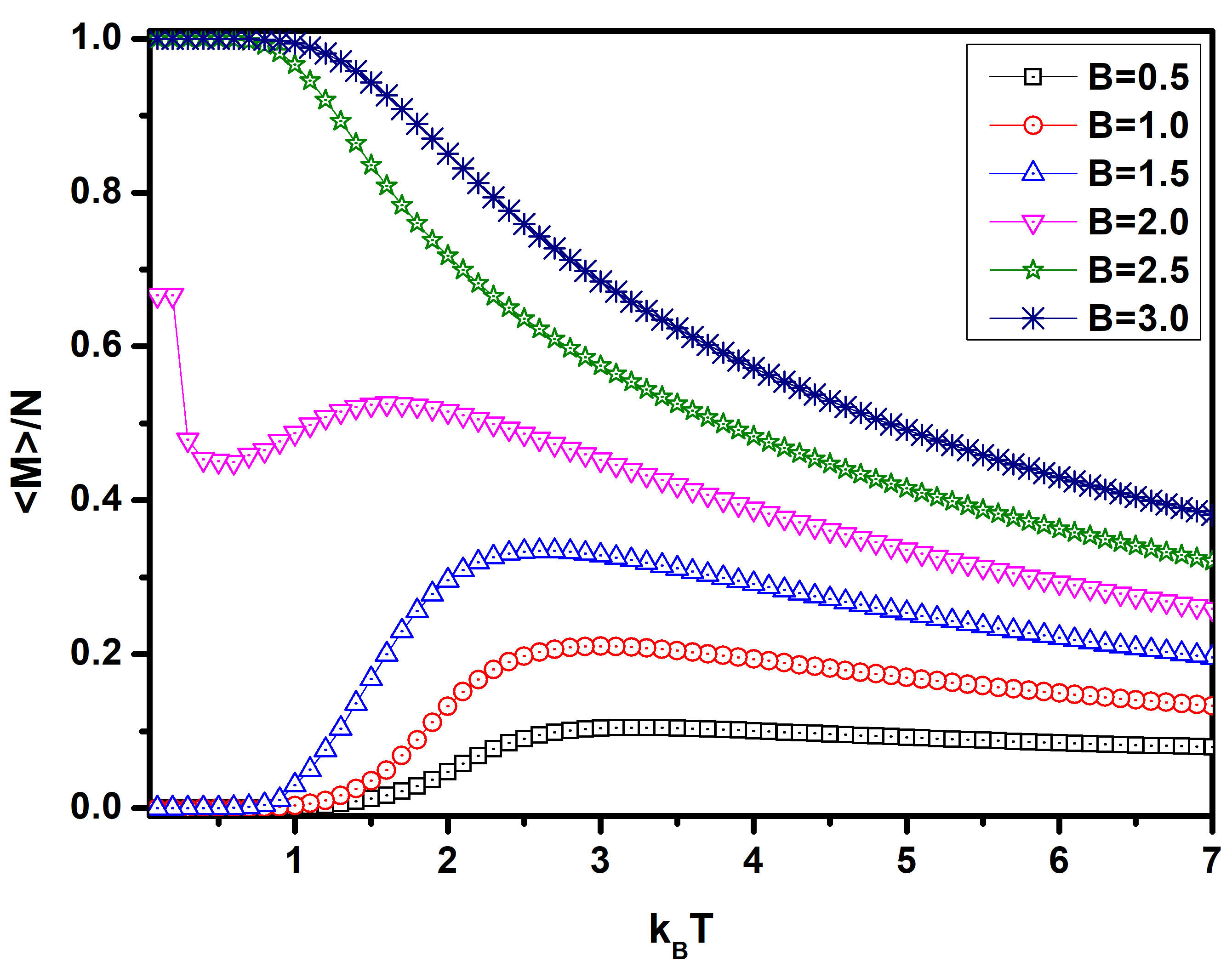}
		\caption{$ $}
		\label{f11a}
	\end{subfigure}
	~
	\begin{subfigure}[b]{0.45\textwidth}
	\includegraphics[width=\textwidth]{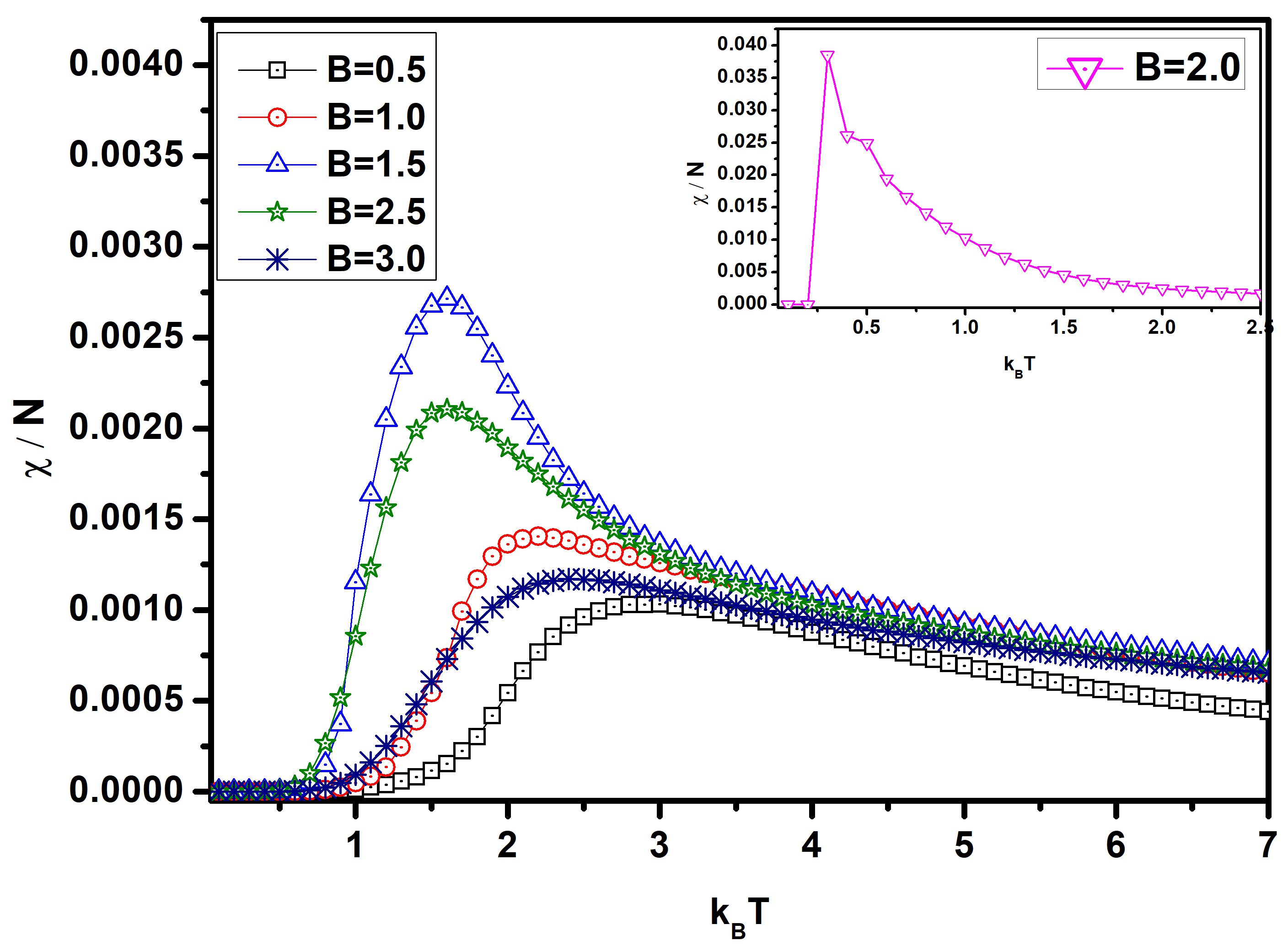}
		\caption{$ $}
		\label{f11b}
	\end{subfigure}
	~
	\caption{Field dependence of observables with increasing magnetic field, (a) average magnetization and (b) susceptibility for the interactions $J_1=+1.0$ and $J_2=-1.0$. The inset plot represents the susceptibility at $B=2.0$.}
	\label{f11}
\end{figure}

The observables for A-type AFM ordered interaction are depicted in Fig. \ref{f11}. As discussed in the preceding section, a small AFM interaction is sufficient to produce an AFM order in a FM system. Hence the system is maintained in A-AFM order even in the presence of FM interaction. However, as the external field is increased, the system loses its AFM behavior and changes to FM behavior. The magnetization plot for various values of $B$ is shown in Fig. \ref{f11a}. The system retains in the A-AFM order when the external field $B$ is less than $2.0$ and it experiences FM order when the external field, $B$ is greater than 2. The transition temperature is observed from the susceptibility plot (Fig. \ref{f11b}). As the external magnetic field is increased, the $T_C$ decreases and the peak value of susceptibility increases. When it is increased beyond $B=2.0$, the $T_C$ increases but the peak value of susceptibility decreases with increase in field strength. The susceptibility plot for $B=2.0$ is highlighted in inset of the Fig. \ref{f11b}.

\begin{figure}[H]
	\centering
	\begin{subfigure}[b]{0.45\textwidth}
		\includegraphics[width=\textwidth]{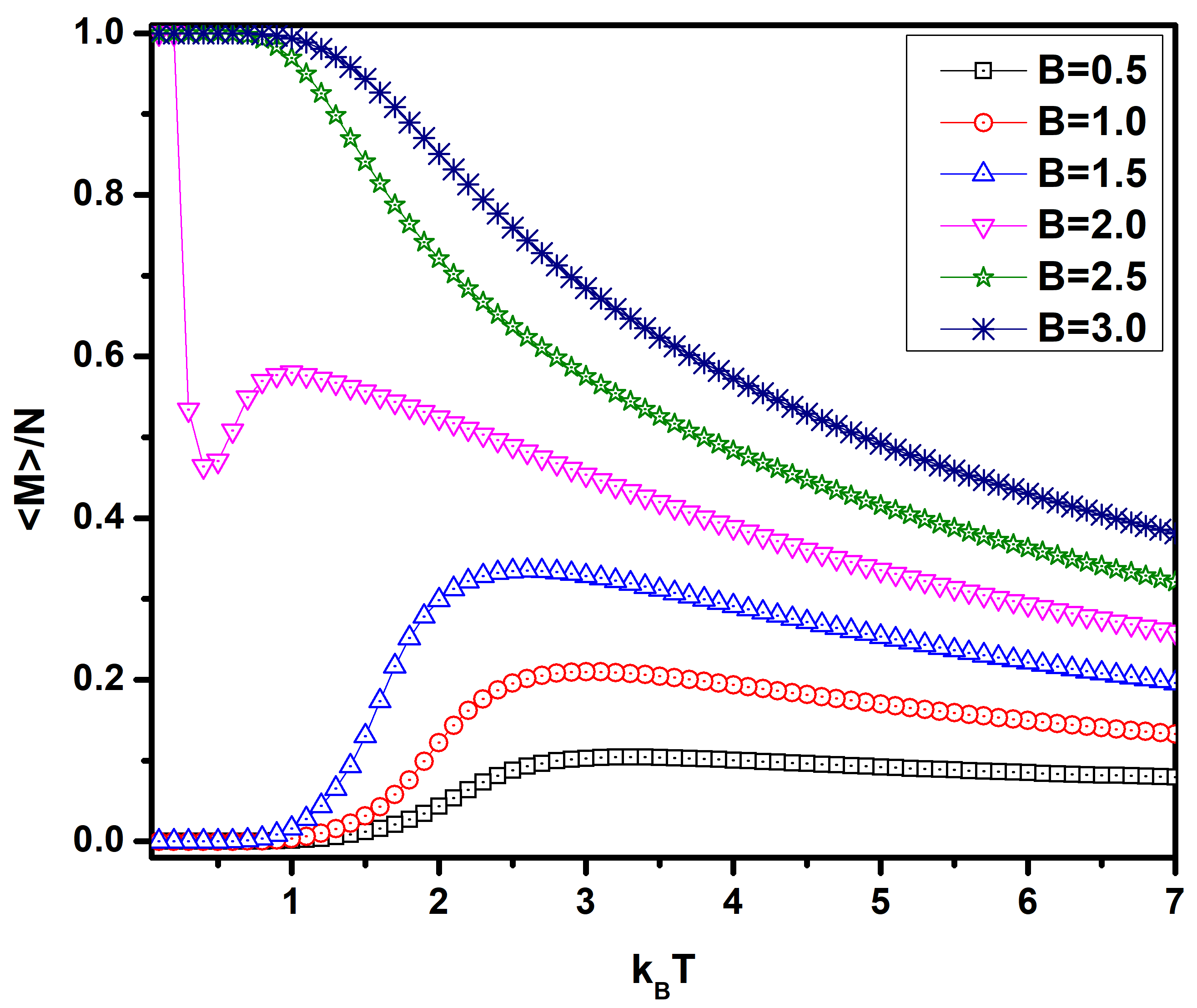}
		\caption{$ $}
		\label{f12a}
	\end{subfigure}
	~
	\begin{subfigure}[b]{0.45\textwidth}
		\includegraphics[width=\textwidth]{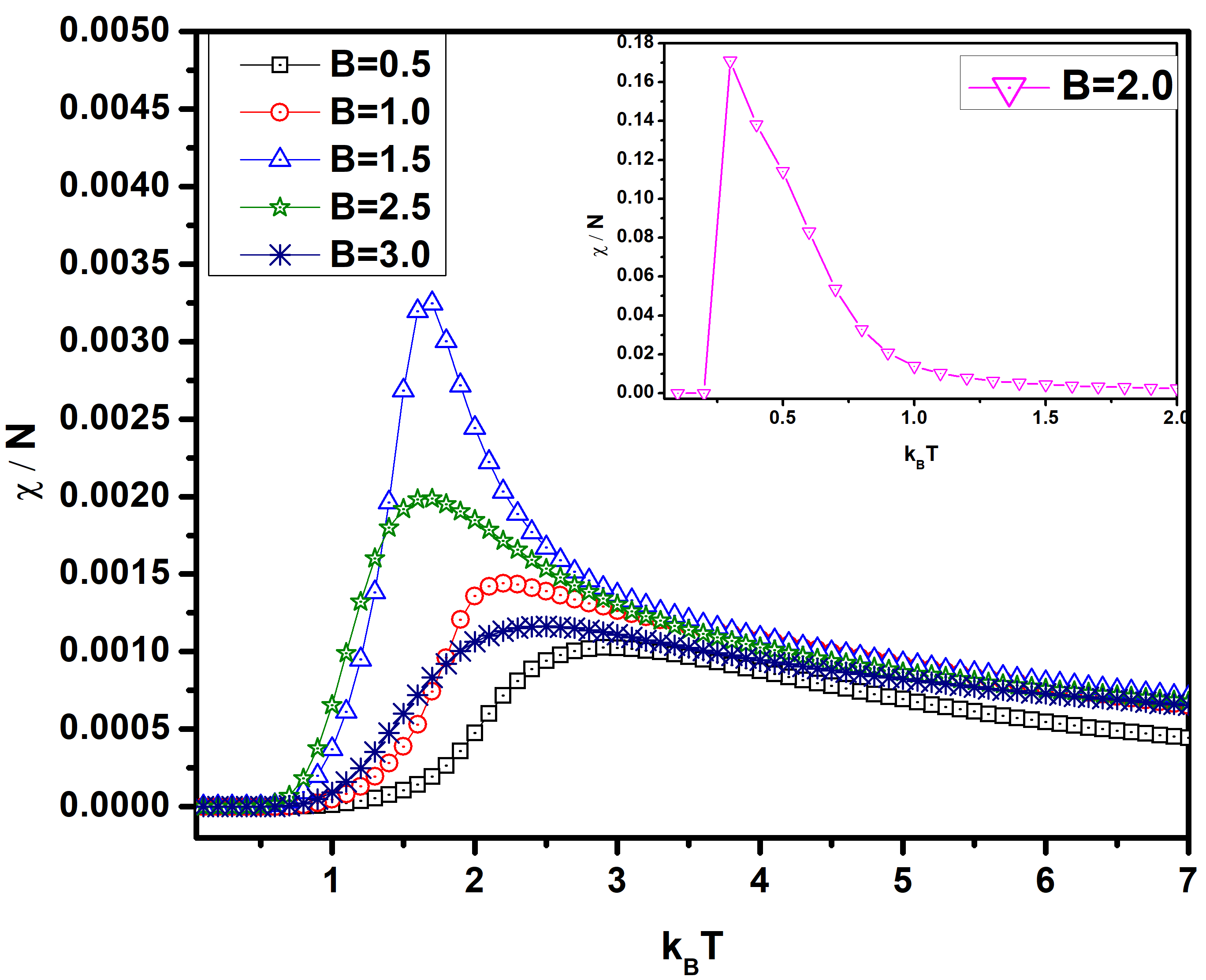}
		\caption{$ $}
		\label{f12b}
	\end{subfigure}
	~
	\caption{Field dependence of observables with increasing magnetic field, (a) average magnetization and (b) susceptibility. The interactions $J_1=-1.0$ and $J_2=+1.0$. The inset plot represents the susceptibility at $B=2.0$.}
	\label{f12}
\end{figure}

Similarly, for the next set of simulations,  interaction strengths of $J_1=-1$ and $J_2=+1$ is applied to the system, which has a C-type magnetic ordering. The results of this interaction are depicted in Fig. \ref{f12}. From the results as shown in Fig. \ref{f12a}, it is clear that, if the magnetic field is less than $2.0$, the system is maintained in C-AFM order and the system turns to FM ordering when the magentic filed is greater than $2.0$ . As the field increases (Fig. \ref {f12b}), the $T_C$ decreases and the maximum value of susceptibility increases until $B=2.0$. Beyond that, the $T_C$ increases and the maximum value of susceptibility decreases with increasing field. The inset in Fig. \ref{f12b} shows the susceptibility curve for $B=2.0$.

\begin{figure}[H]
	\centering
	\begin{subfigure}[b]{0.45\textwidth}
		\includegraphics[width=\textwidth]{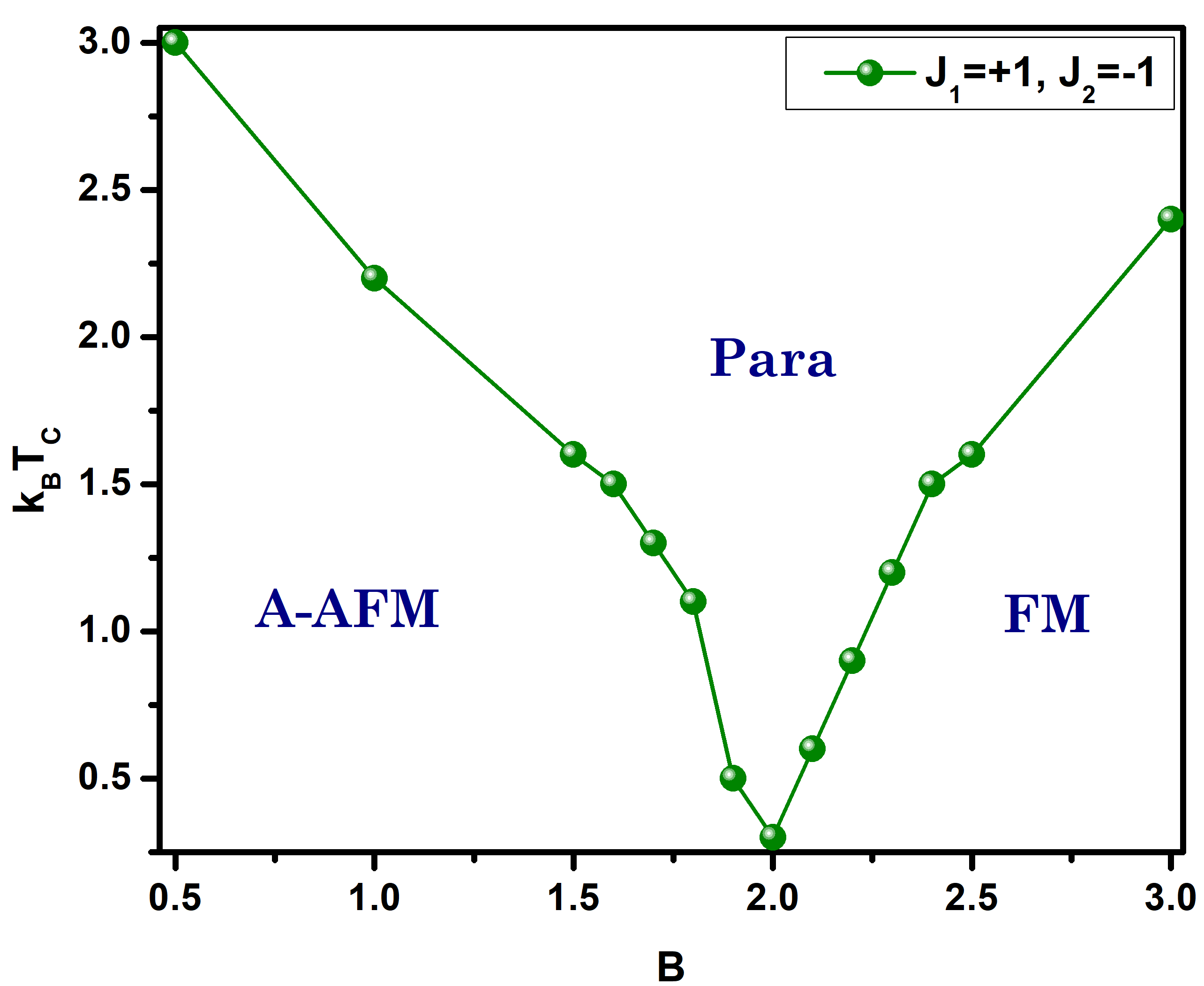}
		\caption{$ 0$}
		\label{f13a}
	\end{subfigure}
	~
	\begin{subfigure}[b]{0.45\textwidth}
		\includegraphics[width=\textwidth]{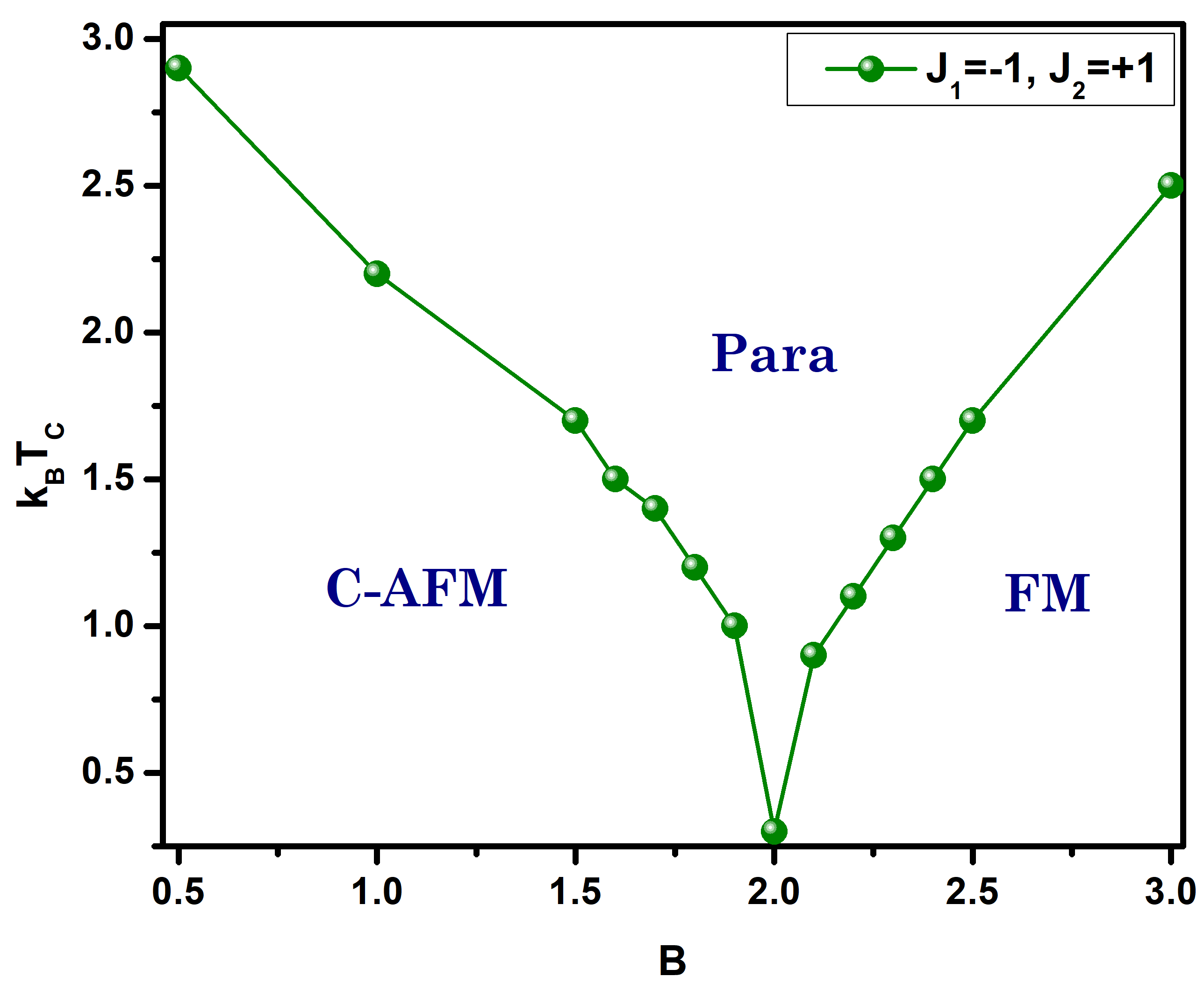}
		\caption{$ $}
		\label{f13b}
	\end{subfigure}
	~
	\caption{Phase diagram: $B$ vs $T_C$. (a) $J_1=+1.0$ \& $J_2=-1.0$; $0.5\leq B\leq 3.0$, and (b) $J_1=-1.0$ \& $J_2=+1.0$; $0.5\leq B\leq 3.0$.}
	\label{f13}
\end{figure}

The transition temperatures for the interactions (A-AFM and C-AFM) against an external magnetic field are plotted in Fig. \ref{f13}.
The phases are well separated in the phase diagram for interaction $J_1=+1,~ J_2=-1$ (Fig. \ref{f13a}). The A-AFM dominates the system below $B=2.0$ and above that FM interaction will follow. Similarly, the C-AFM dominates the system below $B=2.0$, beyond that the FM dominates the system for the interaction $J_1=-1,~ J_2=+1$ (Fig. \ref{f13b}). The evolution of both the systems with interactions $J_1=+1, J_2=-1$ and $J_1=-1, J_2=+1$ possess AFM ordering in low magnetic field ($B<2.0$). As the magnetic field increases, the system changes to FM ordering.

In the subsequent section,  the analysis of hysteresis behavior of the system with the application of the external magnetic field is given in detail.

\subsubsection*{Hysteresis process}

\begin{figure}[H]
	\centering
	\includegraphics[scale=0.4]{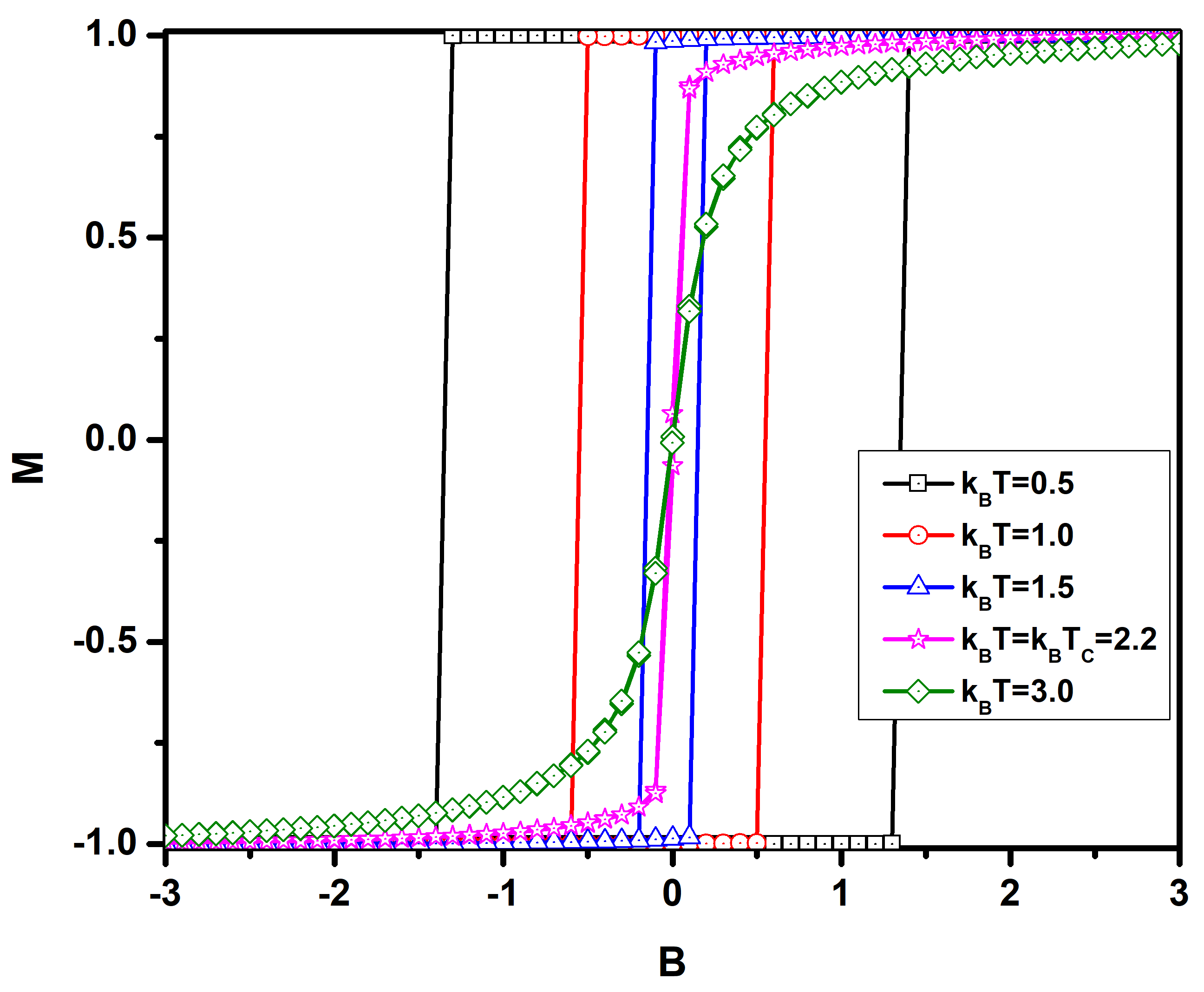}
	\caption{Hysteresis loop for FM interactions at different temperatures.}
	\label{f14}
\end{figure}

\begin{figure}[H]
	\centering
	\begin{subfigure}[b]{0.3\textwidth}
		\includegraphics[width=\textwidth]{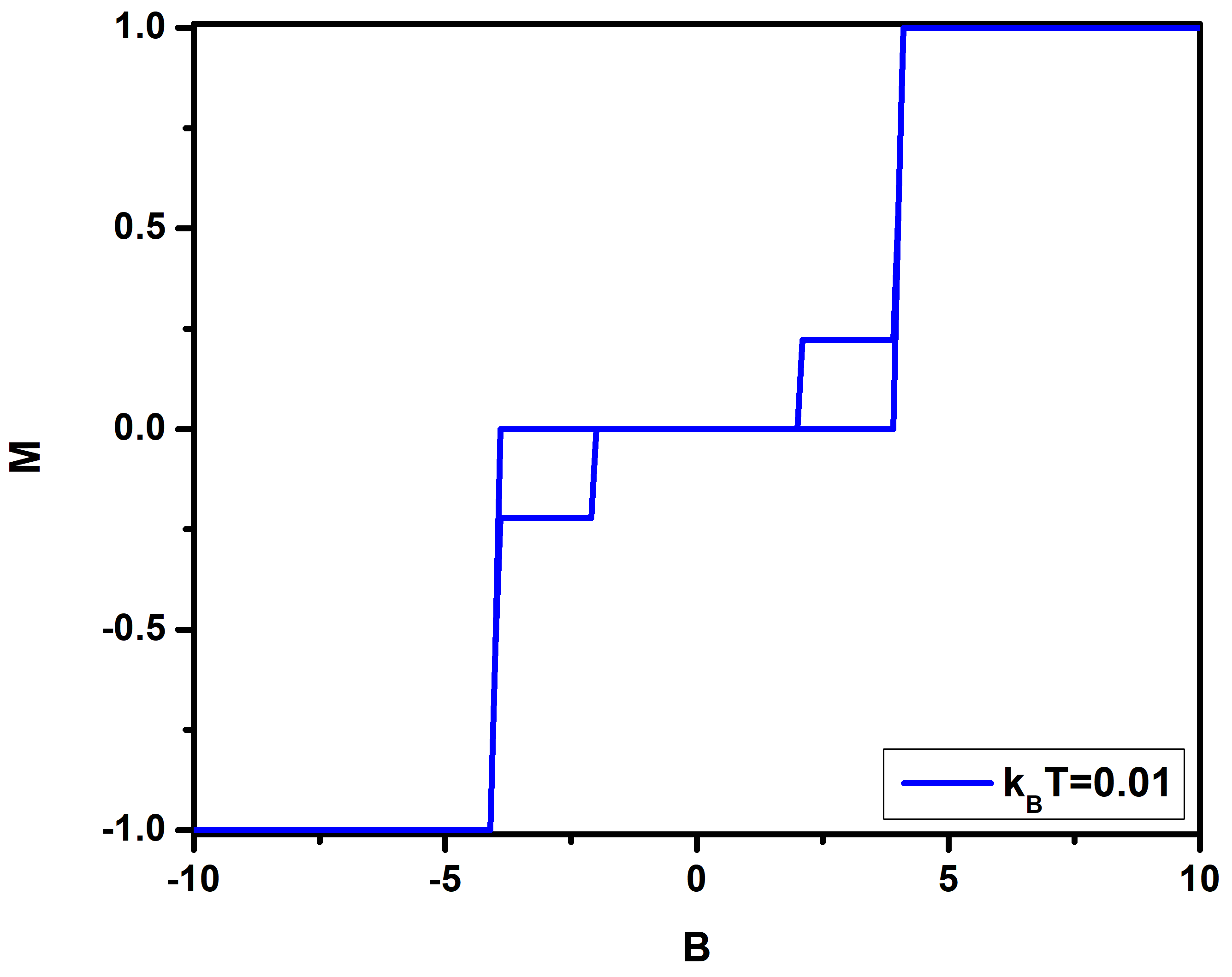}
		\caption{$ $}
		\label{f15a}
	\end{subfigure}
	~
	\begin{subfigure}[b]{0.3\textwidth}
		\includegraphics[width=\textwidth]{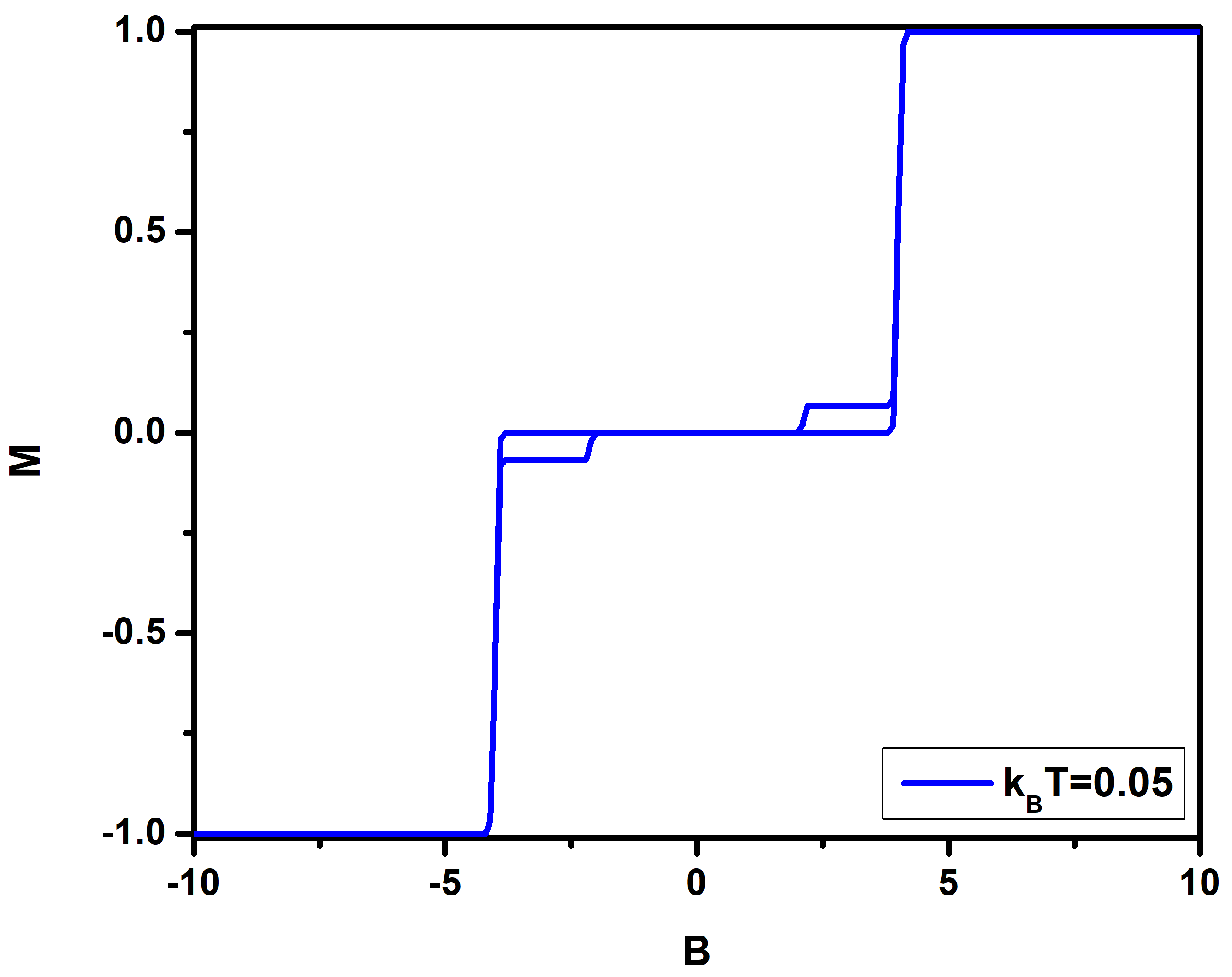}
		\caption{$ $}
		\label{f15b}
	\end{subfigure}
	~
	\begin{subfigure}[b]{0.3\textwidth}
		\includegraphics[width=\textwidth]{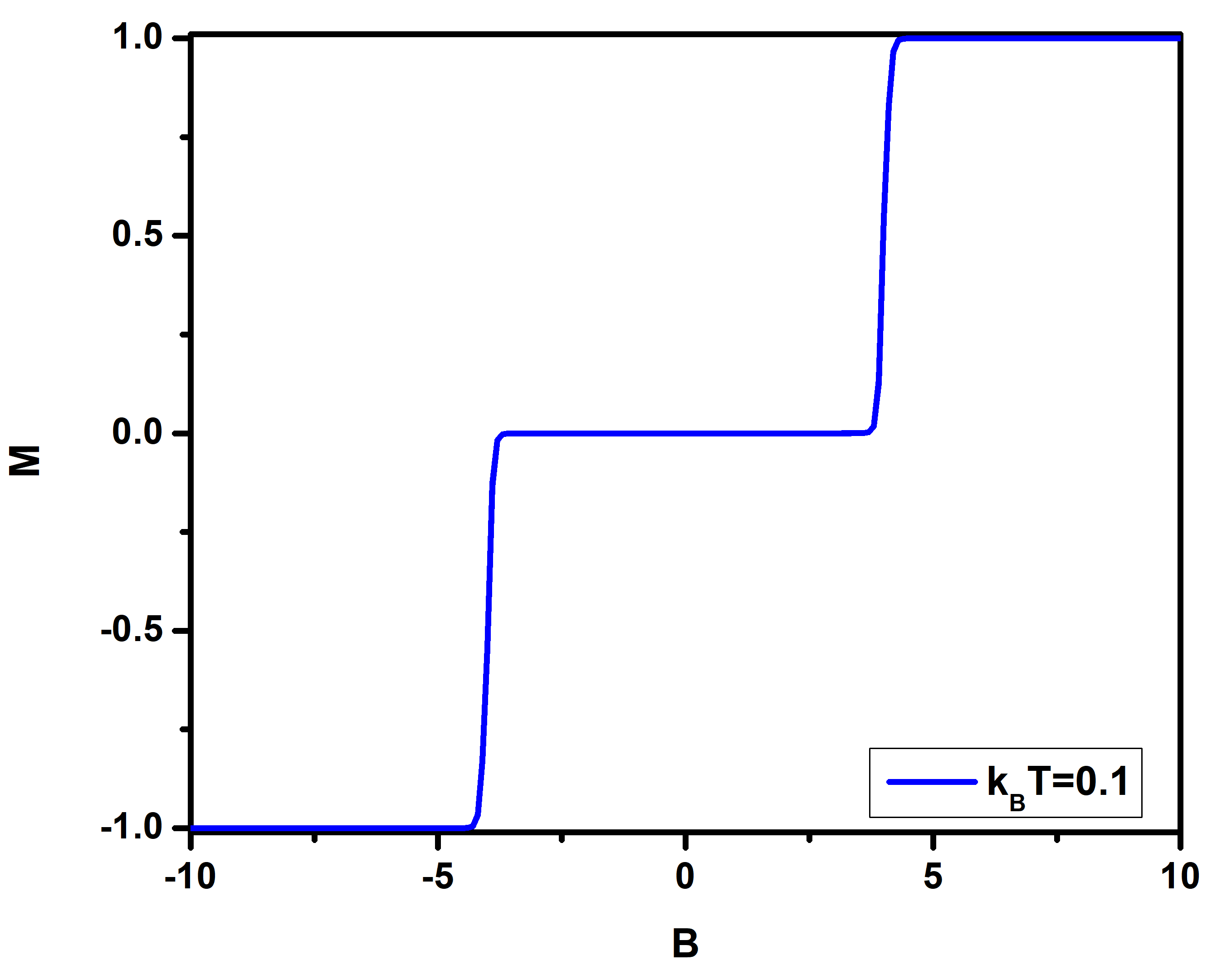}
		\caption{$ $}
		\label{f15c}
	\end{subfigure}
	~
	\begin{subfigure}[b]{0.3\textwidth}
		\includegraphics[width=\textwidth]{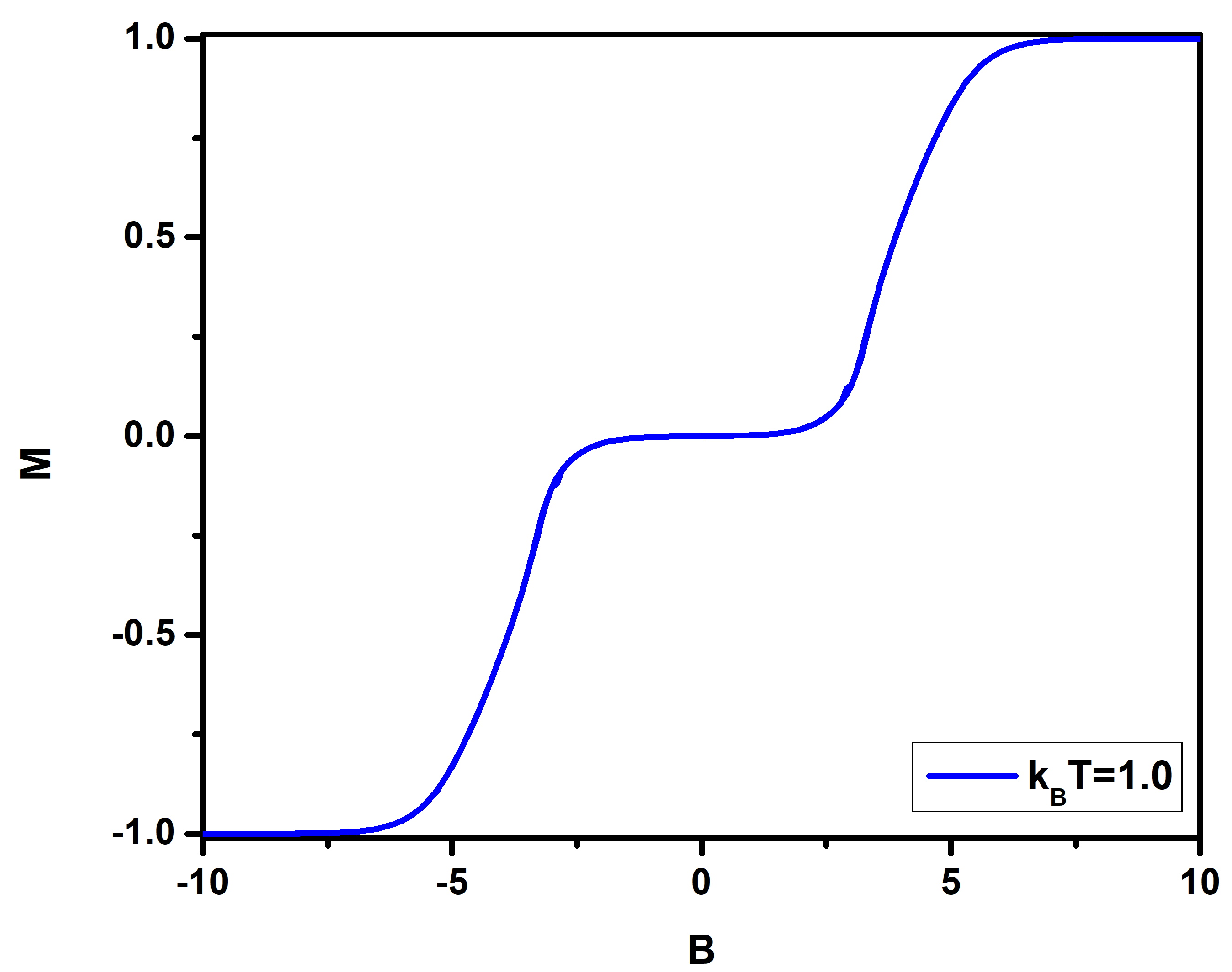}
		\caption{$ $}
		\label{f15d}
	\end{subfigure}
	~	
	\begin{subfigure}[b]{0.3\textwidth}
		\includegraphics[width=\textwidth]{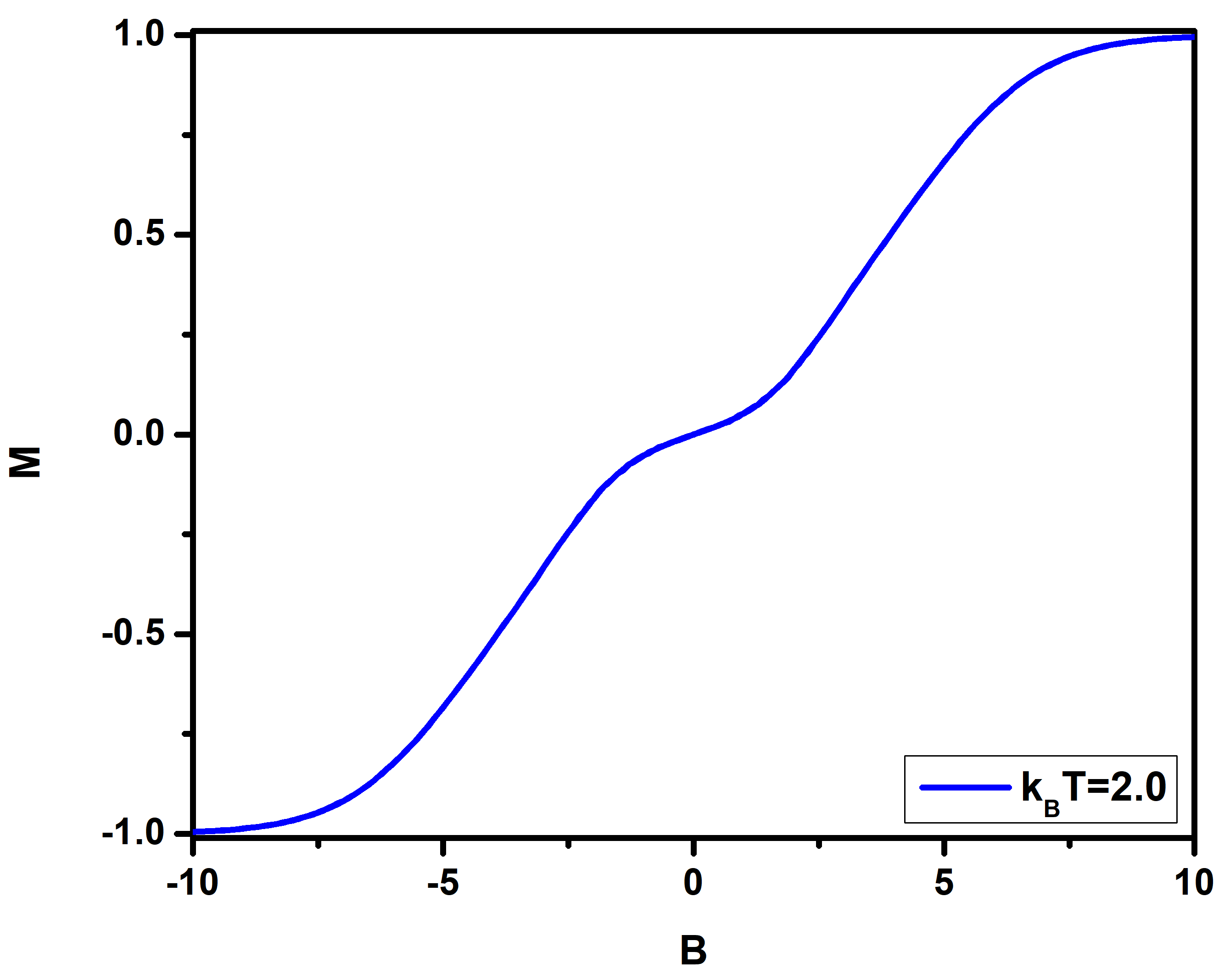}
		\caption{$ $}
		\label{f15e}
	\end{subfigure}
	~
	\begin{subfigure}[b]{0.3\textwidth}
		\includegraphics[width=\textwidth]{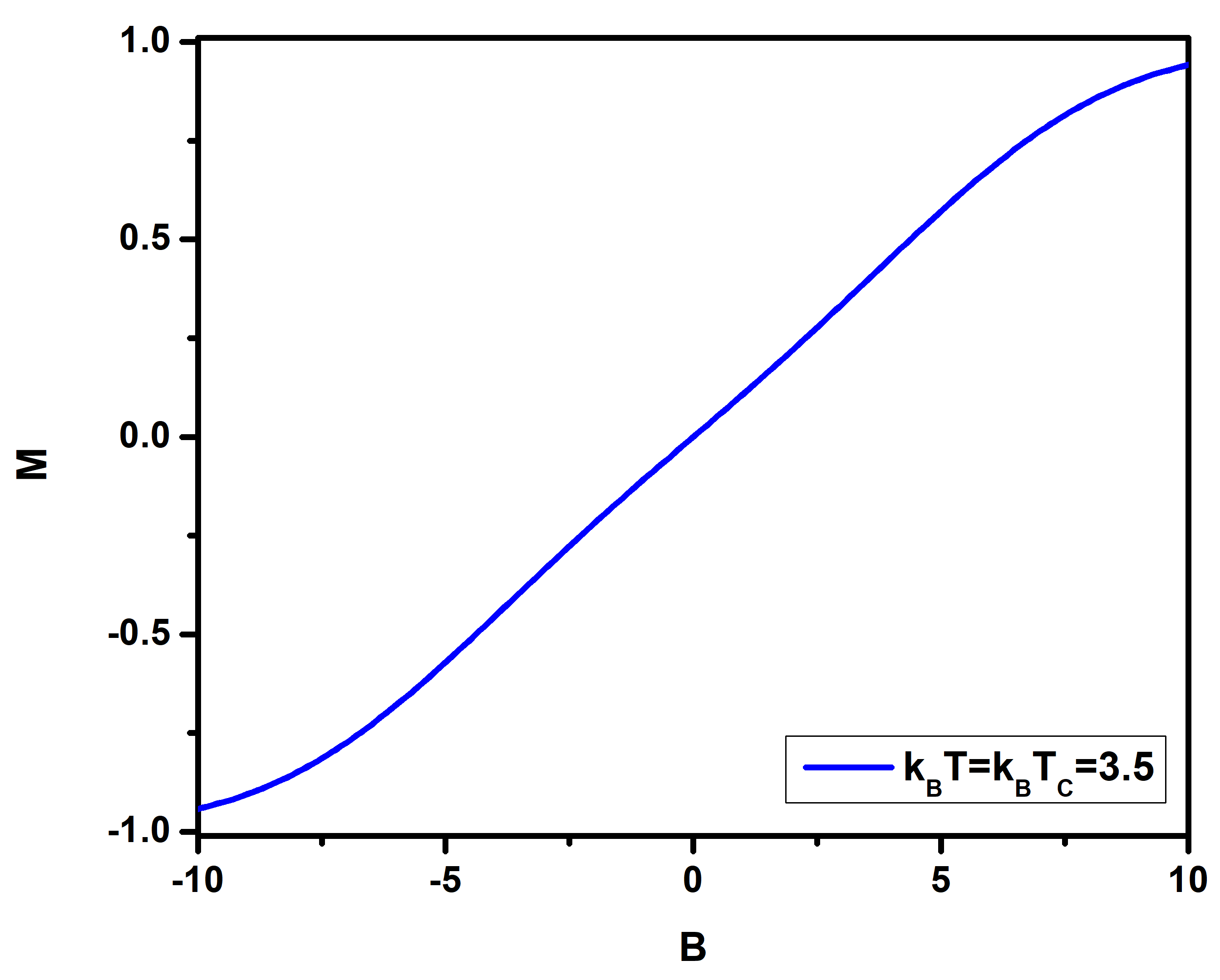}
		\caption{$ $}
		\label{f15f}
	\end{subfigure}
	~		
	\caption{Hysteresis loop for G-AFM order of interaction at different temperatures.}
	\label{f15}
\end{figure}

The magnetization is calculated by increasing and decreasing the external magnetic fields. The saturation limit $M= \pm 1$, is achieved as the system spins are favoured to align with the field. The net magnetization increases with increasing magnetic field, further it decreases in a slightly different path with the decrease in the field strength to form a loop. The hysteresis loops for different temperatures are plotted in Fig. \ref{f14}.  The magnetization shows a sudden shift to its saturation value at lower temperatures. As the temperature increases, the magnetization follows a curved path and gradually reaches the saturation value. The remanent magnetization is observed at zero field and the coercive force is observed within the field values for which the net magnetization is zero. With the increase in temperature, the hysteresis loop area reduces and the coercive force is also found to decrease.  At the transition temperature ($T_C=2.2$), the remanent magnetization gets reduced and the coercive force becomes zero. The forward and reverse paths merge to give a zero hysteresis area at $T_C=2.2$. After the transition temperature, the system is no longer ferromagnetic but transits to a paramagnetic state.
Fig. \ref{f15} shows the hysteresis process for the G-type anti-ferromagnetic ordering. The system exhibits double loop hysteresis for $T=0.01$ (Fig. \ref{f15a}) and $T=0.05$ (Fig. \ref{f15b}). The coercive force is bounded between $B= \pm 4.5$ and is found to decrease with increasing temperature. The double loop gradually vanishes and forms a step hysteresis as shown in  Fig. \ref{f15c} and \ref{f15d}. As the temperature increases, the cycle forms a single line and beyond the transition temperature, $T_C=3.5$, the system is found to be  in the paramagnetic phase (\ref{f15f}). There is no remanent magnetization for all the hysteresis loops.

\begin{figure}[H]
	\centering
	\begin{subfigure}[b]{0.3\textwidth}
		\includegraphics[width=\textwidth]{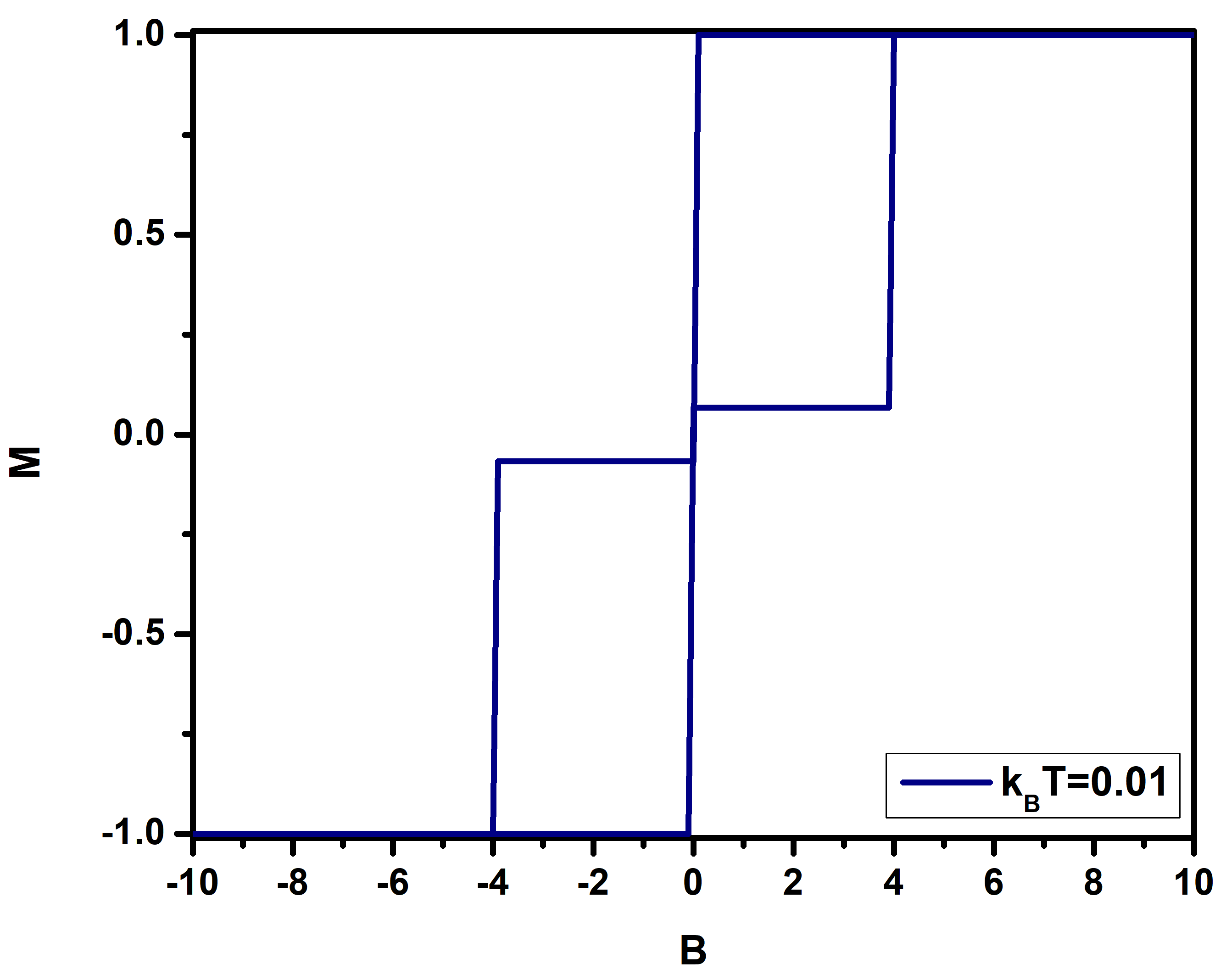}
		\caption{}
		\label{f16a}
	\end{subfigure}
	~
	\begin{subfigure}[b]{0.3\textwidth}
		\includegraphics[width=\textwidth]{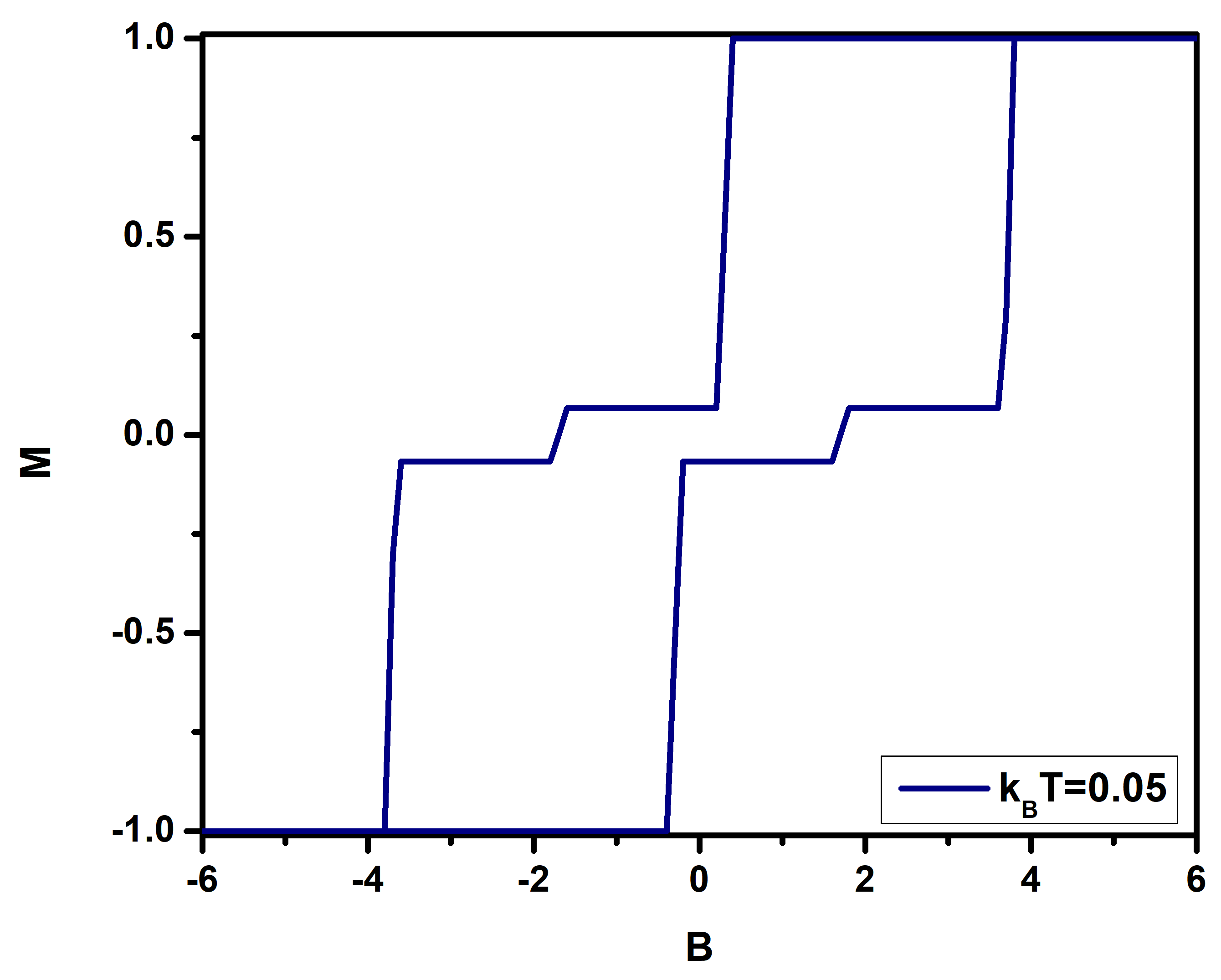} 
		\caption{}
		\label{f16b}
	\end{subfigure}
	~
	\begin{subfigure}[b]{0.3\textwidth}
		\includegraphics[width=\textwidth]{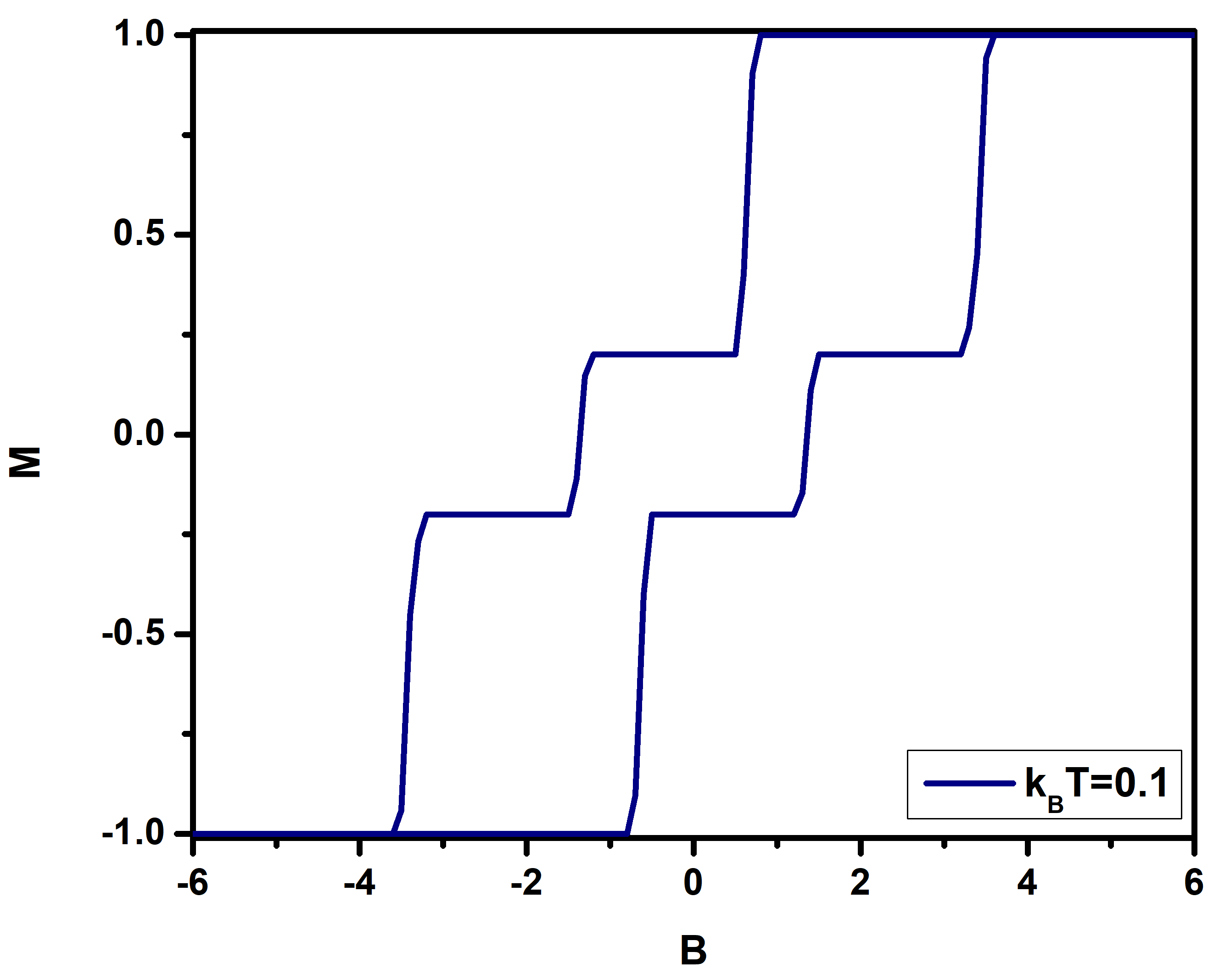}
		\caption{}
		\label{f16c}
	\end{subfigure}
	~
	\begin{subfigure}[b]{0.3\textwidth}
		\includegraphics[width=\textwidth]{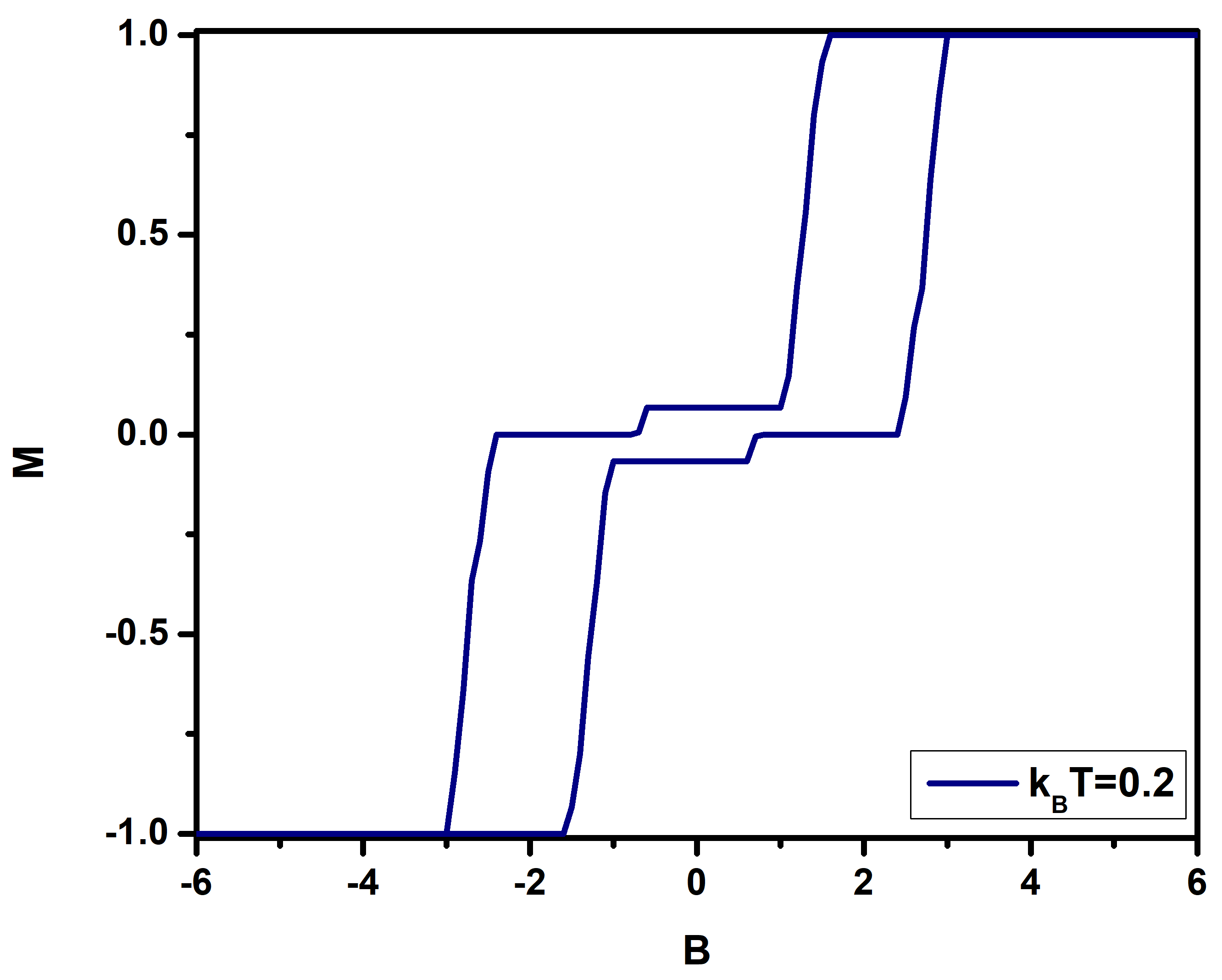}
		\caption{}
		\label{f16d}
	\end{subfigure}
	~
	\begin{subfigure}[b]{0.3\textwidth}
		\includegraphics[width=\textwidth]{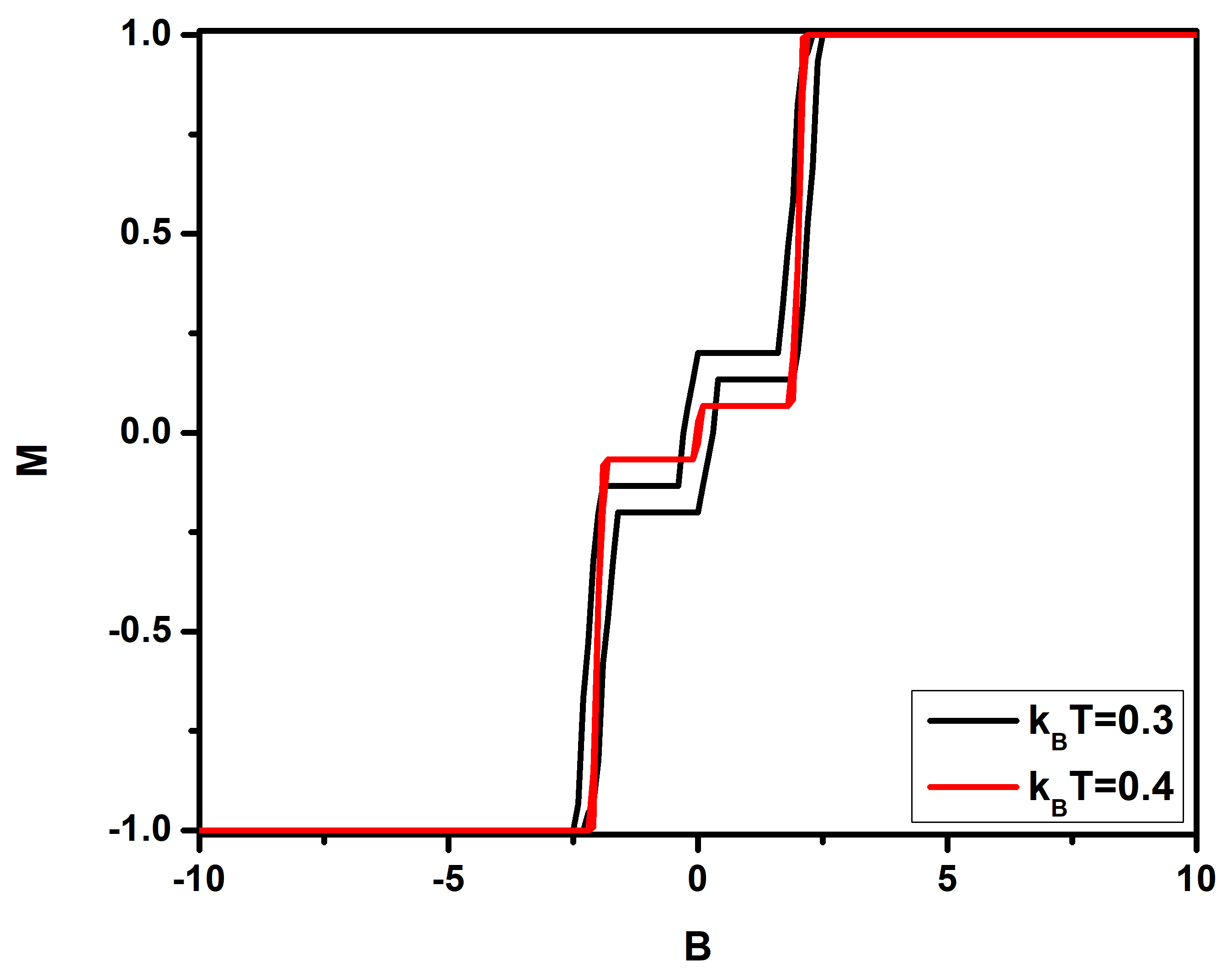}
		\caption{}
		\label{f16e}
	\end{subfigure}
	~
	\begin{subfigure}[b]{0.3\textwidth}
		\includegraphics[width=\textwidth]{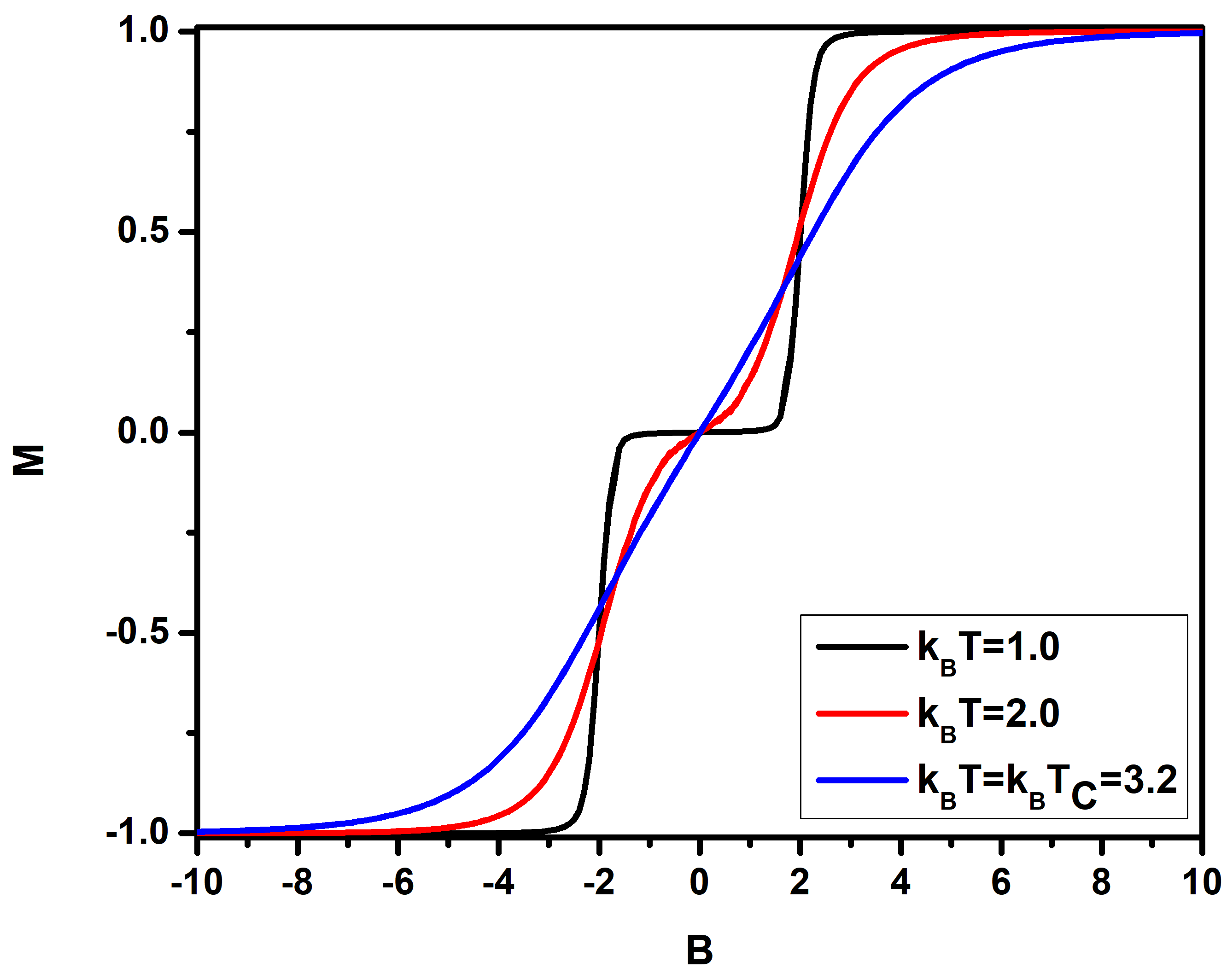}
		\caption{}
		\label{f16f}
	\end{subfigure}	
	~		
	\caption{Hysteresis loop for A-AFM order of interaction at different temperatures.}
	\label{f16}
\end{figure}

Fig. \ref{f16} shows the hysteresis behavior of the system with A-type AFM ordering for various temperatures. The system has a double loop hysteresis at lower temperatures ($T = 0.01$) (Fig. \ref{f16a}). The minimum remanent magnetization and coercive force are observed for the temperatures $T=0.05$ (Figs. \ref{f16b}), $T=0.1$ (Figs. \ref{f16c}) and $T=0.2$ (Figs. \ref{f16d}). The associated coercive force is bounded by $B = \pm 4$. While increasing the temperature, the width of the coercive field is reduced as shown in Figs. \ref{f16b}, \ref{f16c} and \ref{f16d}. Nearly at the temperature, $T=0.4$, the coercive force vanishes and the double loop turns into a single line and is shown in Fig. \ref{f16e}. There is no magnetization and no coercive force above $T=2.0$ and the system enters a paramagnetic phase at $T_C = 3.2$ (Fig. \ref{f16f}).

\begin{figure}[H]
	\centering
	\begin{subfigure}[b]{0.3\textwidth}
		\includegraphics[width=\textwidth]{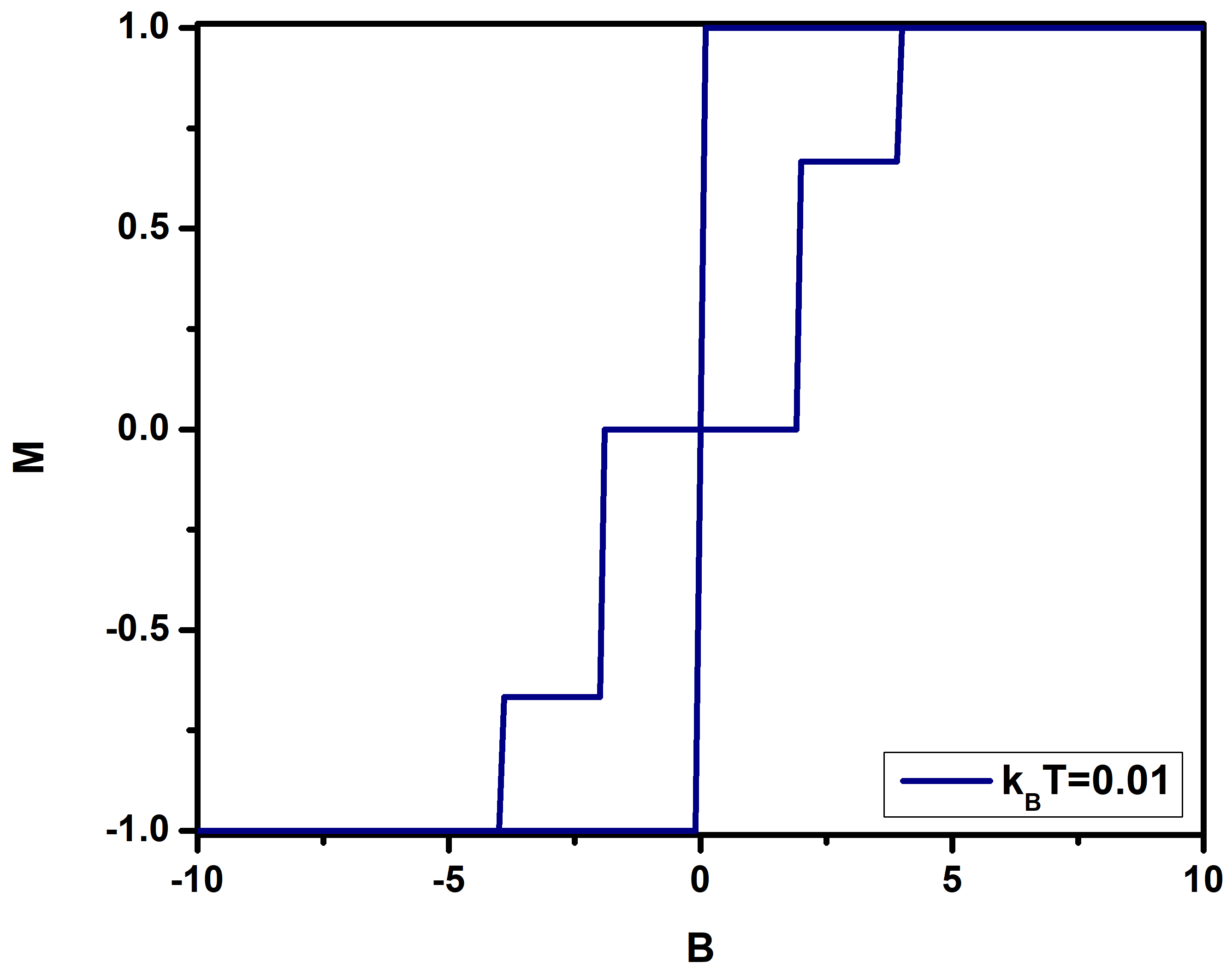}
		\caption{}
		\label{f17a}
	\end{subfigure}
	~
	\begin{subfigure}[b]{0.3\textwidth}
		\includegraphics[width=\textwidth]{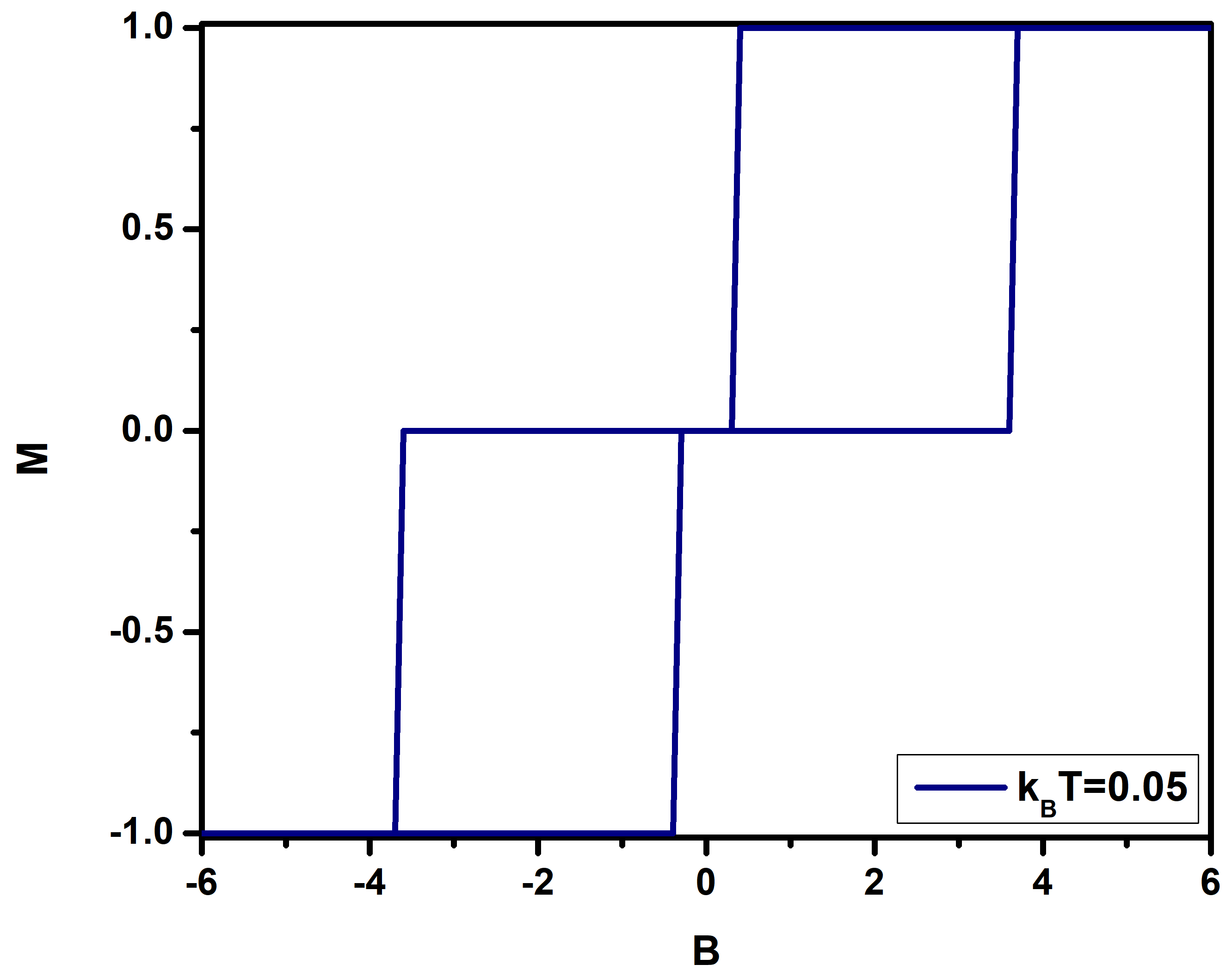} 
		\caption{}
		\label{f17b}
	\end{subfigure}
	~
	\begin{subfigure}[b]{0.3\textwidth}
		\includegraphics[width=\textwidth]{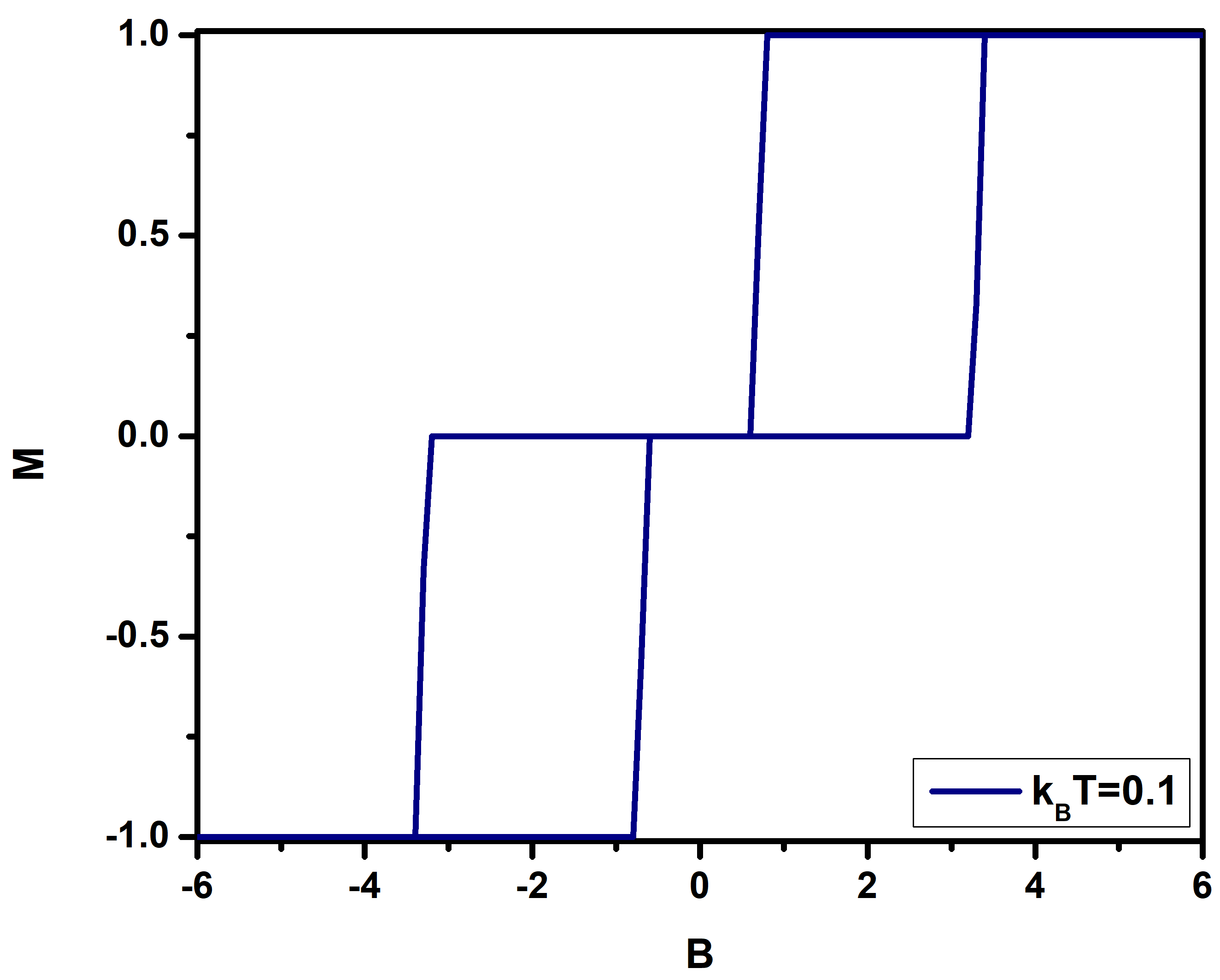}
		\caption{}
		\label{f17c}
	\end{subfigure}
	~
	\begin{subfigure}[b]{0.3\textwidth}
		\includegraphics[width=\textwidth]{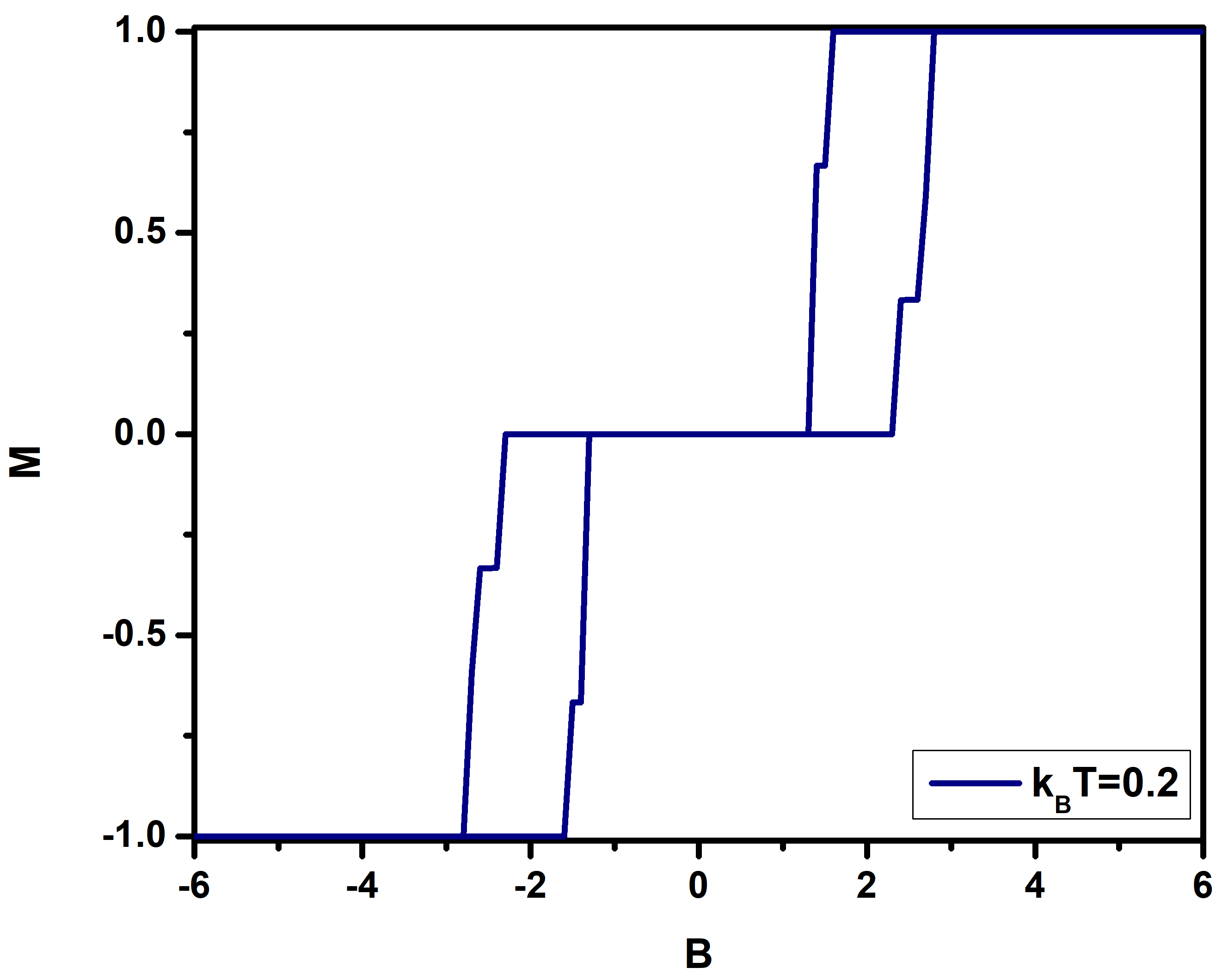}
		\caption{}
		\label{f17d}
	\end{subfigure}
	~
	\begin{subfigure}[b]{0.3\textwidth}
		\includegraphics[width=\textwidth]{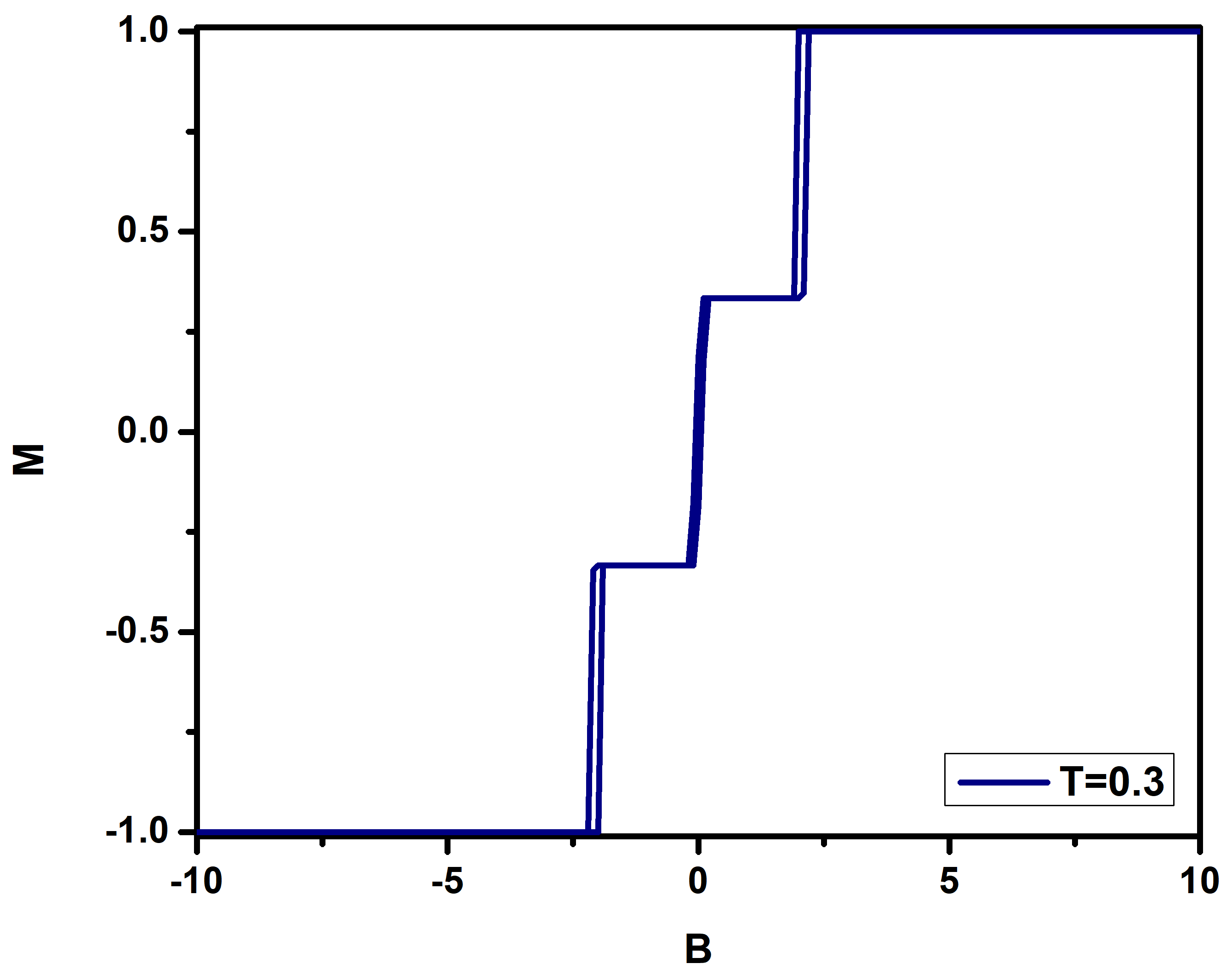}
		\caption{}
		\label{f17e}
	\end{subfigure}
	~
	\begin{subfigure}[b]{0.3\textwidth}
		\includegraphics[width=\textwidth]{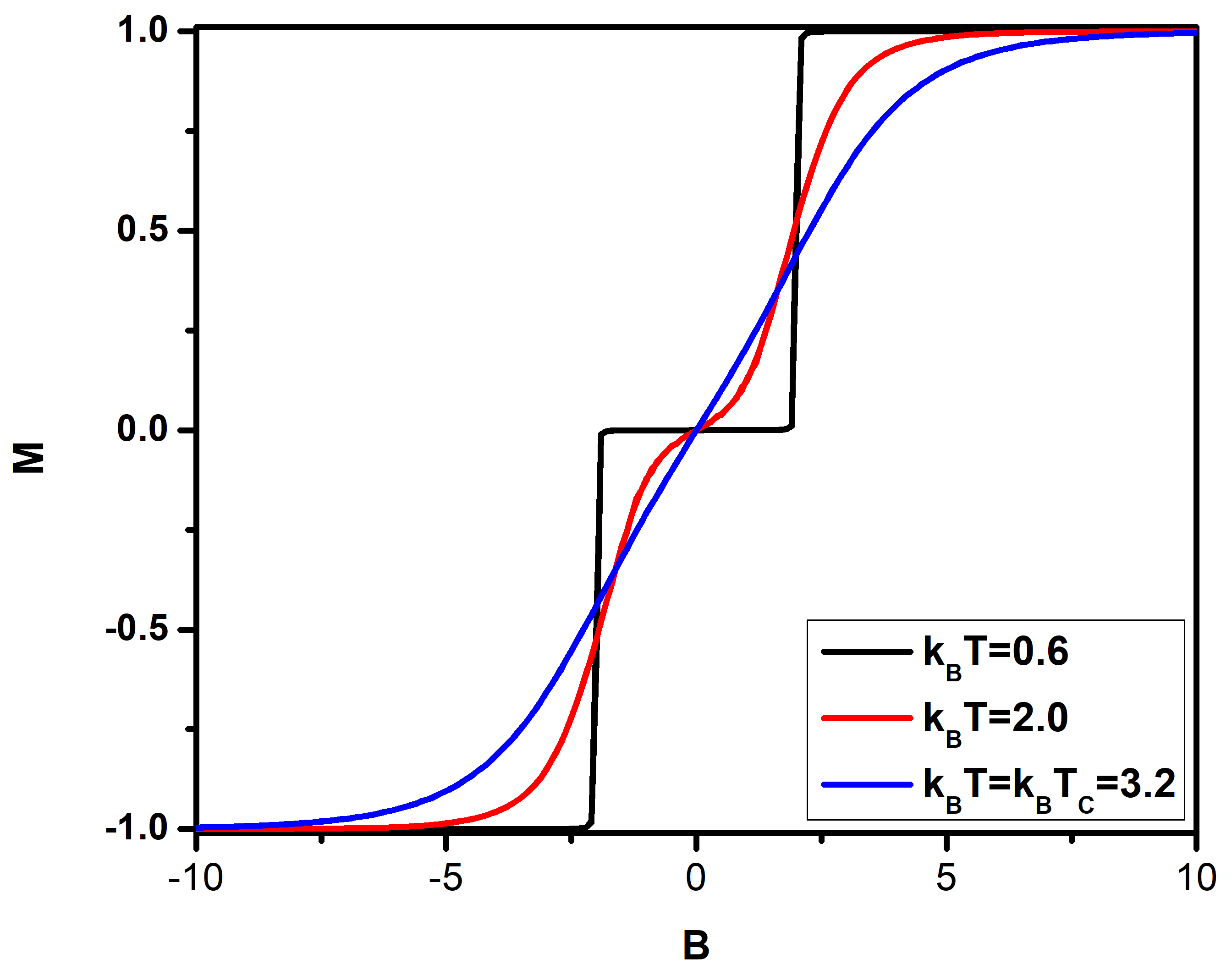}
		\caption{}
		\label{f17f}
	\end{subfigure}	
	~		
	\caption{Hysteresis loop for C-AFM order of interaction at different temperatures.}
	\label{f17}
\end{figure}

Figure \ref{f17} depicts the hysteresis curve for the C-type AFM system ($J_1=-1, J_2=+1$). A step behaviour is observed at lower temperature ( $T= 0.01$) and is shown in Fig. \ref{f17a}. The width of coercive force in the double loop hysteresis decreases as the temperature increases (please see Fig. \ref{f17b}, \ref{f17c} and \ref{f17d}).  At $T=0.3$, the loop becomes a single line and two step hysteresis occurs (Fig. \ref{f17e}). There is a minimum remanent magnetization found at that temperature ($T=0.3$). The Fig. \ref{f17f} shows that the system transits from C-type AFM phase to paramagnetic phase at the $T_C=3.2$.

In the present  work, the canonical ensemble from the Metropolis algorithm are simulated and the mechanical and thermal properties are determined.  However, this method is not suitable for calculating the system's entropy, free energy and the order of the transition \cite{Suman2019Non-}. Hence, to discuss the additional thermal properties and the order of phase transition of the Ising nano tube, WL algorithm is utilized and the following section discusses about them in detail. 

\subsection{Wang-Landau results}
\begin{figure}[H]
	\centering
	\begin{subfigure}[b]{0.35\textwidth}
		\includegraphics[width=\textwidth]{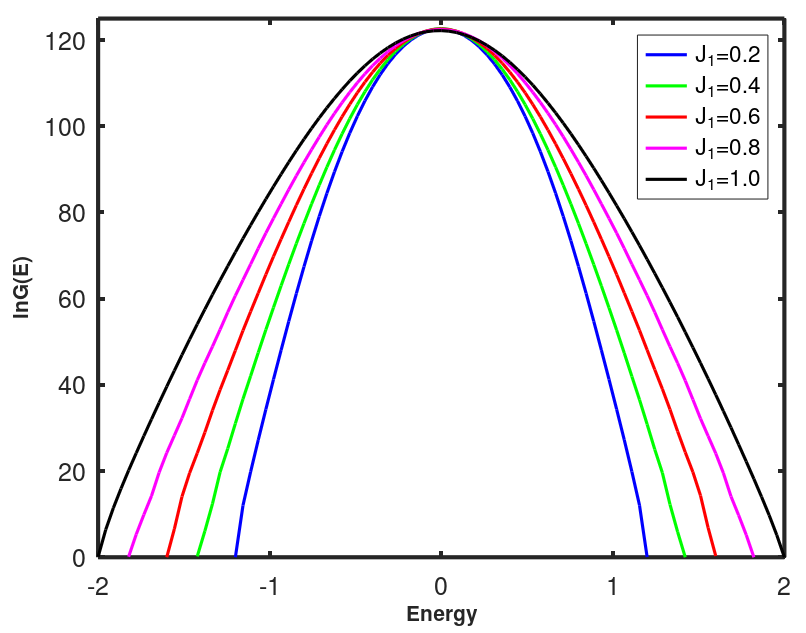}
		\caption{}
		\label{f18a}
	\end{subfigure}
	~
	\begin{subfigure}[b]{0.35\textwidth}
		\includegraphics[width=\textwidth]{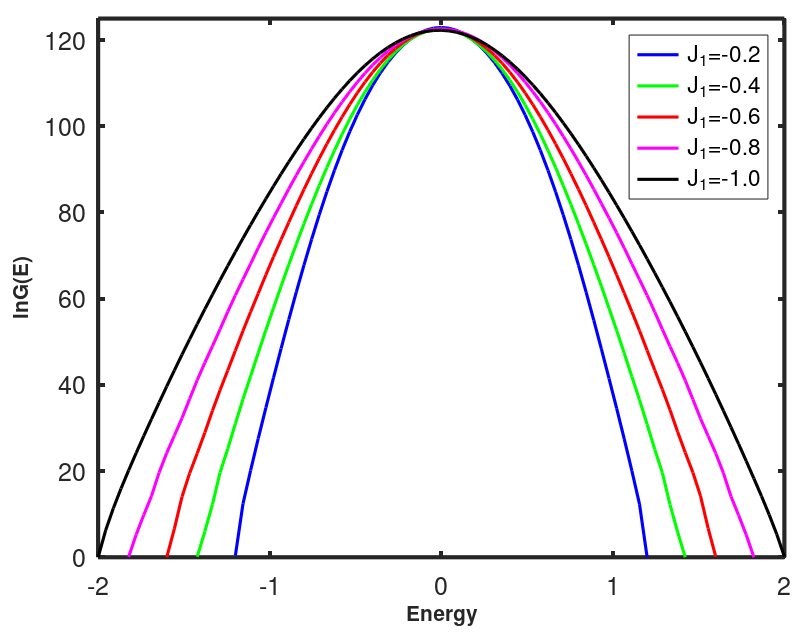} 
		\caption{}
		\label{f18b}
	\end{subfigure}
	~
	\begin{subfigure}[b]{0.35\textwidth}
		\includegraphics[width=\textwidth]{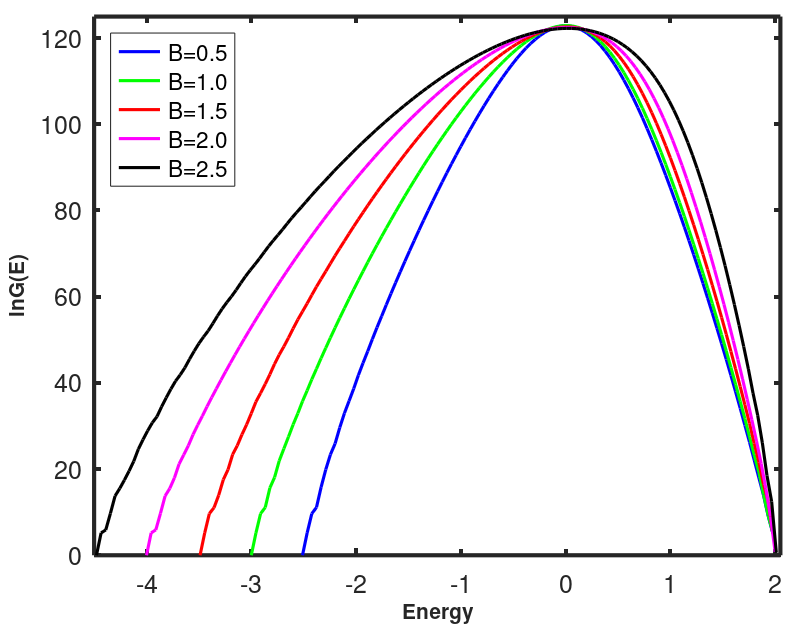}
		\caption{}
		\label{f18c}
	\end{subfigure}
	~
	\begin{subfigure}[b]{0.35\textwidth}
		\includegraphics[width=\textwidth]{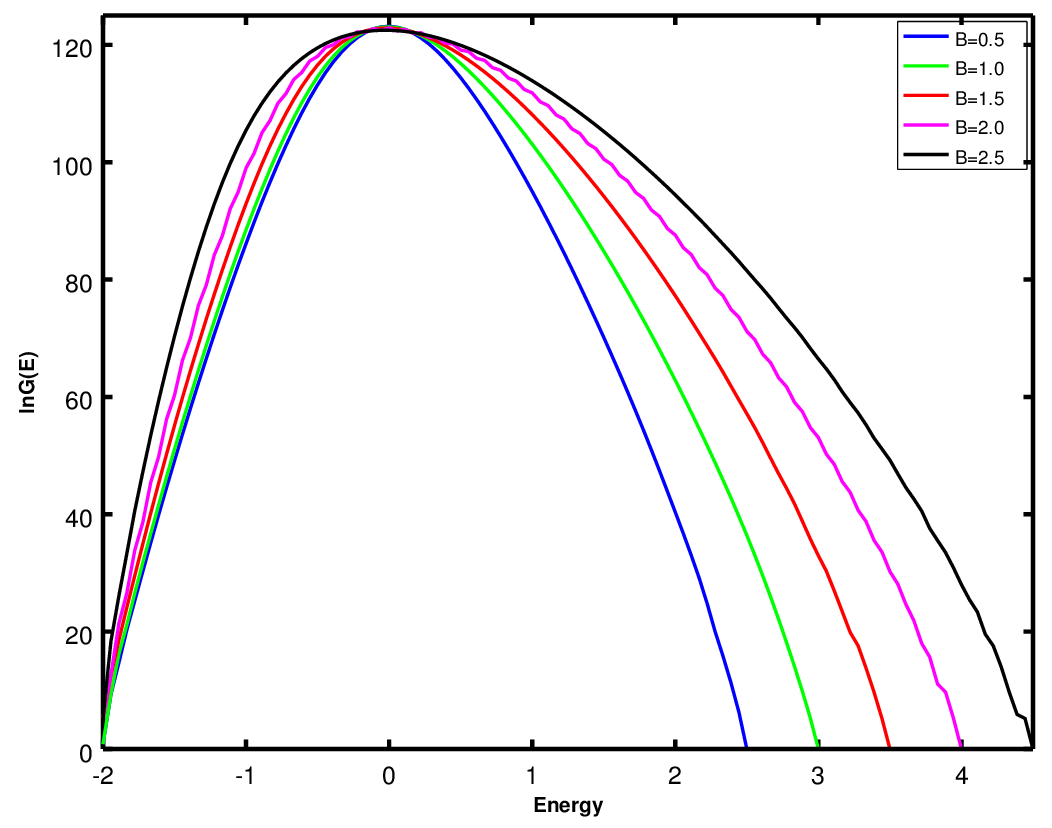}
		\caption{}
		\label{f18d}
	\end{subfigure}
	~		
	\caption{Logarithm of density of states versus energy. (a) $J_1>0$ with $B=0$, (b) $J_1<0$ with $B=0$, (c) $B>0$ for FM interaction, and (d) $B>0$ for G-AFM interaction.}
	\label{f18}
\end{figure}

In this section, the Wang-Landau technique is applied to investigate the system
with FM and G-type AFM order of interaction.  The logarithm of the density of states is calculated for various control parameters and are shown in Fig.\ref{f18}. In the absence of  magnetic field, the span of energy increases for FM interactions (Fig.\ref{f18a}) and decreases for G-AFM interactions (Fig.\ref{f18b}) with increasing interaction strength ${J_1}$. The shape of DOS is symmetric about the zero-energy for FM and AFM interactions without any external field and the shape becomes asymmetric while applying an external magnetic field. 
The logarithm of DOS for FM and G-AFM interactions in the presence of external magnetic field are shown respectively in Fig.\ref{f18c} and Fig.\ref{f18d}. Irrespective of the change in the applied magnetic field,  the highest possible value of DOS remains same in both the cases. However, the DOS  plot for FM interactions show an increase in the range of negative energy values with an increase in the applied magnetic field, whereas the G-AFM interactions show an increase in the range of  positive energy values with an increasing  magnetic field. The maximum value of the DOS is same ($122.2959$) for the energy $E=0$ for all the plots in Fig.\ref{f18}. 

\begin{figure}[H]
	\centering
	\begin{subfigure}[b]{0.4\textwidth}
		\includegraphics[width=\textwidth]{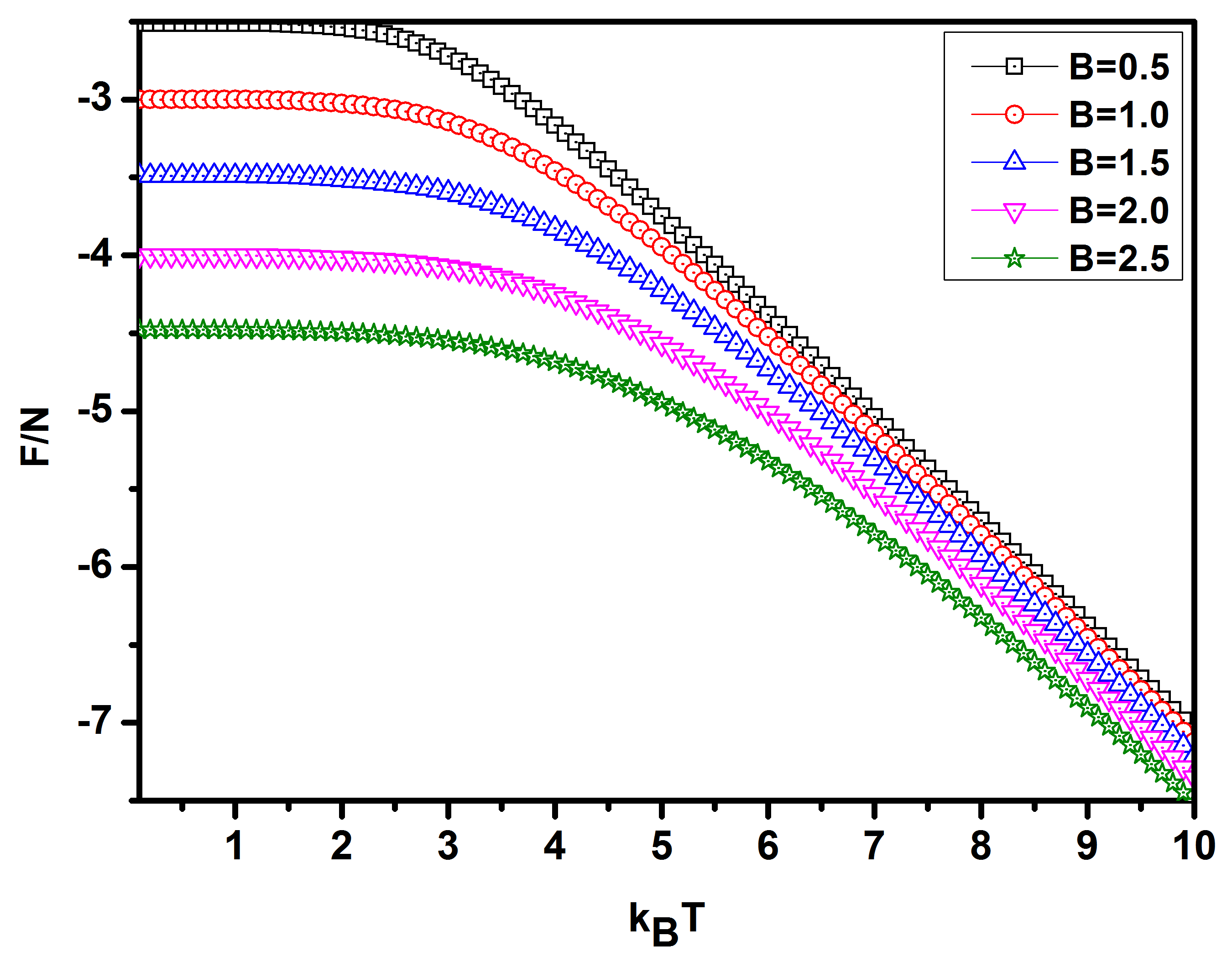}
		\caption{$ $}
		\label{f19a}
	\end{subfigure}
	~
	\begin{subfigure}[b]{0.4\textwidth}
		\includegraphics[width=\textwidth]{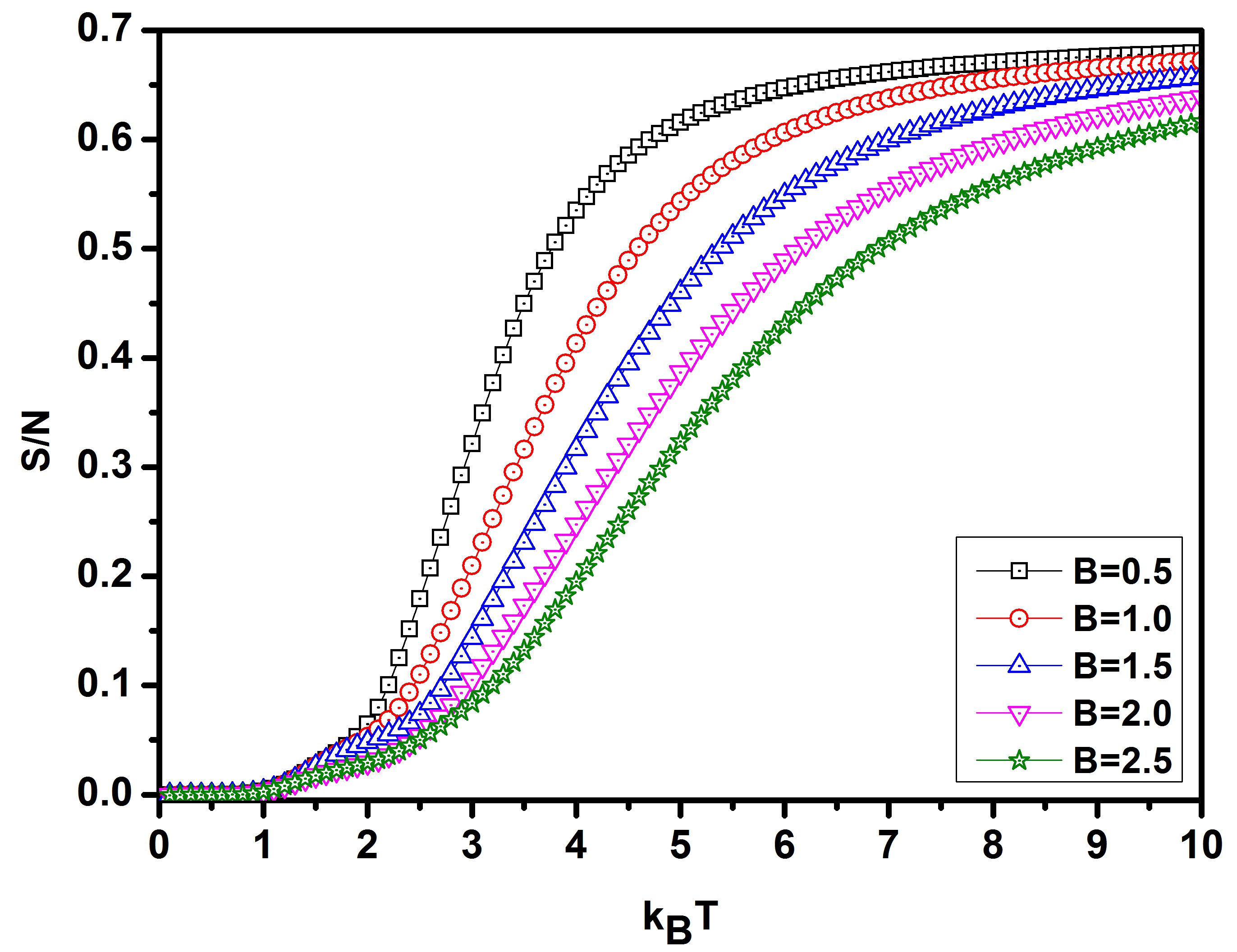}
		\caption{$ $}
		\label{f19b}
	\end{subfigure}
~
	\begin{subfigure}[b]{0.4\textwidth}
		\includegraphics[width=\textwidth]{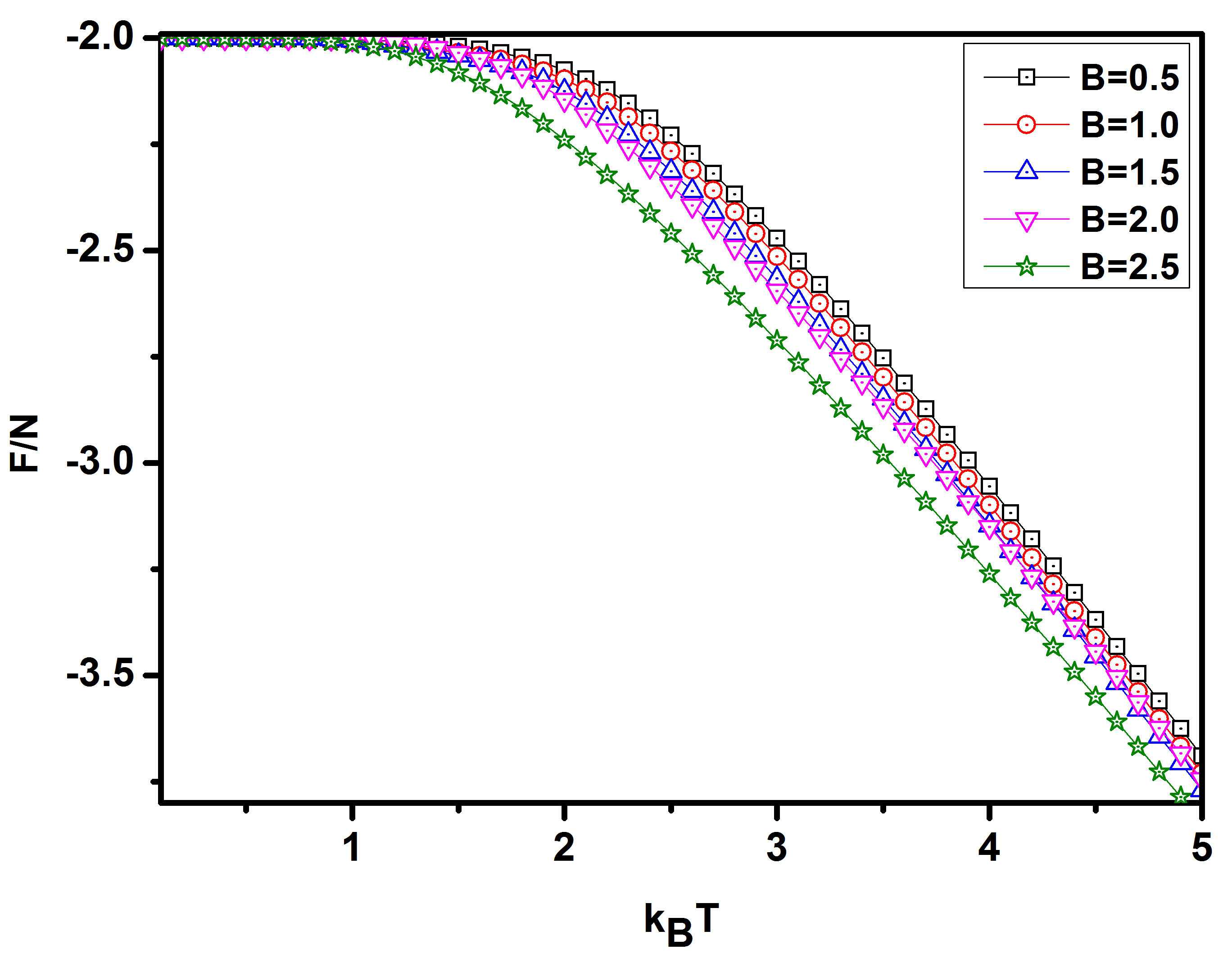}
		\caption{$ $}
		\label{f19c}
	\end{subfigure}
	~
	\begin{subfigure}[b]{0.4\textwidth}
		\includegraphics[width=\textwidth]{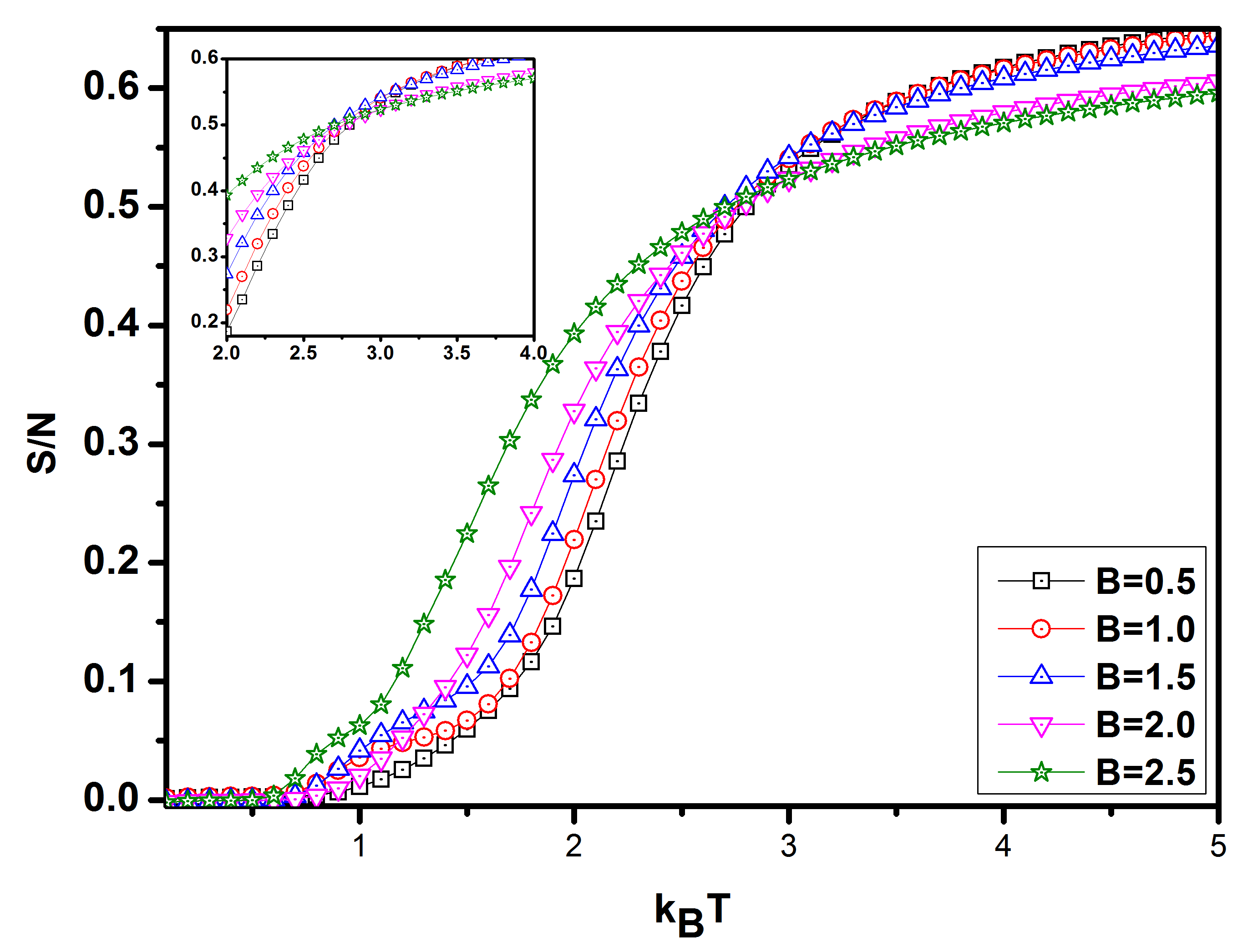}
		\caption{$ $}
		\label{f19d}
	\end{subfigure}
	\caption{Thermodynamic quantities for the Ising nanotube were calculated from DOS with varying magnetic fields. (a) free energy and (b) entropy for FM interaction; (c) free energy, and (d) entropy for G-AFM interaction.}
	\label{f19}
\end{figure}

The DOS of FM and G-AFM are similar to each other in the absence of magnetic field. The DOS shows interesting characteristics in the presence of the magnetic field. Thus, the free energy and entropy for ferromagnetic and anti-ferromagnetic interactions in the presence of the magnetic field are analyzed in detail and are depicted in Fig.\ref{f19}. From the Fig.\ref{f19a}, it is observed that the free energy of the ground state for FM interaction decreases with the increase in the external magnetic field which reflects the characteristics of the DOS as given in Fig.\ref{f18c}. The entropy plot for FM system (Fig.\ref{f19b}) starts from zero entropy for all the external field strengths and it shows the convergence of the system to a single microstate for the lowest energy (ground state). As the field strength increases, the plots follows a decreased trend in entropy values with increasing temperature.

Fig. \ref{f19c} shows the ground state free energy plot for anti-ferromagnetic interation under different magnetic fields and is found to approached at a particular value ($-2.0$). 
The free energy plot for G-AFM interaction shows the decreasing behavior as in the FM interaction with the increasing temperature. The Fig.\ref{f19d} shows a crossing point for all the plots with different magnetic fields at $k_BT=2.8$. This crossing point (inset of Fig.\ref{f19d}) suggests the isentropic behavior of the system with the variation in the magnetic field at a particular temperature, $k_BT=2.8$. The entropy of the system increases for the increase in the magnetic field upto the crossing point, beyond that the entropy decreases. The trend of decreasing free energy and increasing entropy for all the plots in Fig.\ref{f19} shows the phase change from FM (or G-AFM) to paramagnetic order. The phase transition is confirmed to be second-order as the entropy plots are continuous.

\begin{figure}[H]
	\centering
	\includegraphics[scale=0.4]{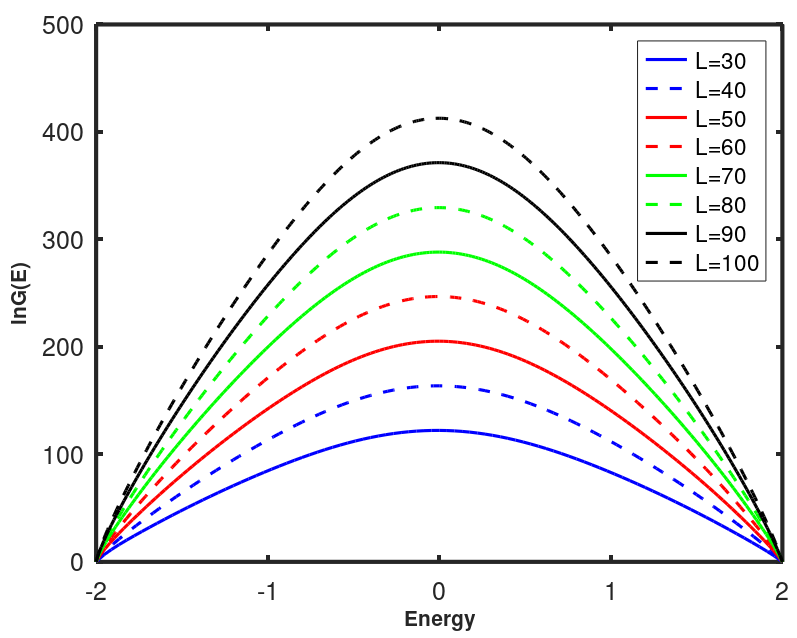}
	\caption{DOS for different number of layer $L$ ($J_1=+1.0, J_2=+1.0$) at $B=0.0$.}
	\label{f20}
\end{figure}
\begin{figure}[H]
	\centering
	\begin{subfigure}[b]{0.4\textwidth}
		\includegraphics[width=\textwidth]{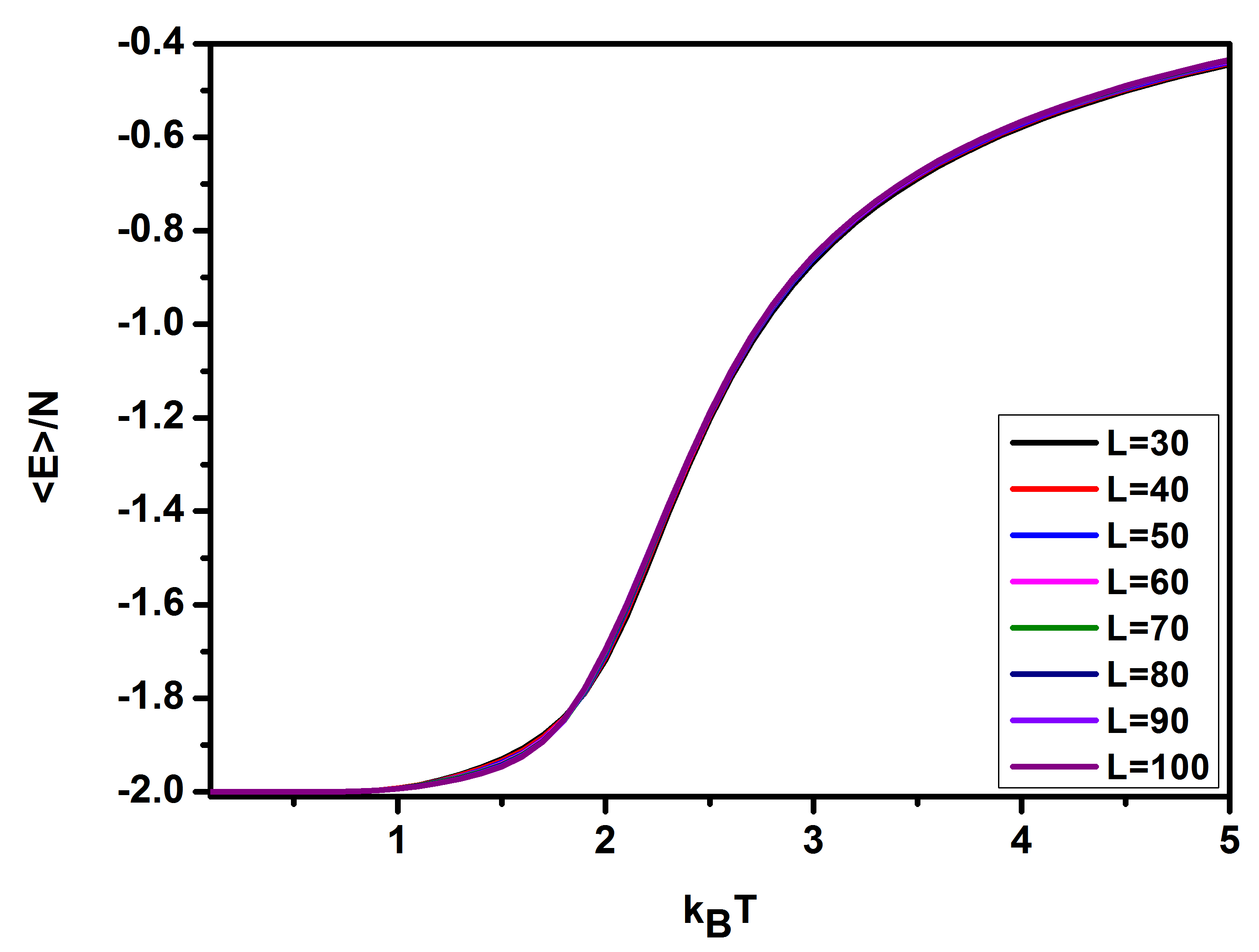}
		\caption{}
		\label{f21a}
	\end{subfigure}
	~
	\begin{subfigure}[b]{0.4\textwidth}
		\includegraphics[width=\textwidth]{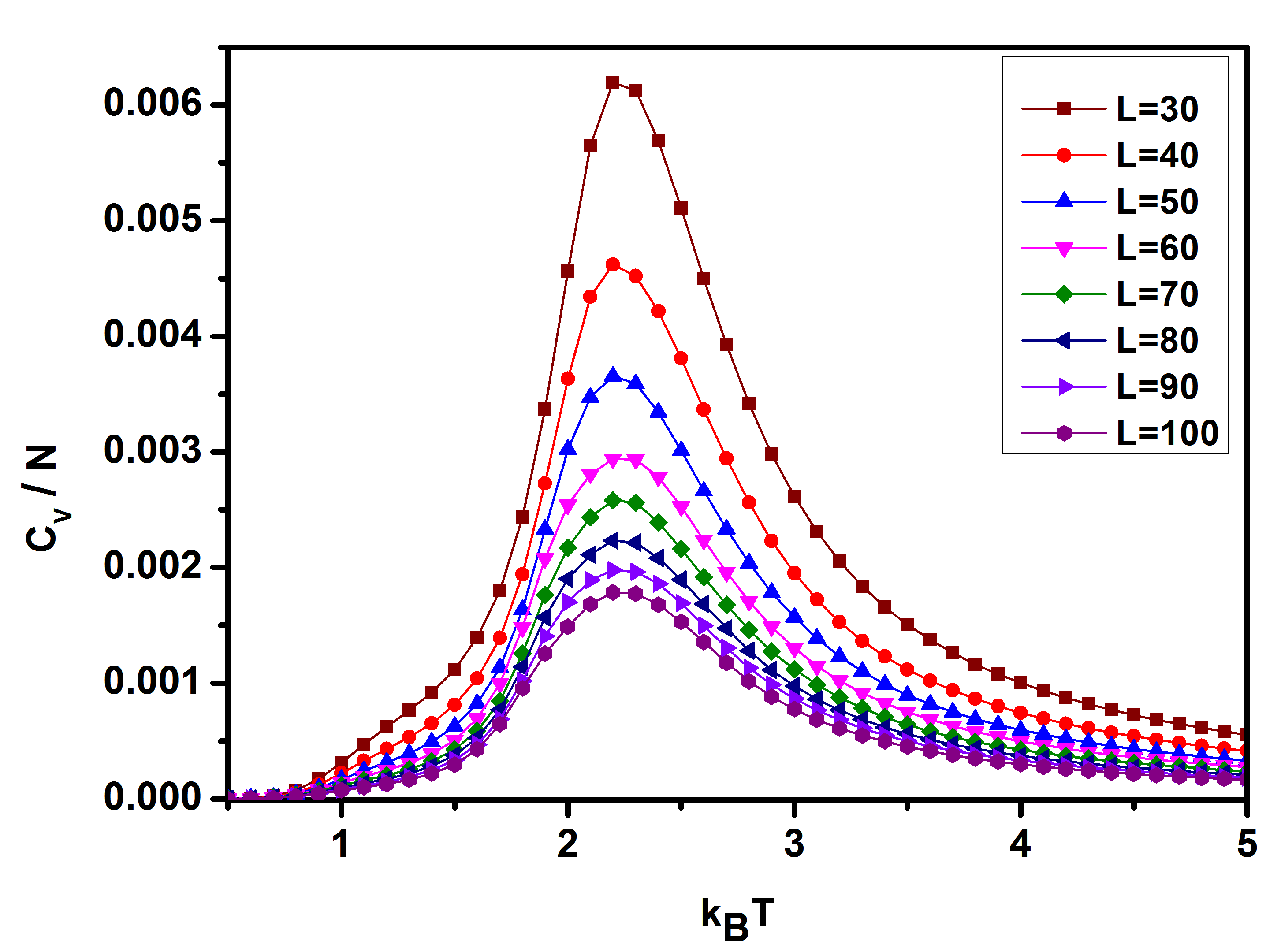}
		\caption{}
		\label{f21b}
	\end{subfigure}
	\caption{Thermodynamic observables  (a) average energy and (b) specific heat capacity for different layer $L$ without applying magnetic field.}
	\label{f21}
\end{figure}

The logarithm of density of states for ferromagnetic interaction in the absence of magnetic field is shown in Fig. \ref{f20} for different number of layers $L$. The DOS is computed by increasing the number of layers $N$ from $30$ to $100$. The energy range is maintained between $-2$ and $+2$. The maximum value of DOS increases in proportion to the number of layers. The average energy is plotted against the temperature (Fig. \ref{f21a}) and the system's stability can be assessed from its profile. The average energy begins in a stable ground state. Then, it rapidly rises near the transition temperature and eventually reaches its maximum value with increasing temperature. This confirms the transition from ferromagnetic to paramagnetic phase. The system under the present study clearly follows a second-order transition since the average energy curve behaves continuously without any jumping. The specific heat capacity as a function of temperature for various number of layers is shown in Fig. \ref{f21b}. The specific heat peak value reduces with increasing the number of layers. For different number of layers, the $T_C$  remained to be the same and the approximate value of $k_B T_C(L)$ is around $2.2$.

\begin{figure}[H]
	\centering
	\begin{subfigure}[b]{0.4\textwidth}
		\includegraphics[width=\textwidth]{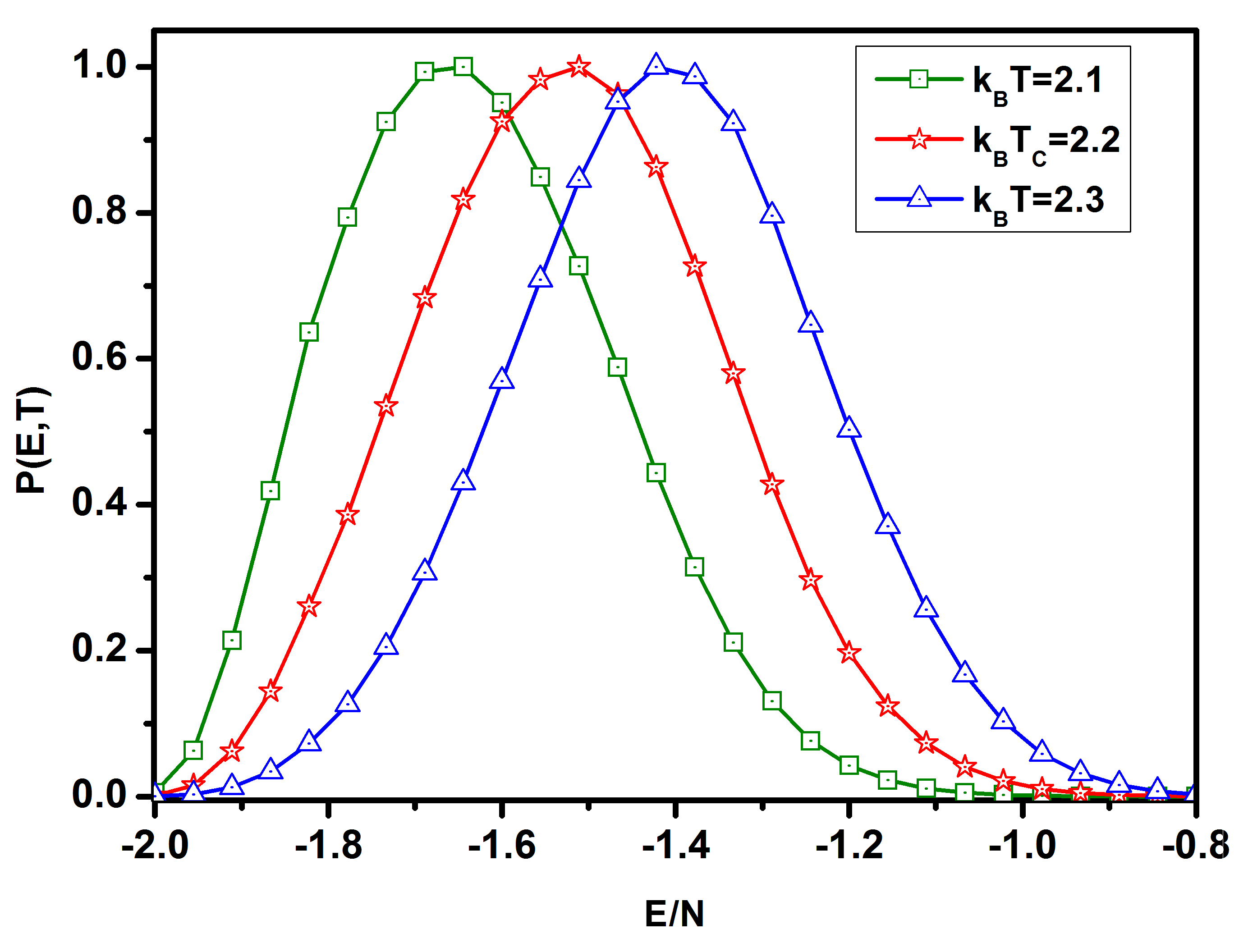}
		\caption{}
		\label{f22a}
	\end{subfigure}
	~
	\begin{subfigure}[b]{0.4\textwidth}
		\includegraphics[width=\textwidth]{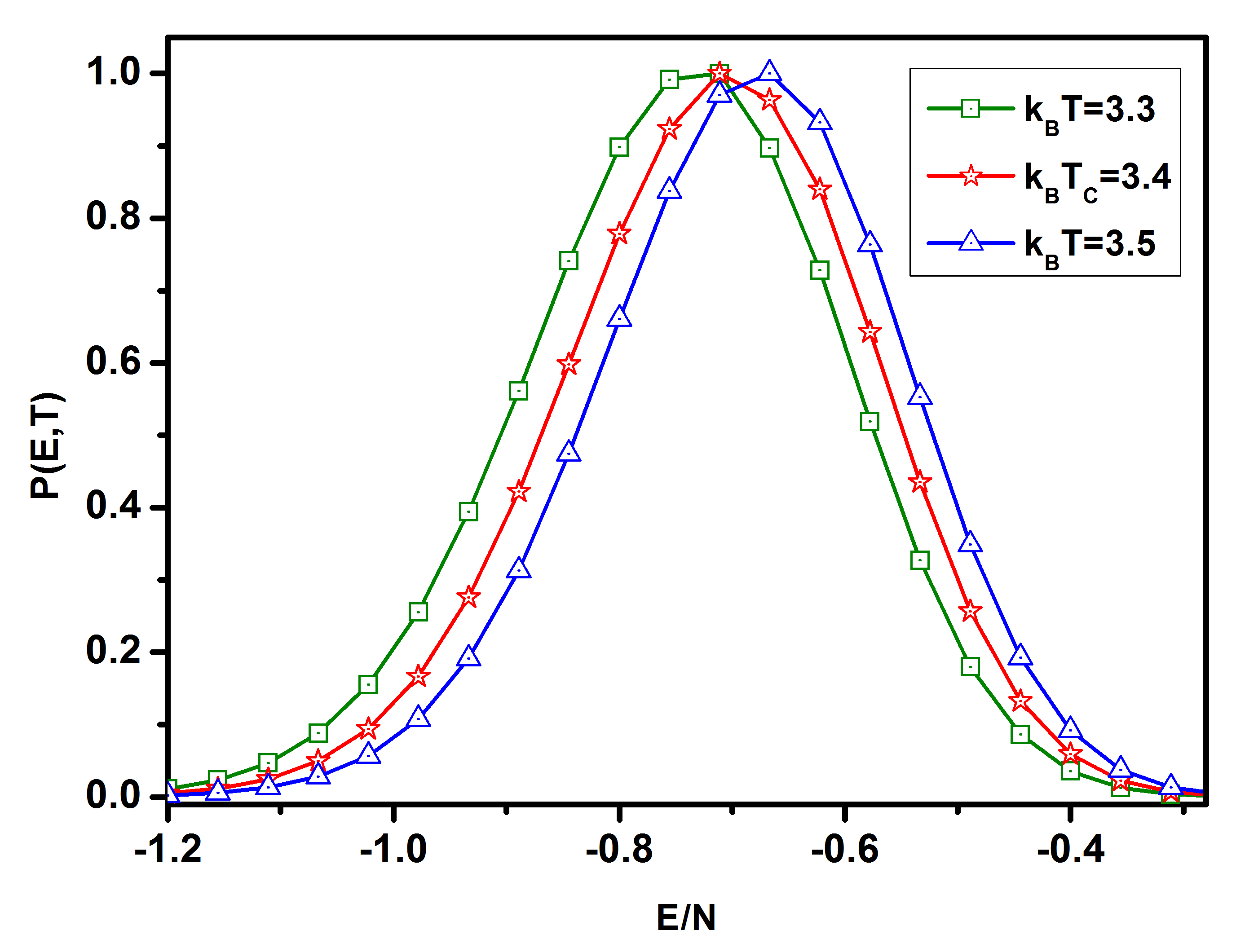}
		\caption{}
		\label{f22b}
	\end{subfigure}
	\caption{Canonical distribution for (a) FM ($J_1=+1,J_2=+1$) and (b) G-AFM ($J_1=-1,J_2=-1$) interaction.}
	\label{f22}
\end{figure}

In order to further confirm the second-order transition, the canonical distribution for FM and G-AFM interaction at the transition temperature is analyzed and is depicted in Fig. \ref{f22}. All the micro-states are chosen based on their weight ($g(E)\exp[-\beta E]$) and  the system shows a single peak for the canonical distribution. Temperatures below and above the vicinity of the transition temperature also show single peak in the distribution. The single peaked form of the probability distribution at the transition temperature is a common indication for the existence of a second-order transition \cite{Landau2004A_New}. It is also confirmed that both the FM and G-AFM interactions show the second-order phase transition.

\section{Conclusion}
The Metropolis and WL algorithms are used to explore the magnetic response of the Ising nanotube.  The obtained magnetic spin orientations of the ground state confirmed the FM and AFM (G, A and C type) ordering about the $z$ axis.  The variations of $T_C$ for different interaction strengths are calculated in the absence of an external magnetic field. At $J_1=0.0$, the C-AFM order will turn to FM ordering and similarly, the A-AFM order changes to G-AFM ordering.  A small disturbance (hump) is found in the magnetization curve for the interaction $J_1=0.0,~ J_2=+1.0$ at lower temperatures and it can be given a special attention in future studies. It is found that a small anti-ferromagnetic interaction is enough to bring the spins in the anti-ferromagnetic ordering, even though a FM interaction is present in the system.

The magnetic response of FM and different AFM interactions is analyzed by including the magnetic field for unit interaction strength. It is found that the transition temperature increases for the FM system and decreases for the G-AFM system when the magnetic field is increased.  Both the alternative interactions, i.e., A and C type AFMs, evolve similarly. In which the transition temperature $T_C$ first decreases then increases with the increasing external magnetic field. 
When the magnetic field, $B$ is less than 2.0, the AFM order dominates the system and the FM order dominates when $B \geq 2.0$. At $B=2.0$, the A-AFM and C-AFM systems changes to FM ordering. 
In general, it is found that the Ising nanotube transits from ferromagnetic (or anti-ferromagnetic) to paramagnetic phase over the finite temperatures. The hysteresis plot shows that the coercive field and the remanent magnetization reduces to zero with an increase in temperature. As a result, the system exhibits a narrow hysteresis loop for higher temperatures. The G, A, and C type AFM interactions exhibit the double step loop hysteresis at lower temperatures and which disappears after the transition temperature.

The DOS is determined using the WL algorithm to obtain the thermal properties of Ising nanotube. The symmetric DOS is observed for FM and G-AFM interaction without an external magnetic field. By applying a magnetic field, the system exhibits the asymmetric DOS pattern. The free energy and entropy are calculated in the presence of a magnetic field from the converged DOS. This study also shows that in the absence of magnetic filed, the difference in the number of layers does not alter the transition temperature. Using Wang Landau algorithm, it is also found that there is no discontinuity in canonical entropy and a single peak in the canonical distribution confirms that the system follows the second-order transition.

\section*{Acknowledgement}
The authors would like to thank the anonymous referees for their insightful comments and valuable suggestions which helped to improve the manuscript.  One of the author A. Arul Anne Elden would like to thank M. Suman Kalyan, S. Siva Nasarayya Chari and Asweel Ahmed A. Jaleel for fruitful discussions, and also thank T. Kiran for useful discussions and critical reading of the manuscript.

\end{document}